\documentclass[aps,reprint,showpacs,superscriptaddress]{revtex4-1}

\usepackage{graphicx}
\usepackage[version=3]{mhchem}
\usepackage{dcolumn}
\usepackage{bm}
\usepackage[usenames, dvipsnames]{xcolor}
\usepackage{threeparttable}
\usepackage{longtable}
\usepackage{afterpage}
\usepackage{multirow}
\AtBeginDocument{\usepackage{booktabs}}
\usepackage{url}
\usepackage[section]{placeins}
\usepackage[colorlinks=true,linkcolor=blue,citecolor=magenta,urlcolor=red]{hyperref}

\definecolor{greenGM}{rgb}{0.4353, 1, 0}
\definecolor{greenmax1}{rgb}{0.7843,0,0.7843}
\definecolor{greenmax2}{rgb}{0,0.6627,0.4118}
\definecolor{greenenv}{rgb}{0,0.8,0}
\definecolor{lightgray}{rgb}{0.65,0.65,0.65}
\definecolor{lightred}{rgb}{1,0.65,0.65}
\definecolor{red2000}{rgb}{1,0,0.2}
\definecolor{red200}{rgb}{1,0,0.4}
\definecolor{red20}{rgb}{1,0,0.6}
\definecolor{lightblue}{rgb}{0., 0.46, 0.8}
\definecolor{grayL}{rgb}{0.3,0.3,0.3}
\definecolor{grayLmk}{rgb}{0.6,0.6,0.6}
\definecolor{redL}{rgb}{1,0,0.3}
\definecolor{redLmk}{rgb}{1,0,0.6}
\definecolor{boxa}{rgb}{1,0.5,0}
\definecolor{boxs}{rgb}{0.5,0.5,0.5}
\definecolor{boxm}{rgb}{0,0.75,0.2}

\newcommand{\blue}[1]{\textcolor{lightblue}{#1}}

\newcommand{\orange}[1]{\textcolor{OrangeRed}{#1}}

\usepackage{varioref}

\begin{document}

\preprint{APS/123-QED}

\title{Computational Investigation of Inverse-Heusler compounds for Spintronics Applications}

\author{Jianhua Ma}
\email{jm9yq@virginia.edu}
\affiliation{Department of Electrical and Computer Engineering, University of Virginia, Charlottesville,VA-22904, USA}%

\author{Jiangang He}

\affiliation{Department of Materials Science and Engineering, Northwestern University, Evanston, IL 60208, USA}

\author{Dipanjan Mazumdar}
\affiliation{Department of Physics, Southern Illinois University, Carbondale, Illinois 62901, USA}

\author{Kamaram Munira}
\affiliation{Center for Materials for Information Technology, University of Alabama, Tuscaloosa, Alabama 35401, USA}%

\author{Sahar Keshavarz}
\affiliation{Center for Materials for Information Technology, University of Alabama, Tuscaloosa, Alabama 35401, USA}%
\affiliation{Department of Physics and Astronomy, University of Alabama, Tuscaloosa, Alabama 35401, USA}%

\author{Tim Lovorn}
\affiliation{Center for Materials for Information Technology, University of Alabama, Tuscaloosa, Alabama 35401, USA}%
\affiliation{Department of Physics and Astronomy, University of Alabama,
Tuscaloosa, Alabama 35401, USA}%

\author{C. Wolverton}
\affiliation{Department of Materials Science and Engineering, Northwestern University, Evanston, IL 60208, USA}

\author{Avik W. Ghosh}
\affiliation{Department of Electrical and Computer Engineering, University of Virginia, Charlottesville,VA-22904, USA}%

\author{William H. Butler}
\email{wbutler@mint.ua.edu}
\affiliation{Center for Materials for Information Technology, University of Alabama, Tuscaloosa, Alabama 35401, USA}%
\affiliation{Department of Physics and Astronomy, University of Alabama, Tuscaloosa, Alabama 35401, USA}%

\date{\today}

\begin{abstract}
First-principles  calculations of the electronic structure, magnetism and structural stability of  inverse-Heusler compounds with the chemical formula \textit{X$_2$YZ} are presented and discussed with a goal of identifying compounds of interest for spintronics.     Compounds for which the number of electrons per atom for \textit{Y} exceed that for \textit{X} and for which  \textit{X} is one of Sc, Ti, V, Cr, Mn, Fe, Co, Ni, or Cu;  \textit{Y} is one of Ti, V, Cr, Mn, Fe, Co, Ni, Cu, or Zn; and \textit{Z} is one of  Al, Ga, In, Si, Ge, Sn, P, As or Sb were considered.   The formation energy per atom of each compound was calculated.    By comparing our calculated formation energies to those calculated for phases in the  Inorganic Crystal Structure Database (ICSD) of observed phases, we estimate that inverse-Heuslers with formation energies within 0.052 eV/atom of the calculated convex hull are reasonably likely to be synthesizable in equilibrium.   The observed  trends in the formation energy and relative structural stability  as the \textit{X}, \textit{Y} and \textit{Z} elements vary are described.     In addition to the Slater-Pauling gap after 12 states per formula unit in one of the spin channels, inverse-Heusler phases often have gaps after 9 states or 14 states.  We describe the origin and occurrence of these gaps.  We identify 14 inverse-Heusler semiconductors, 51 half-metals and 50 near half-metals with negative formation energy.  In addition, our calculations predict 4 half-metals and 6 near half-metals to lie close to the respective convex hull of stable phases, and thus may be experimentally realized under suitable synthesis conditions, resulting in potential candidates for future spintronics applications.\end{abstract}

\maketitle


\section{\label{secI}INTRODUCTION}

During the past two decades, the field of magnetoelectronics and spintronics has experienced significant development\cite{RevModPhys.76.323,RevModPhys.76.323, Takanashi2014Future, Ando2015spintronics}. Half-metals have the potential to play an important role in the continuing development of spintronics because they have a gap in the Density of States (DOS)\ at the Fermi energy for one of the spin-channels, but  not for the other, creating an opportunity to achieve 100\% spin-polarized currents\cite{ANIE:ANIE200601815, RevModPhys.80.315}. A half-Heusler compound inspired the term `half-metal' when, in 1983, de Groot and collaborators calculated the electronic structure of NiMnSb and recognized its unusual electronic structure\cite{PhysRevLett.50.2024}. Since then, many Heusler compounds have been analyzed and have been theoretically predicted to be half-metals or near-half-metals \cite{PhysRevB.66.134428, PhysRevB.66.174429, kandpal2006covalent, galanakis2006electronic,0953-8984-16-18-010,PhysRevB.87.024420}.

Due to the 100\% spin polarization at the Fermi level and the relatively high Curie temperatures shown by some Heusler compounds\cite{0953-8984-18-43-003,PhysRevB.76.024414,Blum2009Co2FeSi}, half-metallic Heusler compounds have been exploited for devices \cite{felser2015basics}, such as electrodes for magnetic tunnel junctions (MTJs) \cite{Ando2009MTJs, Sugimoto2009MTJ, Yamamoto2012MTJ, Yamamoto2015MTJ}, electrodes for giant magnetoresistive spin valves (GMRs)\cite{Takanashi2009GMR,Hono2010GMR, Hono2011GMR, Ando2011GMR} and for injection of spins into semiconductors\cite{PhysRevLett.107.047202}.

In addition to half-metals, magnetic insulators and semiconductors are interesting materials for spintronic applications because of their potential highly spin-polarized tunneling properties\cite{WorledgeandGeballe,LeClairspinfilter}.  In this paper we will discuss our survey of inverse-Heusler compounds.  In addition to the half-metals and semiconductors  that have been predicted for full-Heuslers and half-Heuslers, magnetic semiconductors and a state described as a ``spin-gapless semiconductor(SGS)"\cite{PhysRevLett.110.100401}\ have been predicted for inverse-Heuslers.

The family of Heusler compounds has three subfamilies, the full-Heuslers, the half-Heuslers and the inverse-Heuslers. All three of these subfamilies are based on decorations of atoms and vacancies on an underlying bcc lattice.  The three subfamilies are most easily visualized by beginning with a bcc lattice represented by the familiar cubic cell shown in Figure 1a with all \textit{X}-type atoms.  If the center atom is made different from the atoms at the corners, the structure becomes the $B2$ structure with composition \textit{XY} (Figure 1b).  If, in addition, alternate atoms at the corners in Figure 1b are made different, as shown in the larger cell of Figure 1c, the structure becomes the full-Heusler or $L2_1$ structure with composition \textit{X$_2$YZ}.   The half-Heusler structure (composition \textit{XYZ}) can be visualized by removing half of the  \textit{X} atoms from Figure 1c as shown in Figure 1d.  The inverse Heusler or $XA$ structure (composition \textit{X$_2$YZ}) can be obtained from $L2_1$ by switching one of the \textit{X} atoms in the $L2_1$ structure  with a \textit{Y} or \textit{Z} as shown in Figure 1e. In all three families, the \textit{X} and \textit{Y} atoms are typically transition metals (at least for the systems that are potential half-metals), while the \textit{Z} atom is typically an \textit{s-p} metal.

We have recently completed a survey of the calculated properties of half-Heusler, full-Heusler and inverse Heusler compounds.  The results of these calculations are available in a database \cite{Heusl89:online}.  We have previously reported on our study of the half-Heusler compounds \cite{ma2016computational}.  In this paper we describe the results of calculations of the properties of inverse Heusler compounds with particular emphasis on those inverse Heuslers that may be of use for spintronics applications and which may have a reasonable probability of being synthesized.

One particular feature of the electronic structure of the invese-Heuslers is especially interesting.  The half-Heusler and full-Heusler families often have band gaps after 3 states per atom in the gapped channel.  This type of gap is called a Slater-Pauling gap because Slater and Pauling, but especially Slater, noticed that many bcc based alloys have approximately 3 states per atom in the minority spin channel\cite{PhysRev.94.1498,Slater1937,PhysRev.54.899}.  This ``Slater-Pauling" state has been ascribed to a minimum in the electronic DOS near the center of the $s-d$ band.  This minimum becomes an actual gap after 3 states per atom in one of the spin-channels in some Heusler compounds.    In contrast, although many inverse Heuslers show Slater-Pauling gaps at 3 states per atom, some have gaps after a number of states per atom that differs from three.  

Thus inverse Heuslers often have gaps after 9, 12, and 14 states per formula unit in the gapped channel.  The gap after 12 states corresponds to the Slater-Pauling gap. The gaps after  9 states and 14 states  are additional gaps allowed by the reduced symmetry of the inverse-Heusler compared to the full-Heusler.     Semiconducting states are possible with half-Heuslers and full-Heuslers, but these typically occur when both spin channels have Slater-Pauling gaps, and both spin-channels are identical implying no magnetic moments on any of the atoms.  In contrast, the multiple gaps of the inverse-Heuslers allows the possibility of a magnetic semiconductor, \textit{e.g.}  when the minority channel has a gap after 12 filled states and the majority has a gap after 14 filled states.

The reduced-symmetry-derived gaps after 9 and 14 states are often quite narrow so that a state called a spin gapless semiconductor (SGS), a  special type of  semiconductor in which the Fermi level falls into a relatively large gap in the minority spin-channel and within a narrow gap for the majority\cite{Galanakis2013SGS,Galanakis2014Mn2CoAl}.  
This  band structure allows both electrons and holes to be excited simultaneously carrying 100\% spin polarized current - potentially  with high mobility. Recently,  the existence of some inverse-Heuslers has been confirmed by experiment\cite{Lakshmi2002Mn2CoSn,PhysRevB.84.132405,PhysRevB.83.174448}. Additionally, it has been experimentally verified that Mn$_{2}$CoAl  is an  SGS with high Curie temperature\cite{PhysRevLett.110.100401}.

Interestingly, the half-Heusler $C1_b$ phase has the same space group as the inverse-Heusler $XA$ phase and is also prone to gaps after 9 and 14 states per formula unit in a spin-channel.  However, in this case 9 states per formula unit coincides with the Slater-Pauling gap after 3 states/atom.

We anticipate that a large and complete database of consistently calculated properties of inverse-Heuslers will allow the testing of hypotheses that may explain the occurrence and size of Slater-Pauling band gaps as well as the gaps that occur in some inverse Heusler compounds after 9 and 14 states per formula unit.

Although a large number of inverse-Heulser compounds have been analyzed by first-principles calculations\cite{Xin201710,PhysRevB.87.024420,PhysRevB.77.014424,Galanakis2013SGS,Ahmadian2014243,PhysRevB.91.174439}, a comprehensive study of the structural, electronic and magnetic properties of the inverse-Heusler family is useful, because it is not clear how many  of the inverse-Heusler half-metals  and semiconductors that can be imagined are thermodynamically stable in the $XA$  structure. Thus, a systematic study of the structural stability of the inverse-heusler family can provide guidance for future work.

In Sec.~\ref{sec:compDetail} we describe the details of our computational approach. Sec.~\ref{sec:DFTdetail} describes  the DFT calculations.  In  Sec.~\ref{DetermineStructure}, we discuss how the  equilibrium structures for each compound were determined and  the possibility of multiple solutions in energy and magnetic configuration for some specific compounds. Our approach to estimating the stability of the inverse Heusler compounds is described in Sec.~\ref{Eform}.

Our results are described in Sec.~\ref{Results}.  The relationship between stability and composition is discussed in Sections \ref{FE and HD} and \ref{C&S}.  In Sec.~\ref{SemiHM},  inverse Heusler semiconductors, half-metals and near half-metals with negative formation energy are   listed and discussed in terms of their electronic structure and structural stability. Sec.~\ref{Summary} is a summary of our results and conclusions.

\section{COMPUTATIONAL DETAILS}
\label{sec:compDetail}

\begin{figure*}[t]
        \includegraphics[width=6.5in]{{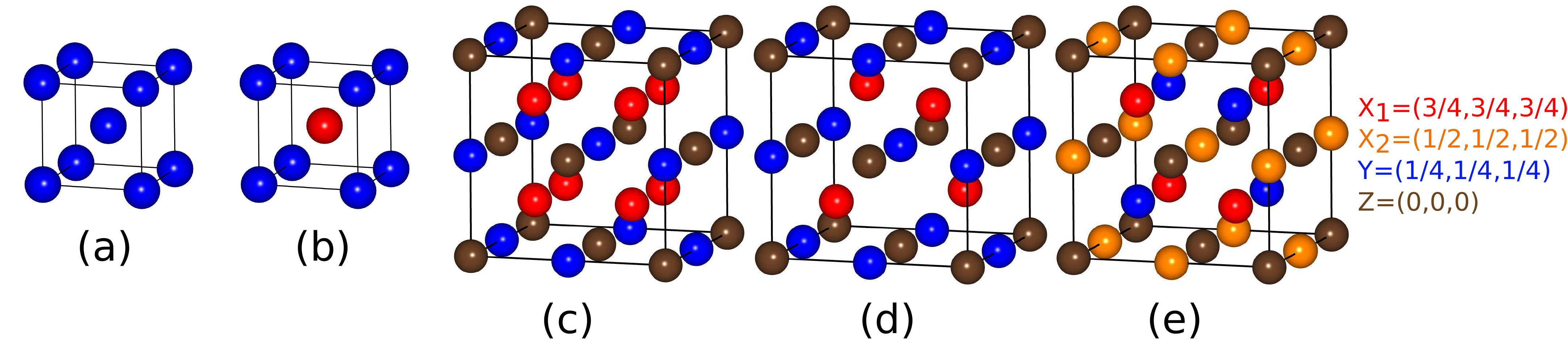}}
        \caption{Schematic representation of full-Heusler $L2_{1}$ structure and inverse-Heusler $XA$ structure. (a) The $bcc$ body-centered cubic structure. (b) The $B2$ cubic structure. (c) The $L2_{1}$ structure consisting of four interpenetrating $fcc$ sublattices with atomic sites $X_{1}$ $\left(\frac{3}{4},\, \frac{3}{4},\, \frac{3}{4}\right)$,  $X_{2}$ $\left(\frac{1}{4},\, \frac{1}{4},\, \frac{1}{4}\right)$, $Y$ $\left(\frac{1}{2},\, \frac{1}{2},\, \frac{1}{2}\right)$, and $Z$ $(0,\, 0,\, 0)$. (d) The  half-Heusler $C1_{b}$ structure consists of three interpenetrating $fcc$ sublattices with atomic sites $X$ $\left(\frac{1}{4},\, \frac{1}{4},\, \frac{1}{4}\right)$, $Y$ $\left(\frac{1}{2},\, \frac{1}{2},\, \frac{1}{2}\right)$, and $Z$ $(0,\, 0,\, 0)$. (e) The $XA$ structure consists of four interpenetrating $fcc$ sublattices with atomic sites $X_{1}$ $\left(\frac{3}{4},\, \frac{3}{4},\, \frac{3}{4}\right)$,  $X_{2}$ $\left(\frac{1}{2},\, \frac{1}{2},\, \frac{1}{2}\right)$, $Y$ $\left(\frac{1}{4},\, \frac{1}{4},\, \frac{1}{4}\right)$, and $Z$ $(0,\, 0,\, 0)$. In the $XA$ structure, $X_{1}$ and $X_{2}$ are the same transition metal element but they have different environments and magnetic moments.}
        \centering
        \label{fig:L21Xa}
\end{figure*}

The inverse-Heusler compound, \textit{X$_{2}$YZ}, crystallizes in the face-centered cubic \textit{XA} structure with four formula units per cubic unit cell (Fig.~\ref{fig:L21Xa}). Its space group is no. 216, $F\bar{4}3m$, (the same as the half-Heusler). The inverse-Heusler structure can be viewed as four interpenetrating fcc sublattices, occupied by the $X$, $Y$ and $Z$ elements, respectively. The $Z$ and $Y$ elements are located at the (0,\, 0,\, 0) and $\left(\frac{1}{4},\, \frac{1}{4},\, \frac{1}{4}\right)$ respectively in Wyckoff coordinates, while the $X_{1}$ and $X_{2}$ elements are at $\left(\frac{3}{4},\, \frac{3}{4},\, \frac{3}{4}\right)$ and $\left(\frac{1}{2},\, \frac{1}{2},\, \frac{1}{2}\right)$, respectively, resulting in two different rock-salt structures, $[X_1Y]$ and $[X_2Z]$ as shown in Fig. \ref{fig:L21Xa}e. We use $X_{1}$ and $X_{2}$ to distinguish these two $X$ atoms sitting at the two inequivalent sites in the $XA$ structure. Since we also calculated the energy for each compound in the $L2_{1}$ structure, we also show the $L2_{1}$ structure in Fig.\ref{fig:L21Xa}c for comparison.

In this study, (a) $X$ is one of 9 elements -- Sc, Ti, V, Cr, Mn, Fe, Co, Ni or Cu, (b) $Y$ is one of 9 elements -- Ti, V, Cr, Mn, Fe, Co, Ni, Cu, or Zn, and (c) $Z$ is one of 9 elements -- Al, Ga, In, Si, Ge, Sn, P, As or Sb. In summary, we  calculated the 405 inverse-Heusler compounds for which the valence (or atomic number since $X$ and $Y$ are both assumed to be 3$d$) of $Y$ is larger than that of $X$.  For each of these 405 potential inverse-Heusler compounds, we calculated its electronic and magnetic structure, stability against structural distortion, and formation energy. We also compared its formation energy to those of  all phases and combinations of phases in the Open Quantum Materials Database (OQMD)~\cite{raey,oqmd_npj_2015}.

\subsection{Density Functional Theory Calculations}
\label{sec:DFTdetail}

All calculations were performed using density-functional theory (DFT) as implemented in the Vienna Ab-initio Simulation Package (VASP)~\cite{Kresse199615} with a plane wave basis set and projector-augmented wave (PAW) potentials~\cite{PhysRevB.50.17953}. The set of PAW potentials for all elements and the plane wave energy cutoff of 520 eV for all calculations were both chosen for consistency with the Open Quantum Materials Database (OQMD)~\cite{raey,oqmd_npj_2015}. The Perdew-Burke-Ernzerhof (PBE) version of the generalized gradient approximation (GGA) to the exchange-correlation functional of DFT was adopted\cite{PhysRevLett.77.3865}. The integrations over the irreducible Brillouin zone (IBZ) used the automatic mesh generation scheme within VASP with the mesh parameter (the number of $k$-points per {\AA}$^{-1}$ along each reciprocal lattice vector) set to 50, which usually generated a $15\times15\times15$ $\Gamma$-centered Monkhorst-Pack grid~\cite{PhysRevB.13.5188}, resulting in 288 $k$-points in the IBZ.The integrations employed the linear tetrahedron method with Bl\"ochl corrections~\cite{PhysRevB.49.16223}. To achieve a higher accuracy with respect to the magnetic moment, the interpolation formula of Vosko, Wilk, and Nusair~\cite{vosko1980accurate} was used in all calculations. Finally, All the crystal structures are fully relaxed until the energy change in two successive ionic steps is less than $1\times10^{-5}$ eV.

\subsection{\label{DetermineStructure}Determination of the Relaxed Structure}

We explain our procedure for obtaining the relaxed structures in some detail in order to make clear that there are many possible structures (in addition to the $XA$ structure) that an \textit{X$_{2}$YZ} compound can assume and that the determination of the ground state can be complicated by the additional degrees of freedom associated with the possible formation of magnetic moments of varying magnitude and relative orientation and by possible distortions of the cell.

We calculated the electronic structure of each compound in the $XA$ phase by minimizing the energy with respect to the lattice constant.  For each system and each lattice constant, we found it to be important to initiate the calculation with multiple initial configurations of the magnetic moments. We considered four kinds of initial moment configurations: (1) moments on the 3 transition metal atoms are parallel; (2) moment on the $Y$ site is antiparallel to the moments on the $X_{1}$ and $X_{2}$; (3) moment on the $X_{1}$ site is antiparallel with the moments on the $X_{2}$ site and $Y$; (4) moment on the $X_{2}$ site is antiparallel with the moments on the X$_{1}$ and $Y$. After the lattice constant and magnetic configuration that determined the minimum energy  for the $XA$ structure was determined we compared this energy to the minimum energy determined in a similar way for the $L2_{1}$ structure.

\textbf{Competition between $XA$ and $L2_1$ }

Comparing the calculated energies of the $XA$ and $L2_1$ phases of the 405 compounds that we considered, we found that 277 had lower energy in the  $L2_{1}$ phase. The remaining 128  preferred to remain in the $XA$ phase. This result may be compared to the conclusion in references \cite{PhysRevB.87.024420,PhysRevB.77.014424,Ozdogan2009L34,AlessthanB-1,AlessthanB-2,AlessthanB-3,AlessthanB-4,AlessthanB-5} that the $XA$ structure is energetically preferred to the $L2_{1}$ structure when the atomic number of the $Y$ element is the larger than the atomic number of $X$.  Although this hypothesis is consistent  with some experimental observations \cite{PhysRevB.77.014424,PhysRevB.83.174448,PhysRevB.84.132405,Klaer,Alijani}, it does not seem to be generally valid because more than two thirds of the systems in our data set (all of which satisfy $N_Y>N_X$) have lower calculated energy as $L2_1$.

After the minimum energy in the cubic $XA$ phase was determined, a full ionic relaxation within an 8-atom tetragonal cell was performed for all of the 405 potential inverse-Heusler compounds and full-Heusler compounds in order to test whether the cubic phase was stable against distortions.  Of the 128 $XA$ compounds 95  were found to remain in the $XA$ structure while 33 relaxed to a tetragonal `$XA$' phase. In contrast to the half-Heuslers which are prone to trigonal distortions\cite{ma2016computational}, we find that the full-Heuslers and inverse-Heuslers tend to retain tetragonal or cubic symmetry during this type of relaxation.  The susceptibility of the half-Heuslers to trigonal distortions is likely related to the vacant site in the cell.

Of the 277 compounds that preferred the  $L2_1$ structure, 136  remained in the $L2_{1}$ structure while 141 relaxed to a tetragonal `$L2_1$' structure. For both $XA$ and $L2_1$ we considered the structure to be cubic if   $\left|c/a-1\right|<0.01$).  All relaxations started from the $XA$ structure or $L2_{1}$ structure adapted to an 8 atom tetragonal cell.

\textbf{Competition between Cubic and Tetragonal $XA$}

Of the 405 potential $XA$ compounds that we considered, 141 were unstable with respect to a tetragonal distortion.  Of the remaining 264 compounds, 136 were calculated to be more stable in the full-Heusler ($L2_1$) phase than in the inverse-Heusler ($XA$) phase.   Few  systems with  $X$=Sc, Ti, or V are calculated to be stable as $XA$. None of the 81 considered compounds with $X$=Sc were found to be $XA$.  10  relaxed into the tetragonal `$L2_{1}$' phase, 63 preferred  the $L2_{1}$ phase and the remaining 8 became tetragonal `$XA$' phase.   Among the 72 compounds with $X$=Ti, only 4 compounds preferred $XA$ phase, while 37  relaxed into the tetragonal `$L2_{1}$' phase, 29 preferred the $L2_{1}$ phase and 2 distorted to be tetragonal `$XA$' phase.   Among the 63 compounds with $X$=V, 43 compounds distorted into tetragonal `$L2_{1}$' phase,  2 preferred the $L2_{1}$ phase and 2 distorted to be tetragonal `$XA$' phase.

\textbf{Competition between magnetic configurations}

For some compounds, we found multiple solutions to the DFT equations at the same or similar lattice constants with different magnetic configurations. These can be categorized into 6 groups:  in the $L2_{1}$ structure; (1) the moments of the $X$ atoms (which are identical by assumption for $L2_1$) are parallel to those for $Y$ ($L2_{1}$ ferromagnetic), or (2) the moments of  $X$ are anti-parallel to those for $Y$($L2_{1}$ ferrimagnetic); in the $XA$ structure,  (3)\ the moments for $Y$ are parallel with those for $X_{1}$ but anti-parallel with those for $X_{2}$ ($XA$ ferrimagnetic A1), or (4) the moments for $Y$ are parallel with those for $X_{2}$ but anti-parallel with those for $X_{1}$ ($XA$ ferrimagnetic A2), or (5) the moments for $X_{1}$ and $X_{2}$ are parallel with each other but anti-parallel with those for Y ($XA$ ferrimagnetic), or (6) the moments for $X_{1}$, $X_{2}$ and Y are parallel ($XA$ ferromagnetic). Magnetic transitions, in both magnitude and orientation as a function of lattice constant are clearly possible and often observed in our calculations.
 We did not consider the possibility of non-collinear moments\cite{PhysRevLett.113.087203_Mn2RhSn}, and because the calculations did not include spin-orbit coupling, we did not obtain information about the orientation of the moments relative to the crystal axes.

\textbf{Example Mn$_2$ZnP}

\begin{figure}[t]
        \includegraphics[width=3.25in]{{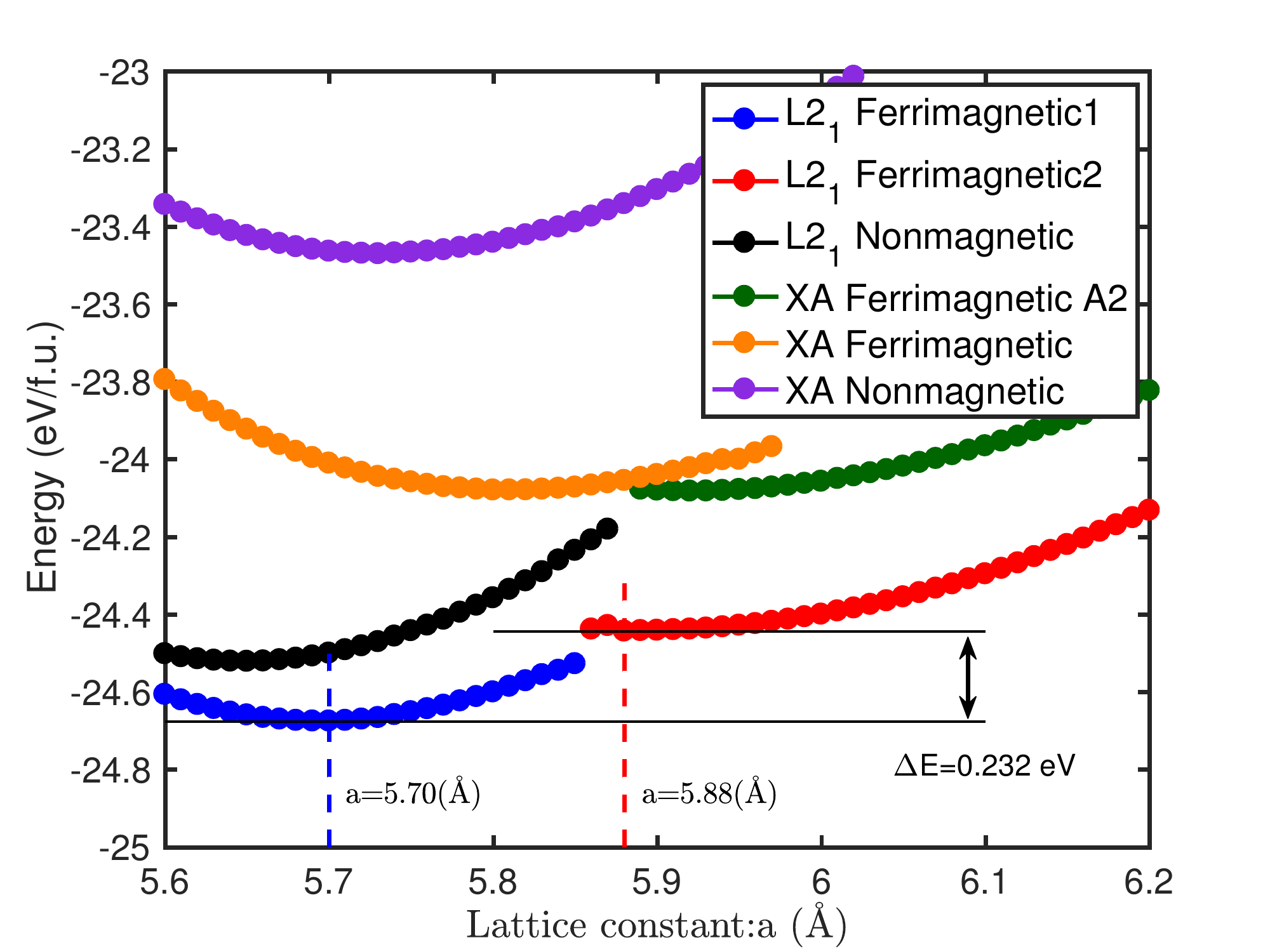}}
        \caption{Calculated total energies of Mn$_{2}$ZnP with full-Heusler $L2_{1}$ structure and inverse-Heusler $XA$ structure as a function of the lattice constant $a$ in ferrimagnetic, ferromagnetic and nonmagnetic states.}
        \label{Mn2ZnP}
\end{figure}

Mn$_{2}$ZnP provides an example of competition between  different magnetic states for both the $L2_1$ and $XA$ structures as shown in Fig. \ref{Mn2ZnP}. In this case, the $L2_1$ phases have lower energy. In addition to a nonmagnetic phase with a lattice constant of 5.645 \AA, two energy minima were found for magnetic phases.  These occur at $a=5.70$ \AA\  and $a=5.88$ \AA. For the solution at $a=5.70$ \AA, the moments within a sphere of radius 1.45 \AA\  surrounding each atom are 1.349 for Mn, -0.012 for Zn and -0.102 for P. For the solution at $a=5.88$ \AA, the compound has a total magnetic moment of 5.358$\mu_{B}$ per unit cell, and the moments within the 1.45 \AA\  spheres in this case are 2.743, -0.031, and -0.137 for Mn, Zn and P, respectively. The solution at $a=5.88$ \AA\  has higher energy than the solution at $5.70$ \AA\  by $\Delta =0.232$ eV per formula unit.

There are also multiple $XA$ magnetic phases.  The two shown are both ferrimagnetic in the sense that the two  Mn moments have opposite signs.  The phase with larger lattice constant actually has smaller net moment because of the closer approximate cancellation of the Mn moments.

To be clear, the energy comparisons of the preceding paragraph only consider the $XA$ and $L2_1$ phases and their possible tetragonal distortions.  Phase stability is discussed more generally in Section \ref{Eform}.

\textbf{Competition between ordered and disordered phases
}
For a few systems, we considered the energy difference between ordered and disordered phases.  As will be discussed in detail in Sec.\ref{FE and HD},  some Mn$_2$CoZ and Mn$_2$RuZ $XA$ compounds, (especially when $Z$ is a larger atom, \textit{e.g.} In, Sn, or Sb,) seem to have disorder in the occupation of the Mn$_1$ and $Y$ sublattice sites. To approximate this disorder in the $XA$ Mn$_2$Co$Z$ ($Z$=In, Sn, and Sb) and Mn$_2$RuSn, special quasi-random structures (SQS) as described by Zunger~\textit{et al.}~\cite{PhysRevLett.65.353} were generated. The SQS are designed to best represent the targeted disordered alloy by a periodic system of a given size. 

Our SQS comprised 32 atom cells for the case  of completely random occupation of the  $X_1$ and $Y$ sites, and 64 atom cells for the case in which the $X_1$ ($Y$) site occupation was 75\%/25\% (25\%/75\%).  The SQS was generated using the Monte Carlo algorithm of Van de Walle \textit{et al.}~\cite{van2013efficient} as implemented in the Alloy Theoretic Automated Toolkit (ATAT)~\cite{van2009multicomponent}. We did not consider the case with smaller $X/Y$ ratios because much larger unit cells would have been required.

\subsection{\label{Eform}Calculation of Energetic Quantities}
\label{ssec:energetic_quantities}

\subsubsection{Formation energy}
\label{sssec:fromation_energy}

We further investigated the structural stability of the 405 potential inverse-Heusler compounds by calculating their formation energies. The formation energy of an inverse-Heusler compound \textit{X$_{2}$YZ} is defined as

\begin{equation}\label{eqn:formation_energy}
\Delta E_{f}\,(X_{2}YZ)=E\,(X_{2}YZ)-\frac{1}{4}\left(2\mu_X+\mu_Y+\mu_Z\right)
\end{equation}

where $E\,(X_{2}YZ)$ is the total energy per atom of the inverse-Heusler compound, and $\mu_i$ is the reference chemical potential of element $i$, chosen to be consistent with those used in the OQMD database.  Typically the reference energy is the total energy per atom in the elemental phase (see Refs. \cite{raey,oqmd_npj_2015} for details). A negative value of $\Delta E_{f}$ indicates that at zero temperature, the compound is more stable than its constituent elements. It is a \textit{necessary but not sufficient} condition for ground state thermodynamic stability. It does not, for example, guarantee the stability of an inverse-Heusler phase over another competing phase or mixture of phases. For that, we must determine the convex hull.

\subsubsection{Distance from the Convex Hull}
\label{sssec:hull_distance}

A compound can be thermodynamically stable only if it lies \textit{on} the convex hull of formation energies of all phases in the respective chemical space. Every phase on the convex hull has a formation energy lower than any other phase or linear combination of phases in the chemical space at that composition. Thus, any phase on the convex hull is, by definition, thermodynamically stable at 0K. Conversely, any phase that does not lie on the convex hull is thermodynamically unstable; \textit{i.e.} there is another phase or combination of phases on the convex hull that is lower in energy.

  The distance from the convex hull $\Delta E_{\rm HD}$ for a phase with formation energy $\Delta E_{f}$ can be calculated as

\begin{equation}\label{eqn:hull_distance}
\Delta E_{\rm HD}=\Delta E_{f} - E_{hull}
\end{equation}

where $E_{hull}$ is the energy of the convex hull at the composition of the phase. The energy of the convex hull at any composition is given by a linear combination of energies of stable phases. Thus the determination of $E_{hull}$ from a database of formation energies is a linear composition-constrained energy minimization problem\cite{ADMA:ADMA200700843, AENM:AENM201200593}, and is available as a look-up feature called ``grand canonical linear programming"(GCLP) on the OQMD website\footnote{\url{http://oqmd.org/analysis/gclp}}. Obviously, if the hull distance $\Delta E_{\rm HD}$ for a phase on the convex hull is 0, there is no other phase or linear combination of phases lower in energy than the phase at that composition. The distance of the formation energy of a phase from the convex hull is an indicator of its thermodynamic stability and the likelihood  of it being fabricated because the greater the distance from the convex hull, the higher is the thermodynamic driving force for  transformation into another phase or decomposition into a combination of phases.

We note that the distance  of a phase from the calculated convex hull depends on the completeness of the set of phases considered in the construction of the convex hull. Ideally, to calculate the convex hull of a system, $X$-$Y$-$Z$, one would calculate the energies all possible compounds that can be formed from elements $X$, $Y$, and $Z$.  Unfortunately, such a comprehensive study  is not currently feasible. 

A practical approach is to construct the convex hull using all the currently reported compounds in the $X$-$Y$-$Z$ phase space. Here, we have limited our set of considered phases to those in the OQMD, which includes many ternary phases that have been reported in the ICSD, and $\sim$ 500,000 hypothetical compounds based on common crystal structures. Thus the calculated formation energy of each \textit{$X_{2}YZ$} inverse-Heusler compound is compared against the calculated formation energies of all phases and all linear combinations of phases with total composition \textit{$X_{2}YZ$} in the OQMD database.

A phase that we calculate to be above the convex hull may still be realized experimentally for three reasons: (a)  There may be errors in the calculation of the energies of the inverse-Heusler compound or in the reference compounds.  These errors  may be intrinsic to PBE-DFT or due to insufficient precision in the calculation or   inability to distinguish the ground state.  We have applied considerable effort to  reduce or eliminate the latter two  possibilities so far as possible.  (b) The compounds are usually synthesized at temperatures well above 0K.     Because entropic effects will differ for different phases, the phase with lowest free energy at the synthesis temperature may be different from the phase with lowest energy at 0K and this high temperature phase may persist to low temperature due to slow kinetics at laboratory time scales. (c) Non-equilibrium processing techniques can sometimes be used to fabricate non-equilibrium metastable phases.

Conversely, an inverse-Heusler phase that we calculate to be on the calculated convex hull may nevertheless be difficult or impossible to fabricate
because of reasons (a)\ or (b)\ or because the
database used to generate the convex hull does not contain all phases so that the true convex hull lies below the calculated one.

\section{\label{Results}RESULTS AND DISCUSSION\label{sec:results_discussion}}

\subsection{Energetics: Formation Energy and Distance from the Convex Hull \label{FE and HD}}
In this section, we systematically investigate the formation energy and thermodynamic stability of  the 405 inverse-Heusler (IH) compounds considered in this paper. The formation energies at 0 K were evaluated by using Eq.~(\ref{eqn:formation_energy}) and the convex hull distances were calculated  using Eq.~(\ref{eqn:hull_distance}). We first explored the relationship between formation energy and convex hull distance for the known synthesized IH compounds, including those collected in the Inorganic Crystal Structure Database (ICSD).    

We compiled a list of the reported IH compounds by extracting IH compounds  from the ICSD and literature including all elements as potential $X$, $Y$ or $Z$ atoms.  From this list we removed the compounds with partially occupied sites.   Some of the ICSD entries were from DFT calculations. These were included except for those with a higher DFT energy than the corresponding full Heusler which were obviously included in the ICSD by mistake.  Finally, we tabulated the formation energy and convex hull distance as calculated in the OQMD. 

A total of 48 distinct IH compounds are reported in ICSD.  Of these, 36  have been  synthesized experimentally. References to six additional synthesized IH compounds, not included in ICSD, (Fe$_2$CoGa, Fe$_2$CoSi, Fe$_2$CoGe, Fe$_2$NiGa, Mn$_2$NiGa, and Fe$_2$RuSi) were found in recent literature ~\cite{yin2015enthalpies,kreiner2014new}. These can be used  for cross validation of our method. Although Mn$_2$NiSn was synthesized by Luo \textit{et al.}~\cite{luo2009effect}, a more recent study shows that it has atomic disorder on the transition metal sites~\cite{kreiner2014new} and our DFT calculation show it is 103 meV/atom above the convex hull. 

\begin{figure}
        \centering
        \includegraphics[width=\linewidth]{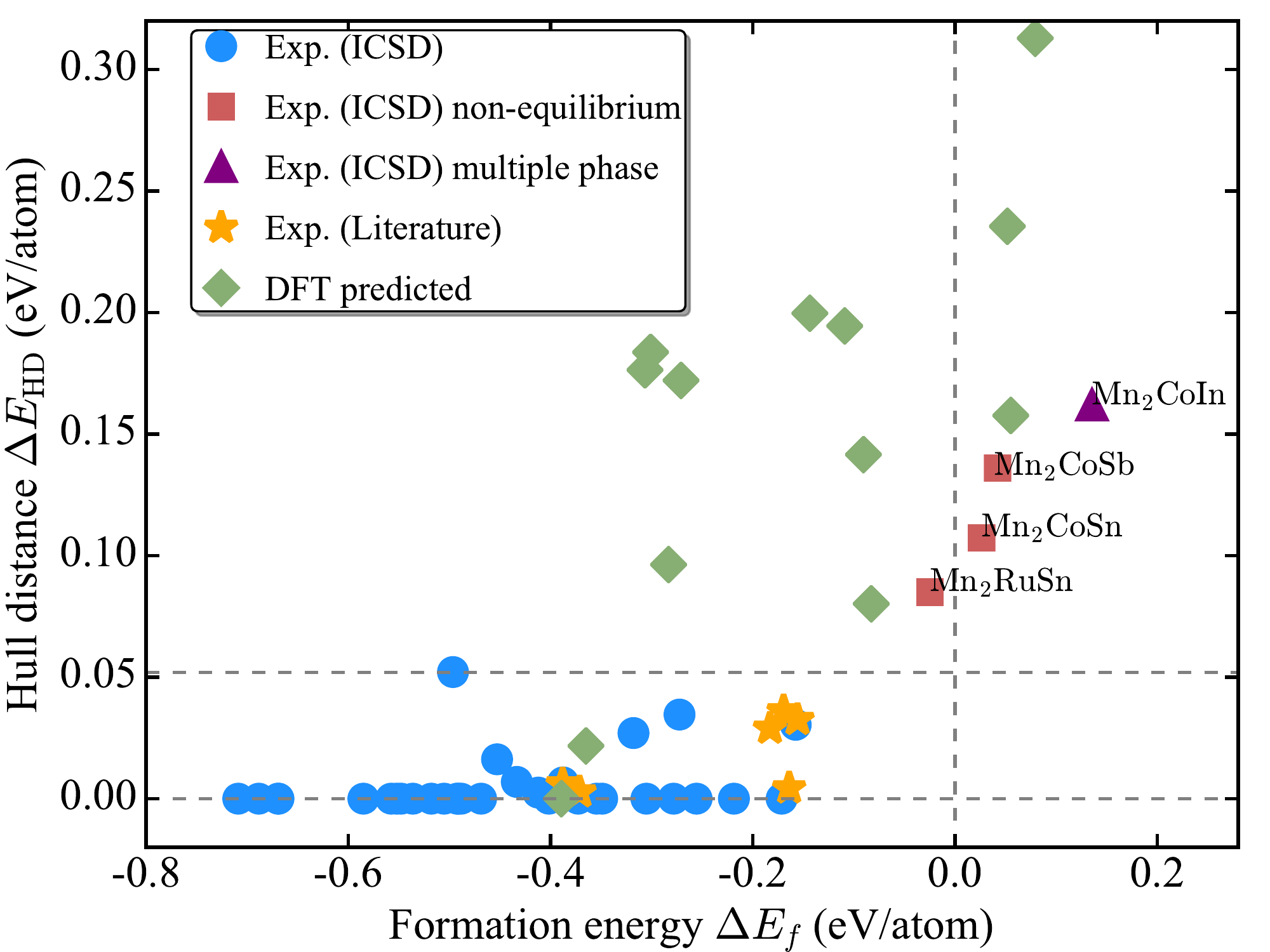}
        \caption{DFT-calculated formation energy vs. convex hull distance of inverse Heuslers reported in ICSD and in recent experimental studies. A hull distance $\Delta{E_{\rm{HD}}}$=0 (on the convex hull) indicates the compound is thermodynamically stable at 0K. Blue circles and red squares represent experimentally synthesized IH through equilibrium and non-equilibrium process, respectively. Purple triangle is synthesized IH coexisting with other phase. Yellow stars represent synthesized compounds reported in literature but not in ICSD. Green diamonds represent IH compounds predicted using DFT.}
        \label{fig:convexhull_icsd}
\end{figure}

These data from ICSD and the recent literature are displayed in Fig.~\ref{fig:convexhull_icsd}, from which it can be seen that most of the synthesized IH compounds  are within 52 meV/atom of the convex hull. The green diamonds represent the IH compounds that have not been synthesized but have been sourced into the ICSD by DFT predictions.

Overall, we find most of the experimentally reported IH compounds (38 of 42) have a convex hull distance less than 52 meV/atom. Thus,  the convex hull distance of the DFT-calculated formation energy at 0K appears to be a good indicator of the likelihood of  synthesis of an IH compound and 52 meV/atom is a reasonable empirical threshold separating  the compounds likely to be synthesized by equilibrium processing from those unlikely to be so synthesized.  

The empirical 52 meV threshold assumed here may be contrasted with the 100 meV threshold that seemed appropriate for the half-Heusler compounds~\cite{ma2016computational}. We speculate that the more open structure of the half-Heusler phase is associated with a high density of states for low frequency phonons which reduces the free energy of the half-Heusler phase relative to competing phases  
at temperatures used for synthesis.

Before proceeding it is important to note that several Mn$_2YZ$ compounds (Mn$_2$CoSn, Mn$_2$CoSb, Mn$_2$CoIn, and Mn$_2$RuSn) appear to significantly violate our empirical hull distance threshold.  They also have vanishing or positive formation energy!    A discussion of our rationale for \textit{not} extending our hull distance threshold to include these systems follows: 

(a) Mn$_2$CoSn: Our calculations predict that $XA$ phase Mn$_2$CoSn lies well above the convex hull. A mixture of phases, CoSn-Mn is predicted to be lower in energy by $\Delta E_{\rm {HD}}$ = 0.107 eV/atom. However, there are reasons to believe that the experimentally reported phase may not be pure $XA$.

The ICSD entry for Mn$_2$CoSn is based on the work of Liu~\textit{et al.} who described synthesis via melt-spinning as well as calculations that employed the FLAPW technique \cite{PhysRevB.77.014424}. They found that the non-equilibrium melt-spinning technique generated a material with a ``clean" XRD pattern consistent with the $XA$ phase.  The measured magnetic moment of 2.98~$\mu_B$ per formula unit was consistent with their calculated value of 2.99 and with their calculated DOS, which indicates that this material is a Slater-Pauling near-half-metal.

Subsequently, Winterlik \textit{et al.} synthesized this compound by  the quenching from 1073K of annealed arc-melted samples\cite{PhysRevB.83.174448}. They obtained similar XRD results to Liu \textit{et al.}, but with a small admixture of a tetragonal MnSn$_2$ phase. They also concluded that the experimental moment was near the Slater-Pauling half-metal value of 3$\mu_B$ per formula unit.  However, their calculations for $XA$-phase Mn$_2$CoSn which also employed FLAPW found, in addition to the near-half-metallic phase found by Liu \textit{et al.}, a different magnetic structure  with  lower energy. 

Our calculations using VASP confirm the Winterlik \textit{et al.} results.  The PBE-DFT-predicted phase for $XA$ Mn$_2$CoSn has a magnetic moment around 1.53 $\mu_B$ rather far from the value of 3$\mu_B$ necessary to place the Fermi energy into the Slater-Pauling gap after 12 states per atom in the minority channel.  There is an interesting competition between two very similar ferrimagnetic states that plays out in many of the 26 and 27 electron Mn$_2$Co$Z$ systems.  The Mn$_2$ and Co moments are aligned and are partially compensated by the Mn$_1$ moment.  For smaller lattice constants, the compensation yields the correct moment for a half-metal or near half-metal (i.e., for Mn$_2$Co(Al,Ga,Si,Ge)), but as the lattice constant increases, (\textit{i.e.,} for Mn$_2$Co(In,Sn)), the Mn$_1$ moment becomes too large in magnitude and the partial compensation fails to yield a half-metal.    

Winterlik \textit{et al.} conclude from XRD, NMR and M\"ossbauer studies that their quenched samples of Mn$_2$CoSn are actually disordered on the MnCo layers.  They modeled this disorder using the KKR-CPA approach and obtained values of the moments in reasonable agreement with magnetometry and XMCD.   They argued that their model with Mn and Co randomly occupying the sites on alternate (001) layers separated by ordered Mn-Sn layers agreed with XRD as well as the $XA$ model because Mn and Co scatter X-rays similarly. This disordered $L2_{1}B$ structure has both Mn and Co at the $\left(\frac{1}{4},\, \frac{1}{4},\, \frac{1}{4}\right)$ and the $\left(\frac{3}{4},\, \frac{3}{4},\, \frac{3}{4}\right)$ site, whereas the $XA$ structure has only Co at $\left(\frac{1}{4},\, \frac{1}{4},\, \frac{1}{4}\right)$ and only Mn at $\left(\frac{3}{4},\, \frac{3}{4},\, \frac{3}{4}\right)$ site. 

We investigated the Winterlik \textit{et al.} hypothesis by comparing the energy of a few cubic 16 atom supercells with differing equiatomic occupations of the MnCo layers to the energy of the $XA$  occupation.  The energies of these different site occupations were all lower than the $XA$ energy, and most of them were half-metallic or near-half metallic.  The decrease in energy for the  non-$XA$ site occupations ranged from 8 meV/atom to 33 meV/atom. These should be considered lower bound estimates of the decrease since we did not optimize the geometries of the model ``disordered" structures.  We also performed calculations for an SQS designed to mimic a system with disorder on the MnCo layers. Our calculation indicates that the SQS with 25\%\ Mn/Co mixing is 31 meV/atom lower in energy than the $XA$ phase. The net magnetic moment of this SQS is about 1.28 $\mu_B$ per f.u., which is lower than that of $XA$ phase (1.53 $\mu_B$) due to the atom disordering.
    
We propose the following hypothesis and rationale for omitting Mn$_2$CoSn from the estimate for the hull energy distance threshold: During synthesis at high temperature, Mn$_2$CoSn is thermodynamically stable in a phase with substantial configurational disorder on the MnCo planes.  The configurational disorder may decrease the free energy by as much as 32 meV/atom at 1073K.  If we add this free energy decrease to the approximately 30 meV/atom difference in total energy between the ordered and disordered systems, we can estimate   a decrease in the relative free energy of formation by approximately 60 meV/atom due to disorder.  This  would yield a negative free energy and a hull distance less than our empirical 52 meV threshold.  

(b) Mn$_2$CoSb:  Mn$_2$CoSb is predicted to lie above the convex hull with a mixture of phases CoMnSb-Mn lower in energy by $\Delta E_{\rm{HD}}$ = 0.133 eV/atom. This phase also entered the ICSD database from the paper of Liu \textit{et al.} who generated their sample by melt-spinning and  also provided FLAPW calculations that predicted $XA$ Mn$_2$CoSb to be a half-metal with a Slater-Pauling moment of 4 $\mu_B$ per formula unit, consistent with their measured value.  

Unlike some Mn$_2$Co$Z$ compounds with smaller $Z$ atoms, synthesis of single $XA$ phase Mn$_2$CoSb is difficult via arc-melting and a high temperature annealing step. Prior X-ray diffraction investigations~\cite{PhysRevB.77.014424} revealed a multi-phase Mn$_2$CoSb in arc-melt samples. Our synthesis supports this finding~\cite{DipanjanMn2CoGa}. On the other hand, a melt-spinning treatment of arc-melt Mn$_2$CoSb ingots  is reported to produce a pure cubic structure~\cite{PhysRevB.77.014424,Dai2006533}. 

We consider it likely that  Mn$_2$CoSb is similar to Mn$_2$CoSn.  A study by Xin \textit{et al.} found that ``disordered'' Mn$_2$CoSb (apparently modeled by replacing the (001)  MnCo layers by  layers of pure Mn and pure Co alternating with the MnSb layers) was lower in energy than $XA$ by 21 meV/atom \cite{Xin201710}. Our SQS calculations show that the 50\% Mn/Co mixture within one rock-salt lattice is 60 meV/atom lower in energy than the $XA$ phases.
Again, our assertion is that the free energy reduction due to configurational entropy together with the lower energy of the disordered configurations may be sufficient to stabilize the $L2_{1}B$ phase as it is rapidly quenched from the melt.

(c) Mn$_2$CoIn: According to Liu \textit{et al.}\cite{PhysRevB.77.014424}, Mn$_2$CoIn appears to be multi-phase in samples prepared by arc-melting. Apart from the cubic phase identified as inverse-Heusler, a second phase described as fcc is also apparent.  Liu \textit{et al.} also report FLAPW calculations performed at a much smaller lattice constant (5.8 \AA) than the experimentally derived lattice constant (6.14 \AA).  These calculations yielded a moment of 1.95 $\mu_B$ in reasonable agreement with experimental value of 1.90 $\mu_B$. A subsequent calculation by Meinert \textit{et al.} found a lattice constant of 6.04 \AA and a moment of 1.95$\mu_B$\cite{AlessthanB-2}.

Our calculations for the $XA$ phase yielded a lattice constant of 6.10 \AA, much closer to the experimental value and a moment of 1.74 $\mu_B$. We also found a surprising and unusual sensitivity of our results to the number of plane waves.  We performed SQS calculations which indicated that the disordered phase is essentially degenerate in energy with the $XA$ phase. The magnetic moment of SQS is about 1.01 $\mu_B$ per f.u..

Given the significant amount of second phase in the experimental sample which would affect the composition of the phase identified as $XA$, together with the possibility of disorder on the MnCo layers we tentatively conclude that it is reasonable to discount the large calculated formation energy and hull distance in setting our threshold for likely fabricability of $XA$ phases .
Additional studies of this putative phase are probably advisable.
 
(d) Mn$_2$RuSn: Mn$_2$RuSn is predicted to lie above the convex hull with a mixture of phases MnSnRu+Mn+Sn$_7$Ru$_3$ by $\Delta E_{\rm {HD}}$ = 0.085 eV/atom. The first reports of fabrication by Endo \textit{et al.}\cite{endo2012magnetic}, were unable to distinguish between the $XA$ and $L2_{1}B$ phases. A later study by Kreiner \textit{et al.} seemed to confidently conclude from powder XRD studies that their sample prepared by alloying through inductive heating had the  $L2_{1}B$ structure\cite{kreiner2014new}. Our calculation of the energy of a SQS for Mn$_2$RuSn with 25\% Mn/Ru mixing gave an energy of the model disordered alloy 18 meV lower than $XA$. The calculated magnetic moment of the SQS is about 0.24 $\mu_B$ per f.u.. A study by Yang \textit{et al.} using the KKR -CPA\ approach estimated 11\% anti-site Mn based on a comparison of the calculated moment to the experimental one\cite{YANG2015247}. 

We tentatively conclude that the four experimental systems in the ICSD with anomalously large calculated hull distances and formation energies are probably not $XA$ phase and exclude them from consideration in determining the threshold for estimating which $XA$ phases are likely to be fabricated by equilibrium approaches. We also remind that an additional potential inverse-Heusler, Mn$_2$NiSn, was excluded from our database of experimental XA phases, because of experimental evidence of similar disorder to that observed in the Mn$_2$CoZ phases.   

We recommend additional studies of the competition between $XA$ and $L2_{1}B$ phases, especially for compounds which have large $Z$ atoms.    Such studies may require careful experimental analysis because of the similarities of the the XRD patterns of the two phases.  Theoretical studies may also be difficult because of the need to accurately calculate the formation energy of the disordered $L2_{1}B$ phase.

\begin{figure}
        \centering
        \includegraphics[width=\linewidth]{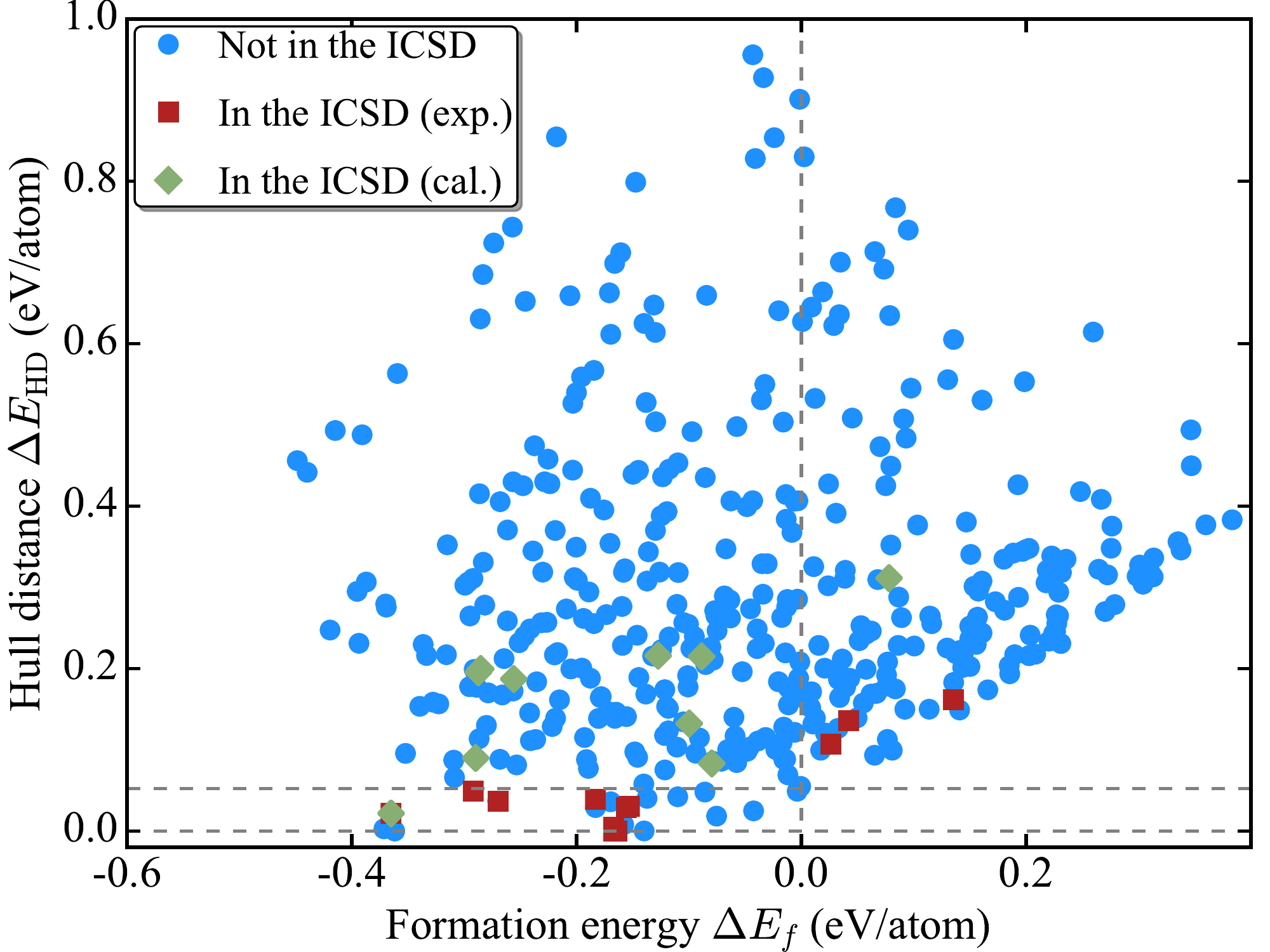}
        \caption{DFT-calculated formation energy vs convex hull distance of all IH considered in this paper. A hull distance $\Delta{E_{\rm{HD}}}$=0 (on the convex hull) indicates the compound is thermodynamically stable at 0K. Red diamonds and green squares represent IH compounds sourced into ICSD by synthesis and DFT computation, respectively. Blue circles represent the IH compounds calculated in this work.}
        \label{fig:convexhull_all}
\end{figure}

We extended the comparison of formation energy and convex hull distance to all the 405 IH compounds considered in this work as shown in Figure \ref{fig:convexhull_all}. It can be seen from Fig.~\ref{fig:convexhull_all} that  the calculated formation energies span a range from -0.45 to 0.38 eV/atom. This may be compared with a range of -1.1 to 0.7 eV/atom that was observed for half-Heuslers~\cite{ma2016computational}.  248 of the Inverse-Heuslers have negative formation energy, indicating thermal stability against decomposition into the  constituent elements at 0 K.   The calculated hull distances vary from 0 to nearly 1 eV/atom.  It is clear from the lack of correlation between formation energy and hull distance in Figure \ref{fig:convexhull_all} that the formation energy is not a reliable indicator of stability.

The IH compounds considered in this work that are included in ICSD are indicated by red squares in Figure \ref{fig:convexhull_all}.  These all lie within 0.052 eV/atom of the convex hull except for  three compounds mentioned previously (Mn$_2$CoSn, Mn$_2$CoIn and Mn$_2$CoSb) that were synthesized by non-equilibrium processing and are likely $L2_{1}B$ rather than $XA$ phase.  Our calculations predict 13 (out of 405) IH compounds to be within 52 meV/atom of the convex hull. The ``success rate'', i.e. the number of predicted potentially stable compounds versus total number of systems investigated is lower for IH (13/405) than for half-Heusler (50/378)~\cite{ma2016computational}, but this result is affected by the different hull distance thresholds and the different choices of potential compounds. 

\subsection{Composition and Stability \label{C&S}}

\begin{figure}
        \includegraphics[width=\columnwidth]{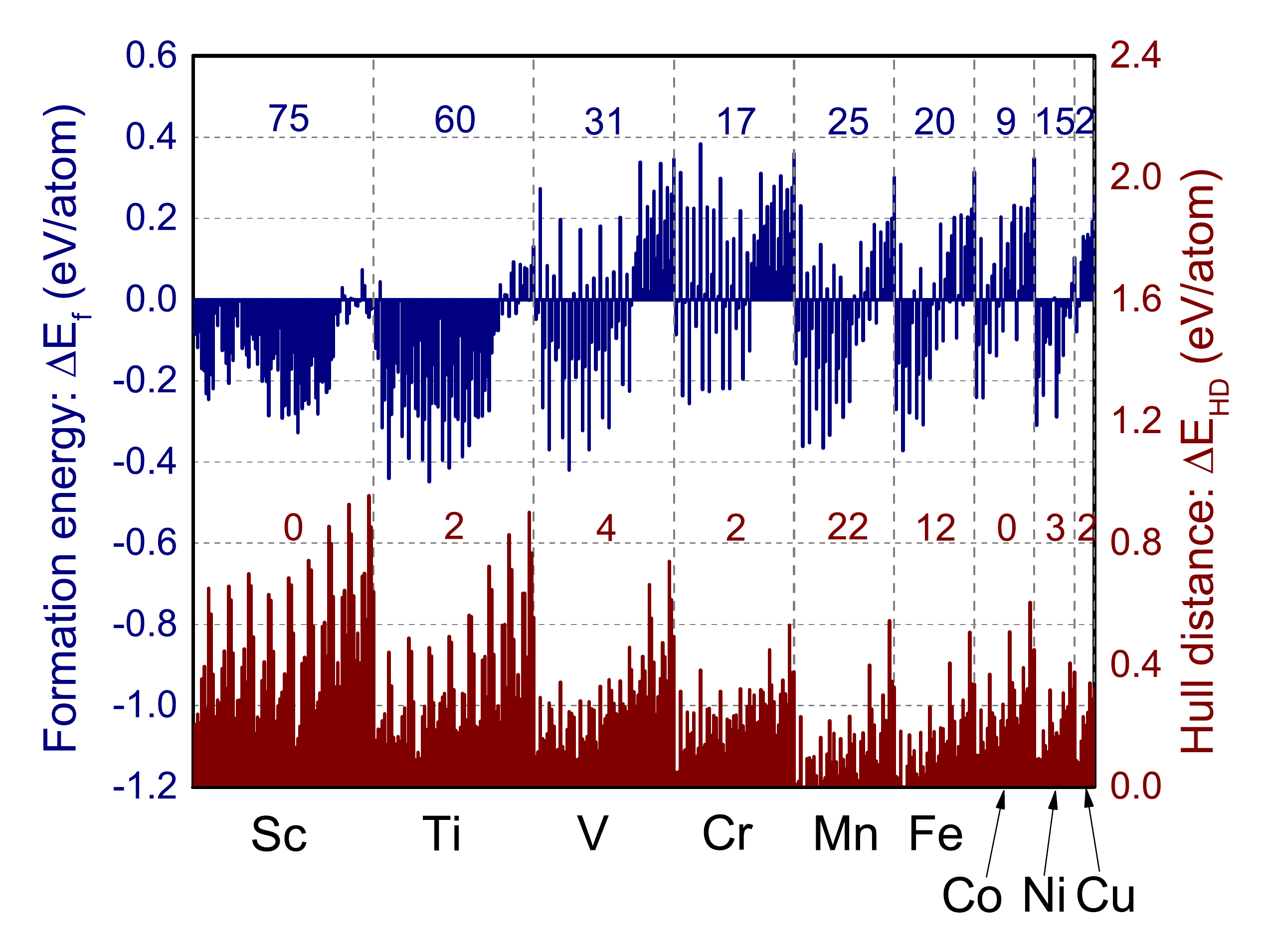}
        \caption{DFT formation energies and hull distances for potential inverse-Heusler compounds grouped by the element on the $X$-site. The numbers near the top (in blue) and center (in brown) of each column denote the number of compounds with negative formation energy $\Delta E_f$ and hull distance $\Delta E_{\rm HD} \leq 0.1$~eV/atom, respectively, in the corresponding $Z$-element group. Within a given $X$-element column, the compounds are ordered first by the element on the $Y$-site (same order as in Fig.~\ref{fig:formation_energy_hull_distance_vs_y}) and then by the element on the $Z$-site (same order as in Fig.~\ref{fig:formation_energy_hull_distance_vs_z}), i.e., $Z$ varies more rapidly than $Y$.}
        \centering
        \label{fig:formation_energy_hull_distance_vs_x}
\end{figure}

\begin{figure}
        \includegraphics[width=\columnwidth]{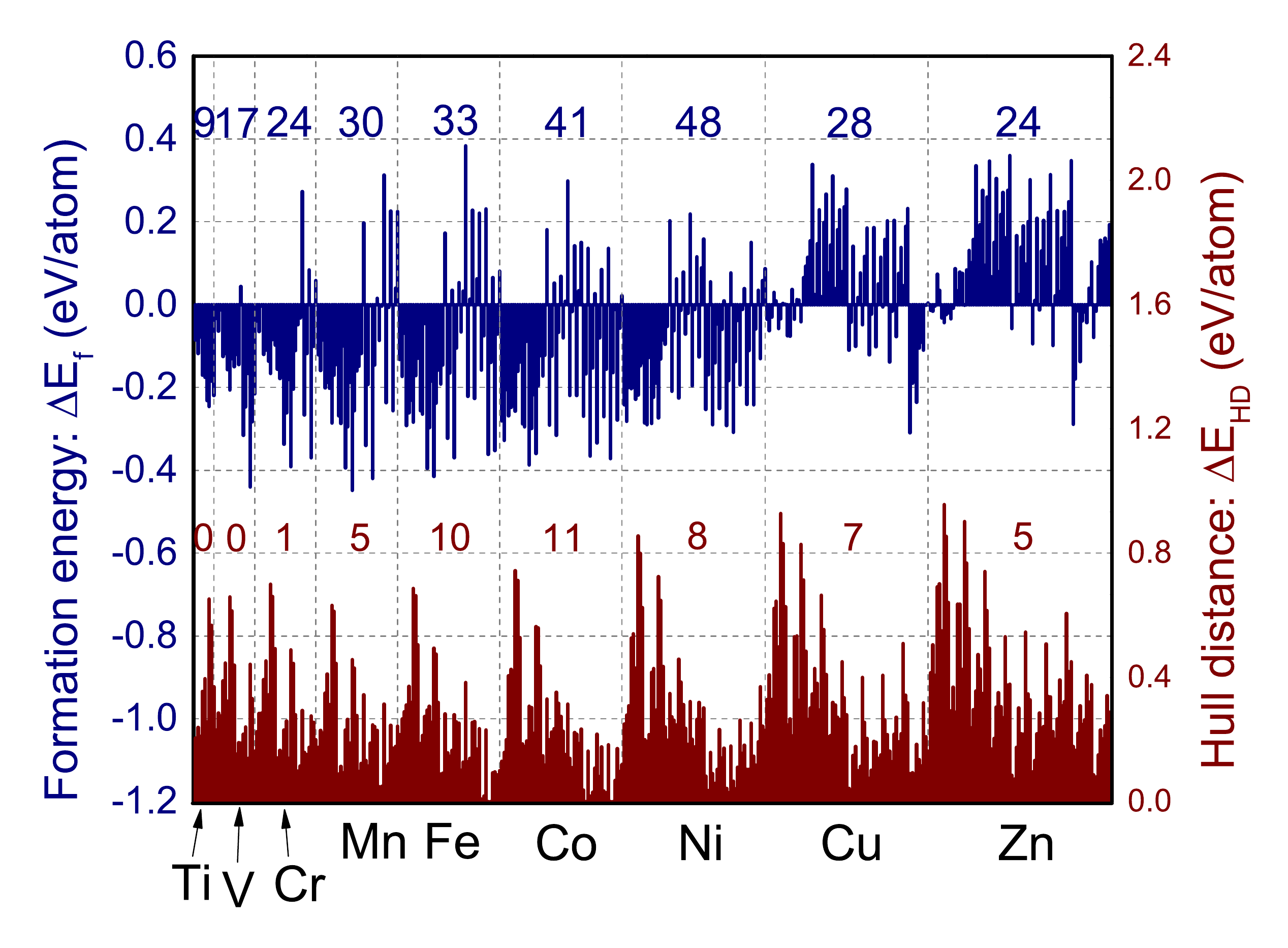}
        \caption{DFT formation energies and hull distances for potential inverse-Heusler compounds grouped by the element on the $Y$-site. The numbers near the top (in blue) and center (in brown) of each column denote the number of compounds with negative formation energy $\Delta E_f$ and hull distance $\Delta E_{\rm HD} \leq 0.1$~eV/atom, respectively, in the corresponding $Y$-element group. Within a given $Y$-element column, the compounds are ordered first by the element on the $X$-site (same order as in Fig.~\ref{fig:formation_energy_hull_distance_vs_x}) and then by the element on the $Z$-site (same order as in Fig.~\ref{fig:formation_energy_hull_distance_vs_z}), i.e., $Z$ varies more rapidly than $X$.}
        \centering
        \label{fig:formation_energy_hull_distance_vs_y}
\end{figure}

\begin{figure}
\includegraphics[width=\columnwidth]{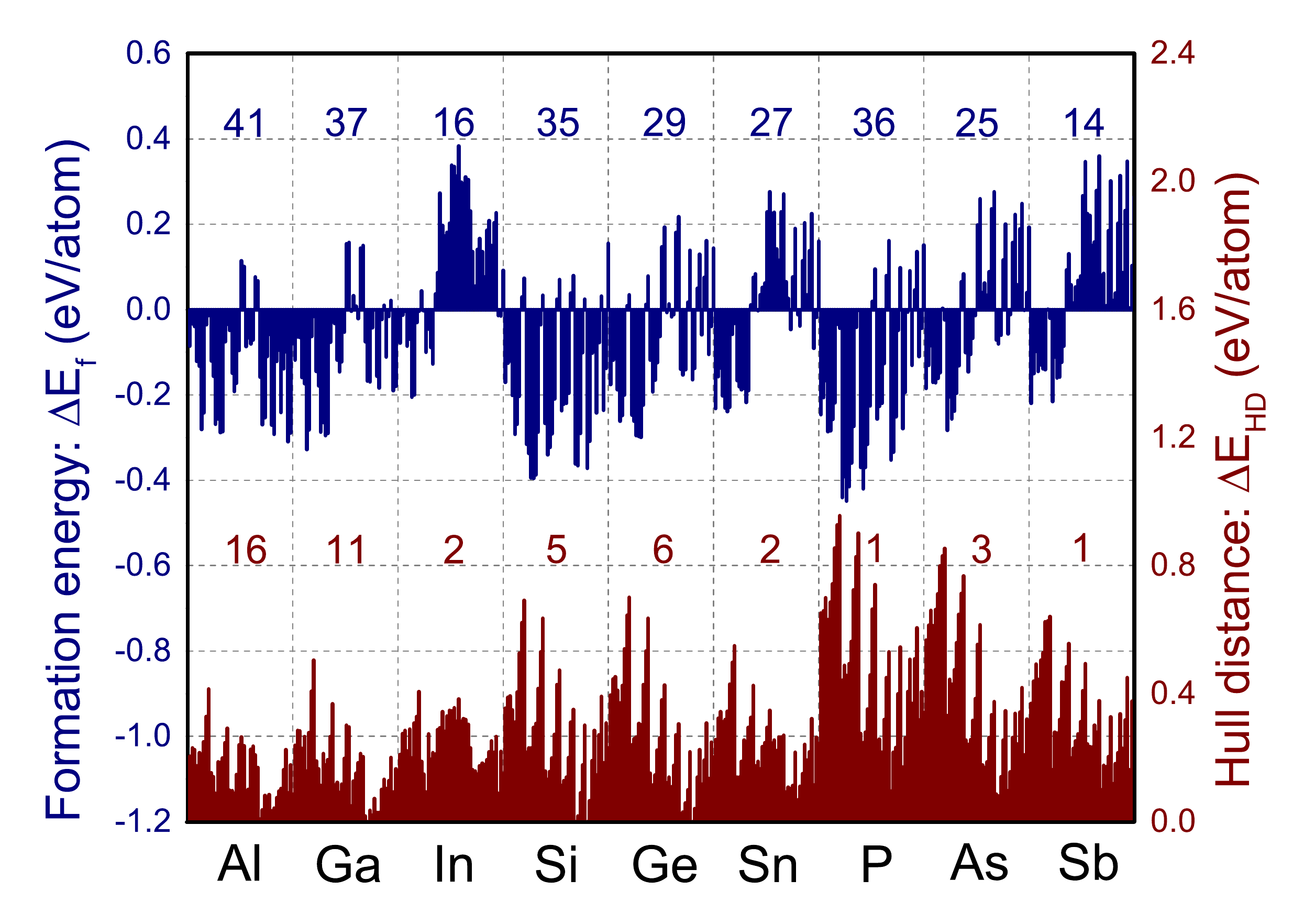}
\caption{DFT formation energies and hull distances for potential inverse-Heusler compounds grouped by the element on the $Z$-site. The numbers near the top (in blue) and center (in brown) of each column denote the number of compounds with negative formation energy $\Delta E_f$ and hull distance $\Delta E_{\rm HD} \leq 0.1$~eV/atom, respectively, in the corresponding $Z$-element group. Within a given $Z$-element column, the compounds are ordered first by the element on the $X$-site (same order as in Fig.~\ref{fig:formation_energy_hull_distance_vs_x}) and then by the element on the $Y$-site (same order as in Fig.~\ref{fig:formation_energy_hull_distance_vs_y}), i.e., $Y$ varies more rapidly than $X$.}
\label{fig:formation_energy_hull_distance_vs_z}
\centering
\end{figure}

\subsubsection{Stability and Composition}

We systematically investigated the formation energies of the 405 inverse-Heusler compounds in the $XA$ phase. We found that many of these  potential inverse-Heusler phases actually have lower energy in other phases,  \textit{e.g.} $L2_{1}$(136), tetragonal `$L2_{1}$' phase (141), or tetragonal $`XA$(33). Here, however, since we are interested in gaining insight into the systematics of the stability of the $XA$ phase, we compare  formation energies as if  all 405 compounds remained in the $XA$ phase. To better illustrate the relation between structural stability and composition, we arranged the formation energy data according to $X$, $Y$ and $Z$ element respectively in Fig. \ref{fig:formation_energy_hull_distance_vs_x}-\ref{fig:formation_energy_hull_distance_vs_z}.

Figure \ref{fig:formation_energy_hull_distance_vs_x} shows the calculated formation energies and hull distances of the 405 calculated $X_2YZ$ compounds arranged by $X$ species, then by $Y$ and finally by $Z$. From Fig. \ref{fig:formation_energy_hull_distance_vs_x}, we can see that Sc, Ti, and V generate more negative formation energy $XA$ compounds as $X$ elements. In terms of the fraction with negative formation energy, we found 75/81 with $X$=Sc, 60/72 with $X$=Ti, 25/45 with $X$=Mn and 20/36 with $X$=Fe, and 15/18 with $X$=Ni.       

 Figure \ref{fig:formation_energy_hull_distance_vs_y} presents the calculated formation energies and hull distances arranged first by $Y$, then by $X$ and finally by $Z$. From Fig. \ref{fig:formation_energy_hull_distance_vs_y}, it can be seen that when $Y$ is Fe, Co or Ni many  $XA$ compounds with negative formation energies result.  However, when one considers the fraction of negative formation energy compounds with a given $Y$ in our database, the highest are 9/9 for $Y$=Ti, 17/18 for $Y$=V and 24/27 for $Y$=Cr.

When the compounds are ordered by $Z$ element as presented in Fig. \ref{fig:formation_energy_hull_distance_vs_z}, we can see that the smaller atoms within a given group ($\it{i.e.}$ $Z$=Al, Si, In) tend to have a larger fraction of negative energy $XA$ phases.  
 The large number of $\Delta E_f<0$  compounds with $Z$=Al is striking in Fig.\ref{fig:formation_energy_hull_distance_vs_z},  but only 14 compounds have negative formation energy for $Z$=Sb. Although the number of negative formation energy compounds with $Z$=Al is greatest, the number among them with formation energies less than -0.2 eV (13) is less than for compounds with $Z$=Si (25) and $Z$=P (24).

It is interesting to contrast the trends in formation energy to those observed by considering the OQMD hull distance.   Although Sc, and Ti tend to generate negative formation energies as $X$ atoms in $XA$ phase $X_2YZ$ compounds, the picture is opposite for the hull distances.  Very few of these compounds have a hull distance near the empirical 0.052 eV/atom range which is indicative of a potentially synthesizable phase. Using a larger hull distance of 0.1 eV as an (arbitrary) stability threshold we find no $X$=Sc phases that meet the criterion and only 2 of 72 $X$=Ti phases.  Sc and Ti are ``active" elements that form numerous low energy phases.  Hull distances tend to be lower when the $X$ atom is Mn or Fe
with 22 of 45 $X$=Mn and 12 of 36 $X$=Fe meeting the criterion of  hull distance  less than 0.1eV/atom. 

While 9/9 $Y$=Ti and 17/18 $Y$=V IH compounds have negative formation energy, none of them meet the relaxed hull distance threshold of 0.1eV/atom.     However, relatively large fractions of $Y$=Fe (10/45), $Y$=Co (11/45) and $Y$=Ni (8/63) meet the criterion.   Thus, there is better correlation between low formation energy and low hull distance when the  $XA$ compounds are grouped by $Y$ element.   By both measures,  $Y$=Ni, Co or Fe tend to generate more stable compounds.  

The trends in stability with $Z$ atom are also somewhat similar when viewed from the perspectives of hull distance and formation energy.  Al, Ga, Si and Ge as $Z$ atom are associated with compounds having relatively low hull distances and negative formation energies.  One big difference, however, is $Z$=P which tends to generate numerous $XA$ compounds with low formation energy, but none with very low hull distances.

\subsubsection{Stability and Gaps near the Fermi Energy}

We calculated the electronic structure of each compound and obtained its spin polarization at the Fermi level, $E_{F}$. The spin polarization $P$ at $E_{F}$ is defined as:

\begin{equation}
P(E_{F})=\frac{N_{\uparrow}(E_{F})-N_{\downarrow}(E_{F})}{N_{\uparrow}(E_{F})+N_{\downarrow}(E_{F})}
\label{eqn:spin_polarization}
\end{equation}

where $N_{\uparrow}$ and  $N_{\downarrow}$ are the densities of states for majority (spin-up) and minority (spin-down) electrons, respectively. We define the compounds with 100\% spin polarization ($P(E_{F})=1$) to be half-metals in this paper.

\begin{figure}
        \centering
        \includegraphics[width=\columnwidth]{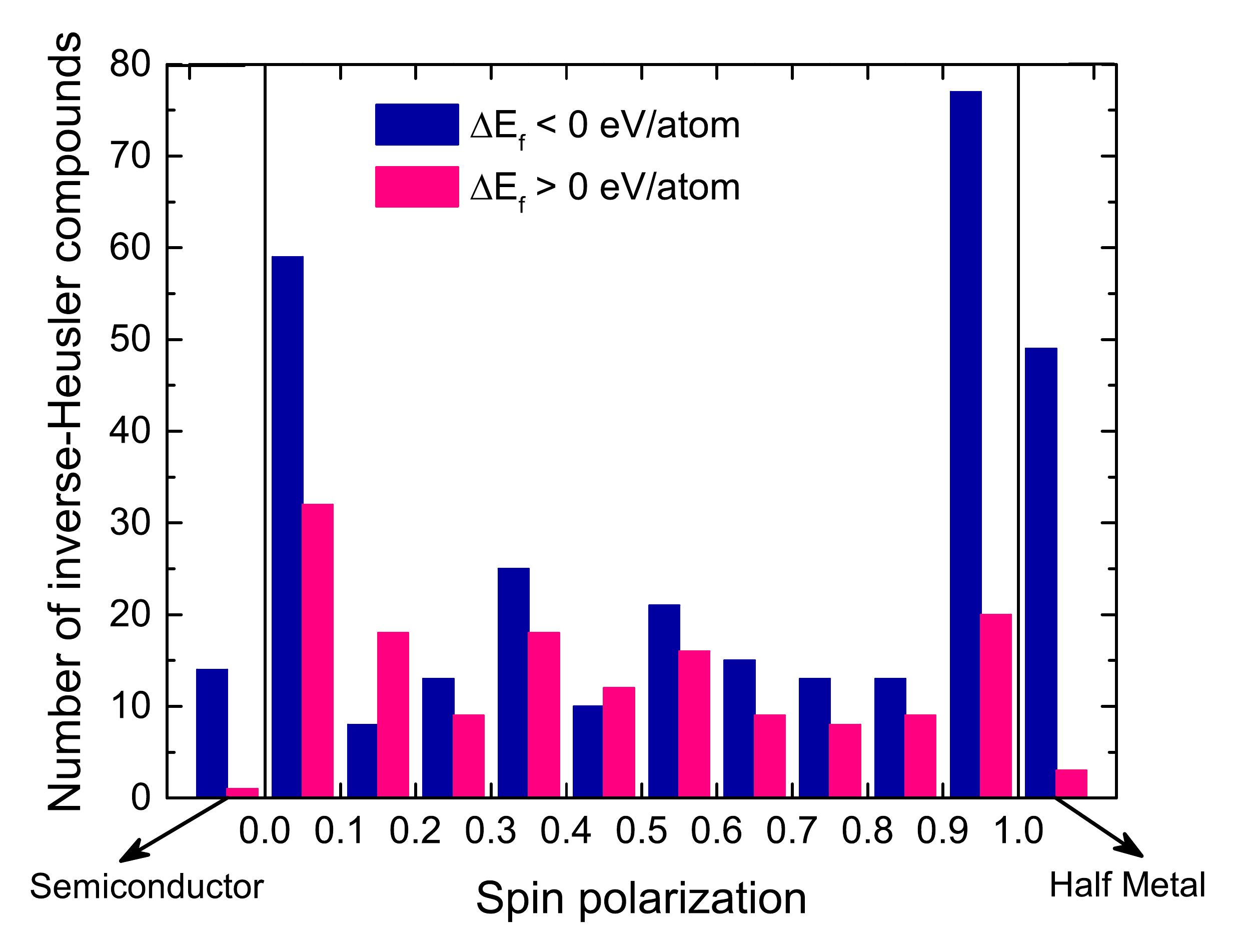}
        \caption{The distribution of the 405 potential inverse-Heusler compounds with negative ($\Delta E_{f} < 0$~eV/atom) and positive ($\Delta E_{f} > 0$~eV/atom) formation energies as a function of spin polarization $P(E_F)$ (given by Eq.~\ref{eqn:spin_polarization}). In the central region, we show the number of inverse-Heusler compounds grouped by 10 percentage points of spin polarization. In an additional region to the left, we show the 15 semiconductors, including 14 compounds with a negative formation energy and 1 with a positive formation energy. In the additional region on the right, we show 52 half-metals, including 50 with negative formation energy and 2 with positive formation energy. }
        \label{fig:formation_energy_vs_polarization}
\end{figure}

    In a study of the half-Heusler compounds gaps at the Fermi energy in one or both spin channels appeared to be associated with low formation energies and low hull distances\cite{ma2016computational}. Figure \ref{fig:formation_energy_vs_polarization} shows how the number of inverse-Heusler compounds with positive and negative formation energy varies with spin-polarization. It can be seen that for semiconductors and half-metals the ratio of the number of negative formation energy compounds to positive formation energy compounds is particularly high.

\begin{figure}
        \includegraphics[width=\columnwidth]{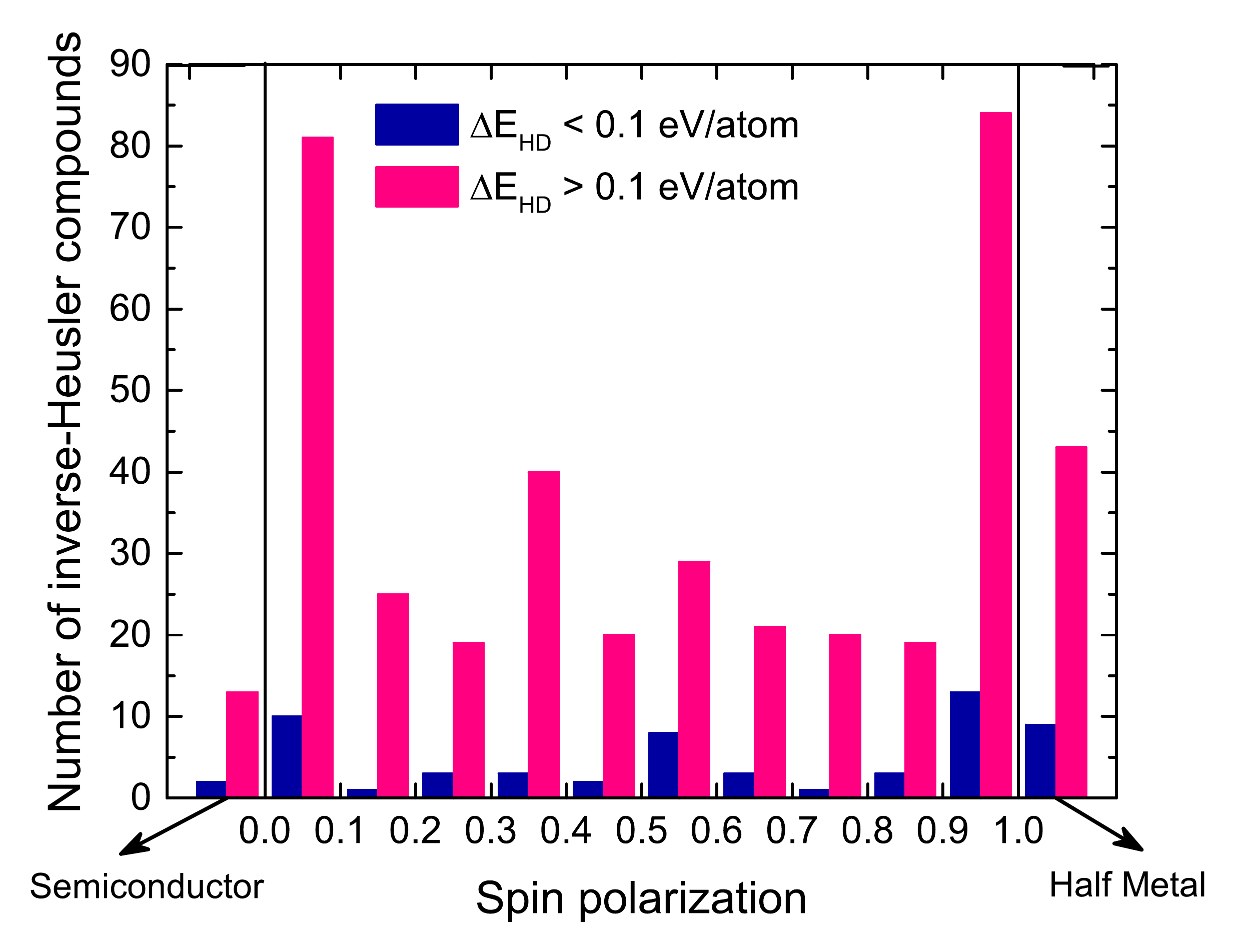}
        \caption{The distribution of the 405 potential inverse-Heusler compounds that lie near ($\Delta E_{\rm HD} < 0.1$~eV/atom) or far ($\Delta E_{\rm HD} > 0.1$~eV/atom) from the convex hull, as a function of spin polarization $P(E_F)$ (given by Eq.~\ref{eqn:spin_polarization}). In the central region, the number of inverse-Heusler compounds grouped by 10 percentage points of spin polarization is shown. In the additional region on the left are shown 15 semiconductors (of which 2 compounds have $E_{\rm HD} < 0.1$~eV/atom). On the right are shown the 52 half-metals (of which 9 compounds have $E_{\rm HD} < 0.1$~eV/atom). }
        \centering
        \label{fig:hull_distance_vs_polarization}
\end{figure}

Figure \ref{fig:hull_distance_vs_polarization} shows the number of inverse-Heusler compounds with positive and negative hull distances less than and greater than 0.1 eV/atom versus  spin polarization.  In contrast to the case of the  half-Heuslers, the plot of hull distance versus spin-polarization does not present an obvious case for a correlation between gaps near the Fermi energy and stability.   As will be discussed in Section \ref{SemiHM}, this is due to the large number of $X$=Sc and $X$=Ti compounds in our database that are semi-conductors or half-metals and although they tend to have negative formation energies, they tend to be less stable than competing compounds.  On the other hand, as will be discussed in Sec. \ref{sec:results_discussion}, only 20 of the IH compounds in our database systems met our 0.052 eV/atom hull distance stability criterion.    Half of these are half-metals or near half-metals.
We consider this to be strong support for the notion that gaps in or near the Fermi energy in one or both spin channels contribute to stability.

{\textbf{}
\subsection{\label{SemiHM}Semiconductor and half-metallic ferromagnetism}

In total we found 14 semiconductors, 51 half-metals, and 50 near half-metals  with negative formation energy among the 405 potential inverse Heusler compounds. We will discuss these  materials in this section.
\subsubsection{Semiconductors \label{SemiSec}}

\begin{figure}[t]
        \includegraphics[width=3.5in]{{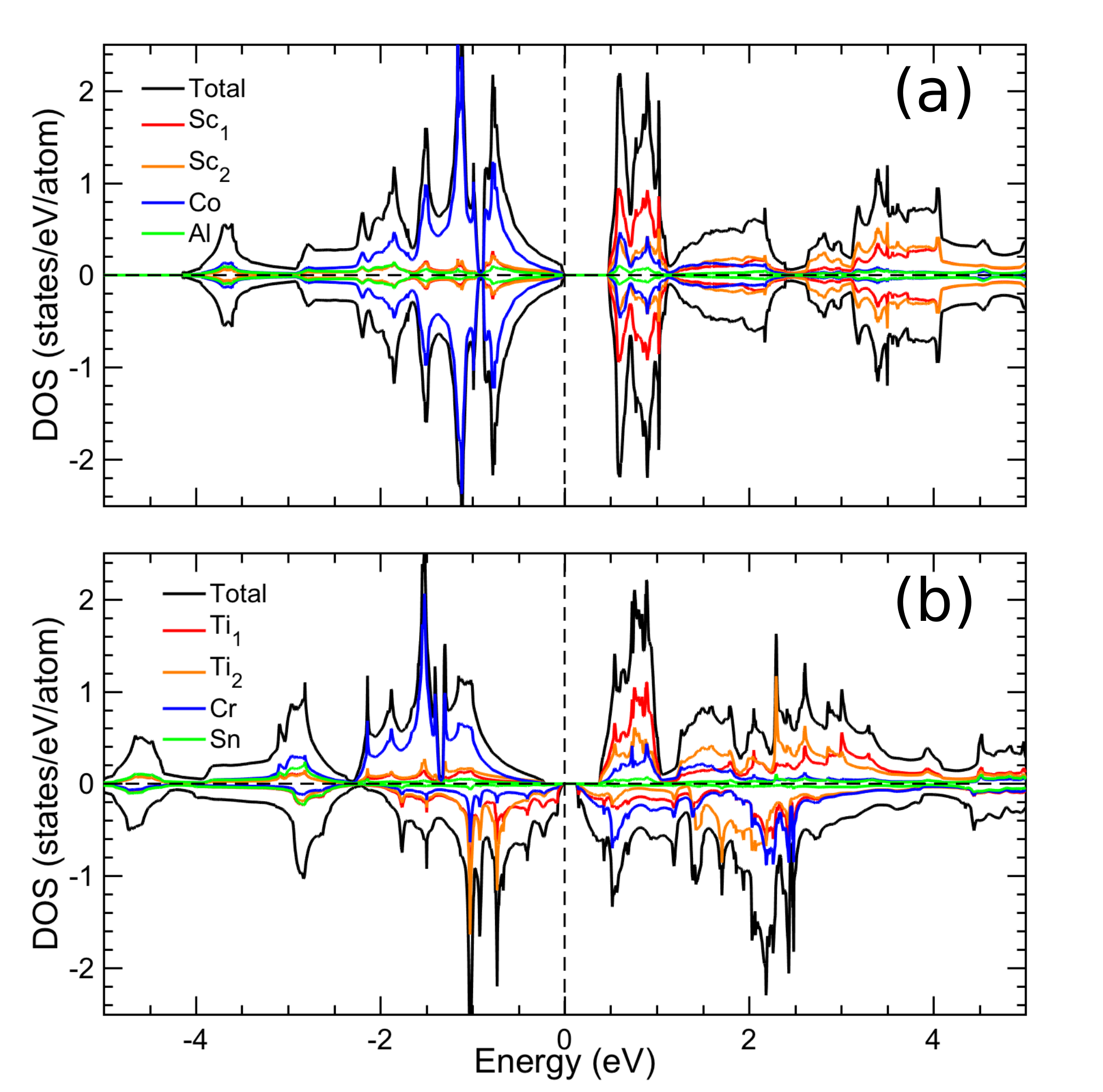}}
        \caption{Density of states curves for the 18-electron semiconductor (a) Sc$_{2}$CoAl and (b) Ti$_{2}$CrSn. The curve presents the total DOS (black) and the projected-DOS contributed by Sc$_{1}$ (Ti$_{1}$) (red), Sc$_{2}$ (Ti$_{2}$) (orange),  Co (Cr) (blue) and Al (Sn) (green). Zero energy corresponds to the Fermi level.}
        \centering
        \label{fig:Ti2CrSn_Sc2CoAl}
\end{figure}

\afterpage{
        \begin{table*}
        \caption{For each of the 14 18-electron \textit{\textit{X$_{2}$YZ}} inverse-Heusler compounds that are non-Slater-Pauling semiconductors with negative formation energy, we list the calculated lattice constant $a$, band gap in two spin channels $E_g^{\uparrow}$, $E_g^{\downarrow}$ within DFT, the type of gap, local moments for atoms on the $X$, $Y$ and $Z$ sites:$m(X_{1})$, $m(X_{2})$, $m(Y)$ and $m(Z)$, the formation energy of the compound in the $XA$ and $L2_{1}$ structures, distance from the convex hull for the $XA$ phase $\Delta E_{\rm HD}^{XA}$, previous experimental reports. ([Legend] Gap type: \orange{D} = direct, \blue{I} = indirect band gap.)}
        \label{tab:SP_semiconductors}
        \begin{tabular}{|l|c|c|c|c|c|c|c|c|c|c|c|c|c|c|}
\toprule
\textit{X$_{2}$YZ} & $a$ & $E_g^{\uparrow}$ & Gap & $E_g^{\downarrow}$ & Gap &  $m(X_{1})$ & $m(X_{2})$ & $m(Y)$ & $m(Z)$    &$\Delta E_f^{XA}$ & $\Delta E_f^{L2_1} $ & $\Delta E_{\rm HD}^{XA}$  \\
 & ({\AA}) & (eV) & type & (eV) & type &  \multicolumn{4}{c|}{($\mu_{B}$)} & \multicolumn{3}{c|}{(eV/atom)}   \\
\midrule
Sc$_{2}$MnP  & 6.18  & 0.052 &  \orange{I} & 0.052 &  \orange{I}  & 0 & 0 & 0 & 0& -0.286 & -0.471  & 0.631   \\ 
Sc$_{2}$FeSi  & 6.255  & 0.347 &  \orange{I} & 0.347 &  \orange{I}  & 0 & 0 & 0 & 0 & -0.292 &  -0.416 & 0.311   \\ 
Sc$_{2}$FeGe  & 6.33  & 0.356 &  \orange{I} & 0.356 &  \orange{I} & 0 & 0 & 0 & 0  & -0.262 & -0.435  & 0.371   \\ 
Sc$_{2}$FeSn & 6.60  & 0.337  &  \orange{I} & 0.337  &  \orange{I} & 0 & 0 & 0 & 0 & -0.230 & -0.358  & 0.319   \\ 
Sc$_{2}$CoAl & 6.42  & 0.475 &  \orange{I} & 0.475 &  \orange{I}  & 0 & 0 & 0 & 0 & -0.280 & -0.350  & 0.130   \\ 
Sc$_{2}$CoGa & 6.36  & 0.532 &  \orange{I} & 0.532 &  \orange{I} & 0 & 0 & 0 & 0  & -0.328 & -0.454  & 0.159   \\ 
Sc$_{2}$CoIn & 6.61  & 0.470 &  \orange{I} & 0.470  &  \orange{I} & 0 & 0 & 0 & 0  & -0.205 & -0.281  & 0.200   \\ 
Ti$_{2}$VSb & 6.47  & 0.427 &  \orange{I} & 0.119 &  \orange{I}  & -1.236 & -0.521 & 1.881 & -0.009 & -0.215 & -0.208  & 0.162   \\ 
Ti$_{2}$CrSi & 6.11  &  0.475 &  \orange{I} & 0.086  &  \orange{I}  & -1.157 & -0.787 & 2.129 & -0.036 & -0.337 &  -0.385 & 0.230   \\ 
Ti$_{2}$CrGe & 6.20  &  0.522 &  \orange{I} &  0.104 &  \orange{I} & -1.254 & -0.883 & 2.348 & -0.022  & -0.262 &  -0.305  & 0.259   \\ 
Ti$_{2}$CrSn & 6.47  & 0.579  &  \orange{I} & 0.133  &  \orange{I}  & -1.396 & -1.018 & 2.727 & -0.016   & -0.181 & -0.154  & 0.139   \\ 
Ti$_{2}$MnAl & 6.23  & 0.561  &  \orange{I} & 0.043 &  \orange{I}   & -1.202 & -1.067 & 2.606 & -0.043 & -0.268 & -0.302   & 0.088   \\ 
Ti$_{2}$MnGa & 6.20  & 0.603 &  \orange{I} & 0.034 &  \orange{I} & -1.201 & -1.115 & 2.611 & -0.020 & -0.287&  -0.341 & 0.114   \\ 
Ti$_{2}$MnIn & 6.47  & 0.394  &  \orange{I} & 0.042 &  \orange{I}  & -1.336 & -1.266 & 3.043 & -0.020  & -0.100 & -0.05 & 0.092   \\   
\bottomrule
\end{tabular}

\end{table*}
}

The 14 $XA$ semiconductors  with negative formation energy that we found are listed in Table \ref{tab:SP_semiconductors} where the DFT-calculated properties (\textit{i.e.} lattice constant, band gap, gap type, local moments, formation energy, hull distance) are presented. Remarkably, all of these  14 systems  are 18 electron \textit{non-Slater-Pauling} semiconductors with gaps after nine electrons per formula unit (2.25 per atom) per spin channel.  Recall that \textit{Slater-Pauling} gaps (at least according to the nomenclature we prefer) occur after 3 electrons per atom per spin.\

From Table \ref{tab:SP_semiconductors}, it can be seen that all of these semiconductors have either Sc or Ti as $X$ atoms.  This is not surprising (given that they all have 18 electrons per formula unit)\ because the choice of systems that comprise our database (see section \ref{sec:compDetail}) implies that if the number of valence electrons on an $X$ atom exceeds 4, the total number of electrons per formula unit will exceed 18.  

Of the 9 $XA$ compounds with $X$=Sc in our database, 7 are calculated to be non-magnetic semiconductors.  The two that are not semiconductors are Sc$_2$MnAs and Sc$_2$MnSb.  The former is  non-magnetic with a very low DOS pseudogap at the Fermi energy. 

 The absence of a gap in Sc$_2$MnAs  can be traced to a reordering of the states at $\Gamma$.  Between Sc$_2$MnP and Sc$_2$MnAs  a singly degenerate $A_1$ state (band 7) switches places with a triply degenerate $T_2$ state (bands 8,9 and 10). This  $A_1$ state (composed of $s-$contributions from all 4 atoms)  decreases in energy relative to the $d-$states as the lattice expands from Sc$_2$MnP to Sc$_2$MnAs to Sc$_2$MnSb. However, Sc$_2$MnSb becomes magnetic as the lattice is expanded before this reordering occurs so it is predicted to be a magnetic near-half-metal with a tiny moment (see Table \ref{NearHalfMetals}). 

 The reason Sc$_2$MnSb is magnetic while Sc$_2$MnP and Sc$_2$MnAs are not is almost certainly due to the larger lattice constant induced by the larger atom (Sb).  For lattice constants less than approximately 6.50 \AA, Sc$_2$MnSb is predicted to be a non-magnetic semiconductor similar to Sc$_2$MnP.

The DOS of the non-magnetic Sc-based semiconductors in Table \ref{tab:SP_semiconductors} can be understood by referring to Figure \ref{fig:Ti2CrSn_Sc2CoAl}a which shows the site decomposed DOS for Sc$_2$CoAl.  Because Sc is an early transition metal  compared to  Co, its $d-$states must lie higher.  Thus, the ordering of the states is Al-$s$ (in this case forming a narrow band well below the Fermi level and not shown in the figure.) followed by Al-$p$, followed by Co-$d$ and finally Sc$_{1}$-$d$ and Sc$_{2}$-$d$.    The gap after 9 electrons per spin channel occurs from the hybridization of the   Co-$d$ and Sc-$d$ states, or, more generally, between the $Y$-$d$ and $X$-$d$ states.

Thus, bonding in these systems appears to have a considerable ionic component with transfer of electrons to the Al and Co atoms from the Sc atoms.It can be seen from Table \ref{tab:SP_semiconductors} that all of the $X$=Sc 18-electron semiconductors are predicted to be substantially more stable in the $L2_1$ phase than in the $XA$ phase.   In addition, the $XA$ hull distances all exceed our 0.052 eV threshold for potentially being stable in equilibrium.

Of the nine 18 electron compounds in our database with $X$=Ti, seven are predicted to be semiconductors and are included in Table \ref{tab:SP_semiconductors}. One, Ti$_2$VP is predicted to be  a near half-metal (see Table \ref{NearHalfMetals}) with a tiny moment.  The final compound, Ti$_2$VAs, is predicted to be a magnetic semiconductor similar to Ti$_2$VSb, but its formation energy is greater than zero and so is omitted from Table \ref{tab:SP_semiconductors}.

In contrast to the non-magnetic 18 electron semiconductors with $X$=Sc, the systems with $X$=Ti develop magnetic moments and have gaps of different size in the majority and minority spin-channels.  The net moment per formula unit is predicted to be zero for the semiconductors, with the Ti moments having the same sign and being balanced by a larger moment of opposite sign on the Y site.  If the sign of the moment on the $Y$ atom is taken to be positive, then the larger gaps are in the majority channel while the smaller gaps are in the minority channel, becoming slightly negative for  Ti$_2$VP.

The larger gap (in the majority channel in Table \ref{tab:SP_semiconductors}) for the Ti$_2YZ$ magnetic semiconductors is very similar to the gaps in both channels for the Sc$_2$$YZ$ non-magnetic semiconductors.  The positive moment on the $Y$ atom shifts its majority $d-$level down while negative moments on the Ti atoms shift their majority $d$-levels up, the larger moment on the Ti$_1$ giving it a slightly larger upward shift.  As a consequence, the state ordering in the majority channel is $Z$-$s$, $Z$-$p$, $Y$-$d$, Ti$_2$-$d$, Ti$_1$-$d$ so that the majority gap after 9 states per formula unit arises from hybridization between $Y$-$d$ and $X_2$-$d$ and  $X_1$-$d$   similar to the case for both channels in the Sc$_2YZ$ 18 electron semiconductors.

The persistence of the very small gaps after 9 states per formula unit in the minority channel is surprising and seems to be aided by avoided band crossings along $\Gamma$ to $K$ and along $L$ to $W$.  We find these small gaps to be common when the $d-$levels of the three transition metals are similar.   The avoided crossings and hence the gaps are likely symmetry related since the non-magnetic compound V$_2$VGa has several allowed band crossings that are avoided when the symmetry between $X_1$, $X_2$ and $Y$ atoms is broken.  These avoided band crossings lead to the small gap after 9 states and are quite common when $X_1$, $X_2$ and $Y$ are similar, but not identical in one of the spin channels.  For example, simple electron counting ignoring charge transfer and using the calculated moments yields for Ti$_2$MnGa, that the number of electrons for the minority channel on each atom are approximately 2.6 for $X_1$ and $X_2$ and 2.2 for $Y$.  The corresponding minority gap is tiny, 0.034eV.

The atom-projected DOS of Ti$_{2}$CrSn is presented in Fig. \ref{fig:Ti2CrSn_Sc2CoAl}. The energy ordering of the atomic orbitals is Sn-$s$ (in this case forming a narrow band more than 10 eV below the Fermi level and not shown in the figure.) followed by Sn-$p$, followed in the majority channel by Cr-$d$ and then the closely spaced Ti$_{2}$-$d$ and Ti$_{1}$-$d$. In the minority channel the ordering of the $d$-states is different, Ti$_1$-$d$, Ti$_2$-$d$, followed by Cr-$d$.  In the majority channel, the hybridization between the lower-lying  Cr-$d$ and the  higher Ti-$d$ states produces a relatively large gap (0.579eV), while in the minority channel, the relatively closely spaced Cr-$d$ and Ti-$d$ states creates a smaller gap (0.133eV). 

Ti$_{2}$CrSn might have interesting spin-dependent transport and tunneling properties if it can be synthesized.  The calculated hull distance for the compound is 0.139eV/atom meaning that the OQMD database contains phases or a combination of phases (in this case a mixture of binaries) that is lower in energy than our 0.052 eV/atom empirical threshold.

We found no  24-electron Slater-Pauling semiconductors among the $XA$ compounds that we investigated.   Of the 405 $XA$ compounds that we studied, 24 had a valence electron count of 24.  None were semiconducting, though 6 were half-metallic with zero moment or near half-metallic with a very small moment.  The reason for the absence of 24-electron $XA$ semiconductors is relatively simple.  As we shall discuss in the next section, the Slater-Pauling gap in the $XA$ phase requires formation of  significant magnetic moments to create a significant difference between  the $d-$onsite energies of the  two $X$ atoms.   When moments form, the two spin channels will generally be different and a semiconductor becomes unlikely. The interesting exception to this rule are the 18-electron semiconductors with moments, but zero net moment in which even very small differences between the $d-$onsite energies of the $X$ atoms can effectuate a small gap.
All of the semiconductors in Table \ref{tab:SP_semiconductors}, have hull distances that exceed our stability threshold criterion of 0.052 eV/atom. 

\subsubsection{Gaps, Half-metals and ``Near'' Half-metals}

\begin{figure*}[t]
        \includegraphics[width=6.95in]{{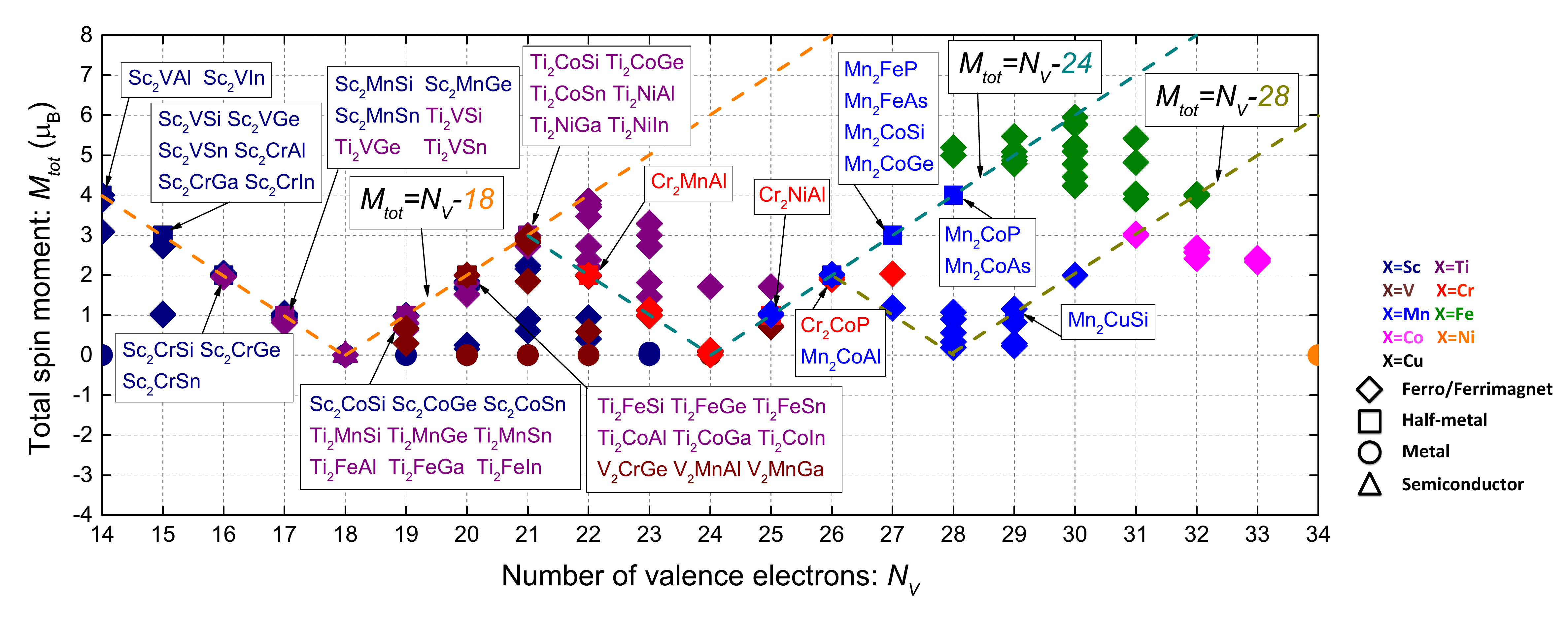}}
        \caption{Calculated total magnetic moment per unit cell as a function of the total valence electron number per unit cell for 254 $XA$ compounds with negative formation energies. The orange dash-dot line represents  $|M_{tot}|=N_{V}-18$, and the purple dash-dot line represents the Slater-Pauling rule: $|M_{tot}|=N_{V}-24$. The compounds from Table \ref{tab:half_metallic_ferromagnets} are listed in the boxes. All of these compounds follow the rules precisely. We used different colors to label different $X$-based inverse-Heusler compounds. We  used diamond, cubic, circle and triangle symbols to label  ferromagnets/ferrimagnets, half-metals, metals and semiconductors, respectively.}
        \centering
        \label{slaterpauling}
\end{figure*}

Although   a large number of $XA$ phase half-metals have been predicted based on theoretical simulations\cite{PhysRevB.87.024420}, we found that most of them do not remain in the $XA$ phase after relaxation, or they have lower energies in the $L2_{1}$ or the tetragonal `$L2_{1}$' phase phase.  However, in order to better understand the $XA$ phase, we have listed in Tables \ref{tab:half_metallic_ferromagnets} and \ref{tab:near_half_metallic_ferromagnets} all of the half-metals and near half-metals found in our survey of 405 systems  for which the calculated formation energy was negative.

In this section we will briefly discuss each of the 405 $XA$ compounds discussing gaps in their DOS, and whether they are half-metals or near half-metals.  The designation of half-metal is binary. If the Fermi energy lies in a gap for one of the spin-channels, but not for the other, then it is a half-metal.  There may be other important quantitative questions such as the distances between the Fermi energy and  the band edges, but the calculated band structure either corresponds to a half-metal or it does not. It should be noted that the determination of whether a given inverse-Heusler has a gap is not always trivial if the gap is small because the Brilliouin zone is sampled at a finite number of points.  We used both the DOS calculated using the linear tetrahedron method and and examination of the bands along symmetry directions to decide difficult cases.

In contrast to the binary choice between half-metal or not, the designation of a ``near-half-metal"  is more qualitative.  There are a couple of ways a calculated DOS could be ``almost" half-metallic; (a) the Fermi energy could fall at an energy for which the DOS for one of the spin channels is very small, i.e. a pseudogap, (b) the DOS could have a gap in one of the spin channels with the Fermi energy lying slightly outside the gap. If we choose criterion (a) we must choose a threshold for ``very small" and if we choose (b) we must choose a threshold for ``slightly".  Criterion (a) is difficult to implement because there are an infinite number of ways the DOS can vary near a pseudogap, e.g. a pseudogap may be wide and only fairly small or it might be narrow but extremely small or some combination.  Another issue with criterion (a) is that a low density of states is often associated with a highly dispersive band that leads to a large conductivity. For these reasons, we choose criterion (b), i.e. we require that there be an actual gap and that the Fermi energy fall ``near" this gap. The (arbitrary) criterion we choose for ``near" is 0.2 electron states per formula unit.  Other criteria could be used, e.g. the distance from the Fermi energy to the nearest band edge.  Our choice of using the number of states has the advantage of being easily related to the magnetic moment.  It has the disadvantage that occasionally a system with a highly dispersive band above or below the Fermi energy with a very small density of states can lead to a situation in which we label a system near-half-metallic when the Fermi energy is as much as 0.3 eV into the band. 

For each of the systems, we note the gaps in the DOS, typically after 9, 12 or 14 states per formula unit per spin channel and whether the Fermi energy falls in one or more of the gaps or close enough  to the gap (0.2 states)  to be designated a near-half-metal. Simple state and electron counting dictates that to make a half-metal by taking advantage of a gap after $N_s$ states when there are $N$ electrons per formula unit, the magnetic moment per formula unit must be $N-2N_s$.

\textbf{13 electron systems:} Our dataset contained 3 systems with 13 valence electrons, Sc$_2$Ti(Al,Ga,In).  All three were magnetic metals with  small  gaps after 9 states per spin channel
in both channels (except for Sc$_2$TiIn minority).  However, the moments were not large enough (a total moment of 5$\mu_B$ per formula unit would have been needed) to move the majority Fermi energy into the gap.

\textbf{14 electrons systems: }Our dataset contained 6 systems with 14 valence electrons, Sc$_2$Ti(Si,Ge,Sn)\ and Sc$_2$V(Al,Ga,In).  These systems require a moment of 4$\mu_B$ to place the Fermi energy in a majority gap after 9 electrons per spin channel.

Sc$_2$Ti(Si,Ge,Sn): Sc$_{2}$TiSi and Sc$_{2}$TiGe are predicted to be non-magnetic metals and to have small gaps after 9 states in both spin channels.
Sc$_2$TiSn is predicted to be magnetic and to have a majority gap after 9 states, but its moment closer to 3 than 4$\mu_B$.   

Sc$_2$V(Al,Ga,In): These all have relatively large gaps after 9 states in the majority channel.   The position of the Fermi energy rises relative to the gap as the atomic size of the $Z$ atom increases the lattice constant, facilitating the formation of large moments on V and even small moments on Sc. For Sc$_2$VAl and Sc$_2$VGa, the moment is not quite large enough to place the Fermi energy in the gap making them near half-metals (Table \ref{tab:near_half_metallic_ferromagnets}). However, the lattice constant of Sc$_2$VIn \textit{is} large enough to support a moment large enough to move the Fermi energy into the gap so it is a half-metal (Table \ref{tab:half_metallic_ferromagnets}).  
 
\textbf{15 electron systems:} Our dataset contained 9 systems with 15 valence electrons, Sc$_2$Ti(P,As,Sb), Sc$_2$V(Si,Ge,Sn)\ and Sc$_{2}$Cr(Al,Ga,In).    A moment of 3$\mu_B$  is needed to place the Fermi energy in a majority gap after 9 states. 

Sc$_2$Ti(P,As,Sb): These systems do not have gaps after 9 states per atom in either channel because of a reordering of the states at $\Gamma.$  A singly degenerate state, usually band 10 at $\Gamma,$ interchanges with a triply degenerate T$_2$ state  so that bands 8, 9, and 10 are associated with the triply degenerate T$_2$ state, precluding a gap. 

Sc$_2$V(Si,Ge,Sn):  These systems have majority gaps after 9 states.  The Fermi energy is in the gap and they are classified as half-metals as described in 
Table \ref{tab:half_metallic_ferromagnets}.

Sc$_{2}$Cr(Al,Ga,In): These systems all have large gaps after 9 states in majority and small gaps or pseudogaps near 9 states in minority.  The Fermi energy falls in the majority gap for all three so they are classified as half-metals and described in 
Table \ref{tab:half_metallic_ferromagnets}.

\textbf{16 electron systems:} Our dataset contained 12 systems with 16 valence electrons per formula unit, Sc$_2$V(P,As,Sb), Sc$_2$Cr(Si,Ge,Sn), Sc$_2$Mn(Al,Ga,In),  \ and Ti$_2$V(Al,Ga,In).  For 16 valence electrons per formula unit and a gap after 9 states in the majority spin channel, a moment of 2$\mu_B$ is needed to place the Fermi energy in the gap to make a half-metal.

Sc$_2$V(P,As,Sb):
Sc$_2$VP and Sc$_2$VAs  have  a region with an extremely low but non-zero DOS at the Fermi energy.  These pseudogaps can be be traced to a band reordering similar to that mentioned previously in which a singly degenerate state at $\Gamma$ drops below a triply degenerate state so that the
triply degenerate state connects to bands 8, 9 and 10, thus precluding a gap.  In the case of Sc$_2$VSb, the singly degenerate state is a few hundredths of an eV above the triply degenerate state
so it is a near half-metal with a tiny gap.   

Sc$_2$Cr(Si,Ge,Sn):
These systems have large majority gaps after 9 states. They generated half-metals as described in Table \ref{tab:half_metallic_ferromagnets}. 

Sc$_2$Mn(Al,Ga,In): The three systems with $X$=Sc and $Y$=Mn have large majority and tiny minority gaps after 9 states.  Their moments are very slightly less than 2 $\mu_B$ so they generated near half metals as described in Table \ref{tab:near_half_metallic_ferromagnets} by placing the Fermi energy at the valence edge of the majority gap.

Ti$_2$V(Al,Ga,In): These systems\ have tiny majority and minority gaps after 9 states per spin channel. All take advantage of the majority gap to generate near half-metals as shown in Table \ref{tab:near_half_metallic_ferromagnets}.   Ti$_2$VIn is not included in the table because it has a positive formation energy.

\textbf{17 electron systems:} Our dataset contained 15 systems that had 17 electrons per formula unit, Sc$_2$Cr(P,As,Sb), Sc$_2$Mn(Si,Ge,Sn), Sc$_2$Fe(Al,Ga,In), Ti$_2$V(Si,Ge,Sn), and Ti$_2$Cr(Al,Ga,In).
 Systems with 17 electrons per formula unit can generate half-metals by taking advantage of a gap after 9 states in majority if their magnetic moment per formula unit is 1$\mu_B$.

Sc$_2$Cr(P,As,Sb):\ Sc$_2$CrP and Sc$_2$CrAs both have majority pseudogaps at the Fermi energy.  The gap after 9 electrons per formula unit is converted into a pseudogap by a band reordering at $\Gamma$  similar to that occurring for  Sc$_2$V(P,As)\ and Sc$_2$Ti(P,As,Sb). Sc$_2$CrSb has a ``normal" band ordering at $\Gamma$ and is a near half-metal similar to Sc$_2$VSb, but with a much larger gap.

Sc$_2$Mn(Si,Ge,Sn): These systems\ have majority gaps after 9 states and are all half-metals as described in Table \ref{tab:half_metallic_ferromagnets}. In addition to the majority gap Sc$_2$MnSi
has a gap after 9 states in minority. A pseudogap near 9 states in minority for Sc$_2$MnSi can also be identified.

Sc$_2$Fe(Al,Ga,In): These systems\ have gaps after 9 states in both majority and minority channels. All three take advantage of the majority gap to generate near-half-metals as described in Table \ref{tab:near_half_metallic_ferromagnets}.
Sc$_2$FeAl also has a minority pseudogap near 12 states and a majority pseudogap near 14 states.

Ti$_2$V(Si,Ge,Sn):  These all have gaps after 9 states in both majority and minority with the majority gap being larger.  All three compounds utilize the majority gap to form half metals as described in Table \ref{tab:half_metallic_ferromagnets}.

Ti$_2$Cr(Al,Ga,In): These  have gaps after 9 states in both spin channels.  They are all near half-metals (Table \ref{tab:near_half_metallic_ferromagnets}), having moments just large enough to satisfy our criterion.  The large moment on the Cr atom opens a large gap in the majority channel (hybridization between Cr and Ti-$d$ and with the Cr $d$-onsite energy well below that of the Ti atoms).  The gap in minority is smaller and is dependent on the difference between the two Ti potentials and avoided band crossings.

\textbf{18 electron systems:} Our dataset contained 18 systems with 18 valence electrons. These were described in Section \ref{SemiSec}.

\textbf{19 electron systems:} Our dataset contained 21 systems with 19 valence electrons per formula unit, Sc$_2$Fe(P,As,Sb), Sc$_2$Co(Si,Ge,Sn), Sc$_2$Ni(Al,Ga,In), Ti$_2$Cr(P,As,Sb), Ti$_2$Mn(Si,Ge,Sn), Ti$_2$Fe(Al,Ga,In), and V$_2$Cr(Al,Ga,In).  Systems with 19 valence electrons need a moment of 1$\mu_B$ to place the Fermi energy in a gap after 9 states in the minority channel. 

Sc$_2$Fe(P,As,Sb): Sc$_2$FeP is non-magnetic with gaps after 9 states in both spin channels. Sc$_2$FeAs is also non-magnetic, but the gaps after 9 states are converted to pseudogaps by the inversion of  singly and triply degenerate states at $\Gamma$ similar to Sc$_2$CrP and Sc$_2$CrAs.  The larger lattice constant of Sc$_2$FeSb allows it to develop a moment.  There are sizeable gaps after 9 states in both the minority and majority channels, but the moment is not quite large enough (0.79$\mu_B$) to meet our criterion for calling it a near-half-metal.
}

  Sc$_2$Co(Si,Ge,Sn): All three of these are half-metals with sizable gaps after 9 states in both spin channels (Table \ref{tab:near_half_metallic_ferromagnets}). The Fermi energy falls near the top of the gap,  in all three cases but especially for Ge.

Sc$_2$Ni(Al,Ga,In): All three have sizable gaps after 9 states in both spin channels and moments as given in Table \ref{tab:near_half_metallic_ferromagnets} that qualify them as near-half-metals.

Ti$_2$Cr(P,As,Sb): All three compounds have gaps in the minority channel after 9 states, but only and Ti$_2$CrAs and Ti$_2$CrSb have sufficiently large moments to be classified as near-half-metals.   (Ti$_2$CrAs only barely made the cut-off.)  Ti$_2$CrSb also has a tiny gap after 9 states in majority.

Ti$_2$Mn(Si,Ge,Sn): All three of these compounds have sizable gaps after 9 states in minority and much smaller gaps in majority.  They also have pseudogaps near 12 states in minority.  For all three, the Fermi energy falls in the gap so they are are included in Table \ref{tab:near_half_metallic_ferromagnets}.

Ti$_2$Fe(Al,Ga,In): All three of these compounds have gaps after 9 states in both channels and peseudogaps in both  channels after 12 states.  The gaps after 9 states are much smaller in majority than in minority.  The Fermi energy for each of the three falls in the minority gap creating three half-metals as described in Table \ref{tab:half_metallic_ferromagnets}.

V$_2$Cr(Al,Ga,In): All three of these systems have gaps after 9 states in both spin channels.  In all cases, however, the magnetic moment is not large enough to meet our criterion for a near-half-metal. The formation energy is small and negative for V$_2$CrAl and V$_2$CrGa.  It is positive for V$_2$CrIn.

\textbf{20 electron systems:} Our dataset contained 24 compounds with 20 electrons per formula unit, Sc$_2$Co(P, As, Sb), Sc$_2$Ni(Si, Ge, Sn), Sc$_2$Cu(Al, Ga, In), Ti$_2$Mn(P, As, Sb), Ti$_2$Fe(Si, Ge, Sn), Ti$_2$Co(Al,Ga,As), V$_2$Cr(Si, Ge, Sn)\ and V$_2$Mn(Al, Ga, In).  

Sc$_2$Co(P,As,Sb): Sc$_2$CoP and Sc$_2$CoAs both have gaps after 9 states per formula unit in both spin channels, but in both cases the net magnetic moment is much too small to pull the Fermi energy into one of the gaps.  Sc$_2$CoSb, on the other hand, due to its larger lattice constant supports a moment sufficiently large (1.999$\mu_B$) that the Fermi energy falls at the top edge of the minority gap.  We classify it as a near half-metal (Table \ref{tab:near_half_metallic_ferromagnets}).

Sc$_2$Ni(Si,Ge,Sn): The compounds, Sc$_2$Ni(Si,Ge,Sn), all have gaps after 9 electrons in both spin channels.  However, the total moment (1.80$\mu_B$, 1.81$\mu_B$, 1.84$\mu_B$ respectively) does not reach the value (2.00$\mu_B$) needed to place the Fermi energy in the gap. Even when the lattice is expanded the magnetic moment of these Sc$_2$Ni(Si,Ge,Sn) compounds hardly increases because the  d-bands of Ni are filled so that Ni only supports a very small moment in these materials.
Nevertheless, these compounds meet (barely)\ our threshold to be called near-half-metals (Table \ref{tab:near_half_metallic_ferromagnets}).

Sc$_2$Cu(Al,Ga,In):
The systems Sc$_2$Cu(Al,Ga,In) all have pseudogaps after 9 states in both spin channels. 

Ti$_2$Mn(P,As,Sb):All three of these compounds have gaps after 9 electrons in both spin-channels.  However only Ti$_2$MnSb (due to its larger lattice constant) has a moment (1.989$\mu_B)$ close enough to 2$\mu_B$ to be classified a ``near-half metal".
Ti$_2$MnAs
(1.7967$\mu_B$)
barely misses the cut-off to be included in Table \ref{tab:near_half_metallic_ferromagnets}.

Ti$_2$Fe(Si, Ge, Sn):
The compounds Ti$_2$Fe(Si,Ge,Sn) all have gaps in both channels after 9 states and pseudogaps after 12 states.  The Fermi energy falls within the gap in all three cases so they are\ predicted to be half-metals as described in Table \ref{tab:half_metallic_ferromagnets}.

 Ti$_2$Co(Al,Ga,As): The compounds  Ti$_2$Co(Al,Ga,In) have large minority gaps after 9 states and somewhat smaller gaps after 9 states in majority.  The Fermi energy falls within the minority gap in all three cases so they are  are predicted to be half-metals as described in Table \ref{tab:half_metallic_ferromagnets}.

 V$_2$Cr(Si,Ge,Sn):V$_2$CrSi has small gaps after 9 states in both spin channels, but is not magnetic.  V$_2$CrGe is magnetic with a moment of 2$\mu_B$.  The Fermi energy falls very close to the top of the gap, but V$_2$CrGe is predicted to be a half-metal as described in Table  \ref{tab:half_metallic_ferromagnets}.  V$_2$CrSn has a gap after 9 states in the majority channel which is not useful for generating a half-metal. There is no gap in minority.

V$_2$Mn(Al, Ga, In): V$_2$MnAl has a tiny majority gap and a sizable minority gap after 9 states.  It also has pseudogaps after 12 states in majority and 14 states in minority.  It takes advantage of the minority gap after 9 states to become a half-metal.  The DOS for V$_2$MnGa is similar.  It is also a half-metal.  Both compounds are included in Table \ref{tab:half_metallic_ferromagnets}. V$_2$MnIn has a gap after 9 states in the majority, but only a pseudogap after 9 states in minority.   Its large moment, in excess of 3$\mu_B,$ places the Fermi energy well below this pseudogap.

\textbf{21 electron systems:} Our database contains 27 systems with 21 valence electrons, Sc$_2$Ni(P,As,Sb), Sc$_2$Cu(Si,Ge,Sn), Sc$_2$Zn(Al,Ga,In), Ti$_2$Fe(P,As,Sb), Ti$_2$Co(Si,Ge,Sn), Ti$_2$Ni(Al,Ga,In), V$_2$Cr(P,As,Sb), V$_2$Mn(Si,Ge,Sn), V$_2$Fe(Al,Ga,In).  Systems with 21 valence electrons are interesting because they can, in principle, place the Fermi energy in a gap after 9 electrons or a gap after 12 electrons or both.
 If a majority gap present after 12 states is present as well as a minority gap after 9 states, a moment of 3$\mu_B$ would yield a magnetic semiconductor.

 Sc$_2$Ni(P,As,Sb): Sc$_2$NiP has tiny gaps in both spin channels after 9 states while Sc$_2$NiAs has pseudogaps after 9 states in both spin channels. Both have very small moments. Sc$_2$NiSb has gaps after 9 states in both spin channels, but its moment (2.15$\mu_B$) is too small to place the Fermi energy in or near the gap. 

Sc$_2$Cu(Si,Ge,Sn):
All three of these compounds have gaps after 9 states in minority and much smaller gaps in majorty.   However their moments are too small to place the Fermi energy in the minority gap.

 Sc$_2$Zn(Al,Ga,In):
These three non-magnetic compounds only have pseudogaps after 9 states in both spin channels because of a highly dispersive band 10 which drops below band 9 along $\Gamma$ to $X$.
 
 None of the 21-electron systems with $X$=Sc  generate half-metals or near half metals because a moment of 3$\mu_B$ would be necessary to take advantage of a gap after 9 states per spin per formula unit or to take advantage of a gap after 12 states per spin per formula unit.  The compounds are not able to do this because Sc is difficult to polarize and Ni, Cu and Zn cannot support a large moment because their d-bands are filled or (in the case of Ni)\ nearly filled.

Ti$_2$Fe(P,As,Sb): These three compounds have large gaps after 9 states in the minority channel and smaller gaps in the majority channel after 9 states. Ti$_2$FeP and  Ti$_2$FeAs have pseudogaps after 12 states in the maority which becomes a very small real gap for  Ti$_2$FeSb.  The moments for the three are 2.72$\mu_b$, 2.94$\mu_b$ and 2.99$\mu_b$ respectively.  Ti$_2$FeAs and Ti$_2$FeSb meet our criterion for near half-metallicity.  If the moment for Ti$_2$FeSb were very slightly larger, we would predict it to be a magnetic semiconductor with a large minority gap and a tiny majority gap, i.e. a 9/12 magnetic semiconductor.  The present prediction is for a magnetic semi-metal that has a pocket of down-spin electrons along $\Gamma$ to $K$ and a $\Gamma$-centered pocket of up-spin holes.  

Ti$_2$Co(Si,Ge,Sn): These three compounds all have gaps in both spin channels after 9 states, with the minority gap being larger. Ti$_2$CoSi also has a majority pseudogap after 12 states.  For all three, the Fermi energy falls within the minority gap giving each a moment of 3$\mu_B$ and making them half-metals (Table \ref{tab:half_metallic_ferromagnets}). For Ti$_2$CoSi
the current at zero K would be carried by up-spin holes at $\Gamma$ and up-spin electrons at $X$.

Ti$_2$Ni(Al,Ga,In): These three compounds all have gaps after 9 states in both spin channels.  The Fermi energy falls well inside the minority gaps yielding half metals with moments of 3$\mu_B$ in each case (Table \ref{tab:half_metallic_ferromagnets}).

V$_2$Cr(P,As,Sb): V$_2$CrP has gaps after 9 states in both spin channels, but is non-magnetic. V$_2$CrP has a gap in the minority channel after 9 states and a pseudogap after 9 states in the majority channel.  Its moment is too small (1.85$\mu_B$) for it to be classified as a near-half-metal. V$_2$CrSb has a pseudogap after 9 states in the majority channel and a sizable gap after  12 states also in majority chanel.  Its moment is large enough (2.89$\mu_B$) that we would classify it as a near-half-metal.  However, its formation energy is positive so it is not included in Table \ref{tab:near_half_metallic_ferromagnets}.      

V$_2$Mn(Si,Ge,Sn): V$_2$MnSi has pseudogaps after 9 states in both spin channels and is non-magnetic. V$_2$MnGe has gaps after 9 states in both spin channels and a pseudogap after 12 states in majority.  Its moment of 2.81$\mu_B$ qualifies it (barely)\ for inclusion with the near half-metals in Table \ref{tab:near_half_metallic_ferromagnets}.  V$_2$MnSn has majority gaps after 9 and 12 states and a pseudogap after 9 states in minority.  It's moment of 2.96$\mu_B$ qualifies it for inclusion in Table \ref{tab:near_half_metallic_ferromagnets} as a Slater-Pauling near-half-metal. 

V$_2$Fe(Al,Ga,In): V$_2$FeAl has gaps after 9 and 14 states in minority and after 12 states in majority.  Its moment of 2.91$\mu_B$ qualifies it as a ``double near-half-metal" since its Fermi energy is near gaps after 9 states in minority and 12 states in majority. V$_2$FeGa has a gap after 9 states in minority and a pseudogap after 12 states in majority.  Its moment of 2.85$\mu_B$
meets the threshold for inclusion in Table  \ref{tab:near_half_metallic_ferromagnets} as a near-half-metal. V$_2$FeIn has gaps in both spin channels after 9 states and a pseudogap after 12 states in majority.  Its moment is large enough for inclusion for inclusion in Table  \ref{tab:near_half_metallic_ferromagnets}, but its formation energy is positive.  

\textbf{22 electron systems:}  Our database contains 27 systems with 22 electrons, Sc$_2$Cu(P,As,Sb), Sc$_2$Zn(Si,Ge,Sn), Ti$_2$Co(P,As,Sb), Ti$_2$Ni(Si,Ge,Sn), Ti$_2$Cu(Al,Ga,In), V$_2$Mn(P,As,Sb), V$_2$Fe(Si,Ge,Sn), V$_2$Co(Al,Ga,As), and Cr$_2$Mn(Al,Ga,In). These systems would need a moment of 4$\mu_B$ to place the Fermi energy into a gap after 9 states in the minority channel or a moment of 2$\mu_B$ to place the Fermi energy into a gap after 12 states in the majority channel.

Sc$_2$Cu(P,As,Sb): The systems with $X$=Sc and $Y$=Cu generate only small moments (0.40, 0.62 and 0.79)$\mu_B$ for $Z$=P, As, and Sb respectively.  They have gaps in both channels after 9 states and no gaps after 12 states.  

Sc$_2$Zn(Si,Ge,Sn):  The systems with $X$=Sc and $Y$=Zn also generate moments that are too small (0.55, 0.92 and 0.93$\mu_B$ for Si, Ge, and Sn respectively). These systems have pseudogaps after 9 states in both spin channels, but no gaps after 12 states. 

Ti$_2$Co(P,As,Sb): These three systems have gaps after 9 states in both spin channels, but their moments (2.20, 2.38, and 2.73)\ are much smaller than the 4$\mu_B$ needed to place the Fermi energy in the minority gap.

 Ti$_2$Ni(Si,Ge,Sn): These three systems have gaps after 9 states in both spin channels.  Their moments are larger than those of the Ti$_2$Co(P,As,Sb) systems (3.47, 3.69 and 3.86) for Si, Ge and Sn respectively. Thus, Ti$_2$NiSn meets our threshold for inclusion in Table \ref{tab:near_half_metallic_ferromagnets} as a near-half-metal.

Ti$_2$Cu(Al,Ga,In):
Ti$_2$CuAl has gaps after 9 states in both spin channels, but its moment is too low (3.72$\mu_B$) to meet our criterion for near-half-metal status. Ti$_2$CuGa and Ti$_2$CuIn have gaps after 9 states in majority, but only pseduogaps after 9 states in minority.

The 22 electron systems with $X$=Ti and $Y$=Co, Ni or Cu tend to have gaps after 9 states per spin channel per formula unit in both spin channels. They do not have gaps after 12 states per spin channel, presumably because insufficient contrast can be generated between the $d$-onsite energies of the two Ti atoms.   In most cases the moments are too small to place the Fermi energy in the minority gap after 9 states.   The moments are largely on the Ti atoms and the moments tend to  increase with the atomic number of the $Y$ atom, increasing from Co to Ni to Cu, and with the size of the $Z$ atom.

V$_2$Mn(P,As,Sb): V$_2$MnP has gaps after 9 states in both spin channels but a very small moment (0.59$\mu_B$) so it cannot take advantage of the minority gap to make a half-metal. V$_2$MnAs also has gaps after 9 states in both channels and a pseudogap after 12 states in the majority.  Its moment of 1.97$\mu_B$ places the Fermi energy close to the majority pseudogap after 12 states, but the absence of a real gap precludes its designation as a near-half-metal. V$_2$MnSb has a majority gap after 9 states, a minority pseudogap after 9 states and a majority gap after 12 states.  Its moment of 1.9987$\mu_B$ places the Fermi energy just below the majority gap after 12 states.  Only its positive formation energy precludes its inclusion in Table \ref{tab:near_half_metallic_ferromagnets}. 

V$_2$Fe(Si,Ge,Sn): All three compounds have gaps after 9 states in both spin channels.  V$_2$FeGe also has a majority  gap after
12 states and V$_2$FeSn has a pseudogap after 12 states in the majority channel.  The moment of V$_2$FeGe (1.98$\mu_B$) places its Fermi energy quite close to the majority gap so it is a near-half-metal as described in Table \ref{tab:near_half_metallic_ferromagnets}.

V$_2$Co(Al,Ga,As): All three of these compounds have gaps in both channels after 9 states with the minority gap being much larger than the small majority gap.  All three also have pseudogaps after 12 states in the majority. The pseudogap for V$_2$CoAl is particularly pronounced.  The moment is slightly less than 2$\mu_B$ for all three compounds, but according to our definition, they are not near-half-metals.

Cr$_2$Mn(Al,Ga,In):Cr$_2$MnAl and Cr$_2$MnGa both have gaps after 12 states in the majority channel. Cr$_2$MnIn only has a pseudogap there. Cr$_2$MnAl is a half metal as described in Table \ref{tab:half_metallic_ferromagnets} and Cr$_2$MnGa is a near-half-metal as described in Table \ref{tab:near_half_metallic_ferromagnets}.
All three of these compounds are predicted to have ferrimagnetic states in which the moments of Cr$_1$ and Mn are aligned, and the moment of Cr$_2$ is opposite.

\textbf{23 electron systems:} Our database contains 27 systems with 23 electrons, Sc$_2$Zn(P,As,Sb), Ti$_2$Ni(P,As,Sb), Ti$_2$Cu(Si,Ge,Sn), Ti$_2$Zn(Al,Ga,In), V$_2$Fe(P,As,Sb), V$_2$Co(Si,Ge,Sn), V$_2$Ni(Al,Ga,In), Cr$_2$Mn(Si,Ge,Sn), and Cr$_2$Fe(Al,Ga,In). These compounds can take advantage of a gap after 9 states in minority to make a half-metal, but a moment of 5$\mu_B$ would be needed.  Alternatively, they could take advantage of a gap after 12 states in majority which would only require a moment of 1$\mu_B$.
 A few of these systems have pseudogaps after 14 states.  In principle, one could have a 23 electron inverse-Heusler magnetic semiconductor with a moment of 5$\mu_B$ that uses gaps after 9 states in minority and 14 states in majority; however such solutions did not appear in our database.

Sc$_2$Zn(P,As,Sb):
The three systems Sc$_2$Zn(P,As,Sb) have pseudogaps after 9 states in both channels. They are all non-magnetic and have no gaps after 12 sates.  

 Ti$_2$Ni(P,As,Sb): These compounds all have gaps after 9 states in both spin-channels, but their moments are much too small to make half-metals or near-half-metals.

Ti$_2$Cu(Si,Ge,Sn):
The three systems Ti$_2$Cu(Si,Ge,Sn)\ have gaps  in both channels after 9 and pseudogaps in majority after 14 states.  Their  moments, however (3.00, 3.16, and 3.33)$\mu_B$ \ are much less than the 5$\mu_B$ that would be needed to place the Fermi energy in the
minority gap.

Ti$_2$Zn(Al,Ga,In):
These compounds all have pseudogaps near 9 states per formula unit in the minority channel but no gaps at 9 or 12 states nor is the Fermi energy near the pseudogap.
Ti$_2$ZnAl and Ti$_2$ZnGa also have majority gaps after 9 states and Ti$_2$ZnAl
also has a pseudogap after 14 states in majority.

V$_2$Fe(P,As,Sb):
These compounds   have gaps in both spin channels after 9 states. V$_2$FeP and V$_2$FeAs have pseudogaps in the majority channel after 12 states.   V$_2$FeSb, however does have a majority gap after 12 states and the magnetic moment (0.9996$\mu_B$)
places the Fermi energy at the lower edge of this gap. V$_2$FeSb is not included in Table \ref{tab:near_half_metallic_ferromagnets} because it has a positive formation energy. 

V$_2$Co(Si,Ge,Sn): These compounds have  gaps after 9 states in both channels, but only pseudogaps after 12 states except for V$_2$CoSi which has a tiny gap after 12 states. The Fermi energy falls just below this gap (moment = 0.997$\mu_B$).
(See Table \ref{tab:near_half_metallic_ferromagnets}.) 

V$_2$Ni(Al,Ga,In):
These compounds have sizable gaps after 9 states in both spin channels, but the moments are small ($\approx 1 \mu_B$) and there are no gaps after 12 states in the majority channel, hence no half-metals or near half-metals.

Cr$_2$Mn(Si,Ge,Sn):
Cr$_2$MnSi has a gap after 9 states in the majority channel and a pseudogap after 9 states in minority.  Cr$_2$MnGe has a pseudogap after 9 states in majority and a sizable gap after 12 states in majority.  Its moment of 0.99$\mu_B$  allows it to use this gap to become a near-half-metal (Table \ref{tab:near_half_metallic_ferromagnets}).  Cr$_2$MnSn also has a majority gap after 12 states and a moment of approximately 1.00$\mu_B$ so it is a near half-metal, but it is not included in Table \ref{tab:near_half_metallic_ferromagnets} because its formation energy is positive.

Cr$_2$Fe(Al,Ga,In): Cr$_2$FeAl has minority gaps after 9 states and 14 states, and a majority pseudogap near 12 states. Its moment of approximately 1.00$\mu_B$  places the Fermi energy into this peseudogap.  Cr$_2$FeGa has a minority pseudogap after 9 states and a majority gap after 12 states.  Its moment of 0.98$\mu_B$ makes it a near-half-metal, but it is not included in Table \ref{tab:near_half_metallic_ferromagnets} because its formation energy is positive.
Cr$_2$FeGa has a majority pseudogap near 12 states.

\textbf{24 atom systems:} Our database contained 24 systems with 24 valence electrons, Ti$_2$Cu(P,As,Sb), Ti$_2$Zn(Si,Ge,Sn), V$_2$Co(P,As,Sb), V$_2$Ni(Si,Ge,Sn), V$_2$Cu(Al,Ga,In), Cr2$_2$(P,As,Sb), Cr$_2$Fe(Si,Ge,Sn), Cr$_2$Co(Al,Ga,In).  Based on electron count alone, these systems would offer the opportunity for non-magnetic semiconductors or zero net moment half-metals arising from Slater-Pauling gaps after 12 states per formula unit in one or both spin-channels.  In practice, we find that only four of the 24  systems display Slater-Pauling gaps (and only in one spin-channel).   In all four of these systems the Fermi energy falls at the edge of the gap but just outside so that they would be classified as near half-metals.  Three of these four near half-metals have positive formation energy.

The gaps after 9 states, on the other hand, seem to be more robust.
Only Cr$_2$MnSb failed to show a gap or a pseudogap after 9 states in at least one of the spin channels.  Most systems showed gaps after 9 states in both spin channels.  However, none of the systems were able to generate a moment of 6$\mu_B$ that would have been necessary to use one of the gaps after 9 states to generate a half-metal.

Ti$_2$Cu(P,As,Sb): These systems  showed gaps in both channels after 9 states, but their moments were much too small
to place the Fermi energy near the minority gap.

Ti$_2$Zn(Si,Ge,Sn): These systems  showed pseudogaps in both channels near 9 states per formula unit. The gaps were converted to pseudopgaps by a highly dispersive band 10 along $\Gamma$ to $X$\ dropping down below the maximum of bands 7-9 at $\Gamma$.

The 9 systems in our database with $X$=V all showed gaps after 9 states in both spin channels.  In all cases, the moments were quite small so that the Fermi energy could not fall into these gaps.  Six of the 9 systems, V$_2$Co(P,As,Sb) and V$_2$Ni(Si,Ge,Sn)\ have pseudogaps around 12 states per atom in at least one spin channel.

V$_2$Co(P,As,Sb): These systems show gaps after 9 states and pseudogaps after 12 states in both spin channels.  The moments are too small (V$_2$CoSb) or zero precluding half-metals.

V$_2$Ni(Si,Ge,Sn): V$_2$NiSi has gaps in both spin channels after 9 states and pseudogaps after both spin channels after 12 states. V$_2$NiGe has gaps in both spin channels after 9 states and a marginal pseudogap after 12 states in majority. V$_2$NiSn has gaps after 9 states in both spin channels and pseudogaps after 12 states in minority and after 14 states in majority.  The lack of real gaps after 12 states and their very small moments preclude half-metals.

V$_2$Cu(Al,Ga,In): These compounds have gaps after 9 states in both spin channels and minority gaps after 14 states.  V$_2$CuIn also has a majority gap after 14 states while V$_2$CuAl and V$_2$CuGa have pseudogaps in majority near 14 states.  Their small moments and the absence of gaps after 12 states preclude half-metals. 

Cr$_2$Mn(P,As,Sb): In the minority channel, Cr$_2$MnP has a gap after 9 states and a pseudogap near 12 states.  It also has a pseudogap near 9 states in majority.  It has a small moment placing the Fermi energy in the minority pseudogap near 12 states. Cr$_2$MnAs has minority gaps after 9 and 12 states.  The very small moment (0.005$\mu_B$)places the Fermi energy just below the gap at 12 states.  Thus Cr$_2$MnAs is a near half-metal, but it is not included in Table \ref{tab:near_half_metallic_ferromagnets} because its calculated formation energy is positive. Cr$_2$MnSb has a minority gap after 12 states. Its tiny moment (0.0002$\mu_B$) makes it a near half-metal, but it is not included in Table \ref{tab:near_half_metallic_ferromagnets} because its formation energy is positive.

Cr$_2$Fe(Si,Ge,Sn): Cr$_2$FeSi is non-magnetic and has gaps after 9 states in both channels.  Cr$_2$FeGe has gaps in both spin channels after 9 states and a minority gap after 12 states. Its small moment makes it a near-half-metal, but we do not list it in Table \ref{tab:near_half_metallic_ferromagnets} because of its positive formation energy.  Cr$_2$FeSn has a majority gap after 9 states and a minority gap after 12 states.  Its small moment places the Fermi energy somewhat below this gap yielding a near-half-metal.  It is not listed in Table \ref{tab:near_half_metallic_ferromagnets} because of its positive formation energy. 

Cr$_2$Co(Al,Ga,In): These systems all have gaps after nine states in the majority channel and all have gaps (Al,Ga)\ or pseudogaps (In) in the minority channel
after 12 states. All three have tiny moments. Cr$_2$CoAl
and Cr$_2$CoGa are near half-metals, but only the former has a negative formation energy.  
Cr$_2$CoAl is the only Slater-Pauling 24 electron  near half-metal with negative formation energy in Table \ref{tab:near_half_metallic_ferromagnets}.

\textbf{25 electron systems:} Ti$_2$Zn(P,As,Sb), V$_2$Ni(P,As,Sb), V$_2$Cu(Si,Ge,Sn) V$_2$Zn(Al,Ga,In), Cr$_2$Fe(P,As,Sb), Cr$_2$Co(Si,Ge,Sn), Cr$_2$Ni(Al,Ga,As), Mn$_2$Fe(Al,Ga,As). 
Many of the 25 electron inverse-Heusler systems show gaps after 9 states per formula in one or both spin-channels. None of these systems, however, are able to take advantage of these gaps to make a half-metal or near half metal by moving the Fermi energy to the gap, because that would require a relatively large moment (7$\mu_B$ per formula unit). This large moment would be difficult to produce in systems whose transition metal elements come from the early or late part of the transition metal series, e.g. $X$=Ti or V and $Y$=Ni, Cu or Zn.  The systems that could produce such large moments, e.g. $X$=Cr or Mn and $Y$=Fe or Co seem to prefer smaller net moments closer to 1 $\mu_B$ which allows them to take advantage of gaps and pseudogaps after 12 states per formula unit. Several systems show gaps after 14 states, but none of these have the correct moment (3$\mu_B$) to make a half-metal.

Ti$_2$Zn(P,As,Sb):
All three compounds have gaps after 9 states in both spin channels, but their moments are too small to move the Fermi energy near these gaps.

V$_2$Ni(P,As,Sb):
All three compounds have gaps after 9 states in both spin channels, but their moments are too small to move the Fermi energy near these gaps.

V$_2$Cu(Si,Ge,Sn):
All three compounds have gaps after 9 states in both spin channels. V$_2$CuSn also has gaps after 14 electrons in both spin channels. In all three cases, the moments are too small to move the Fermi energy near the minority  gap after 9 states or the majority gap after 14 states.

V$_2$Zn(Al,Ga,Sb):
V$_2$ZnAl has a gap after 14 states in majority and a pseudogap after 14 states in minority.  V$_2$ZnGa has a pseudogap after 14 states in majority and a gap after 14 states in minority.  V$_2$ZnIn has a gap in both spin channels after 14 states.  Their small moments preclude half-metals or near half-metals. 

Cr$_2$Fe(P,As,Sb): Cr$_2$FeP has gaps in both spin channels after 9 states and a pseudogap in minority after 12 states. Cr$_2$FeAs has gaps in both spin channels after 9 states and a sizable gap in minority after 12 states. Cr$_2$FeSb has a gap after 9 states in majority and a pseudogap after 9 states in minority. It also has a gap after 12 states in minority.  All three have moments of approximately 1.01$\mu_B$ which places the Fermi energy near the gap or pseudogap at 12 states.  Cr$_2$FeAs and Cr$_2$FeSb are near half-metals, but are not included in Table \ref{tab:near_half_metallic_ferromagnets} because of their positive formation energies.

Cr$_2$Co(Si,Ge,Sn): Cr$_2$CoSi has gaps in both spin channels after 9 states and a minority gap after 12.  Cr$_2$CoGe has a majority gap after 9 states and a minority gap after 12 states.  Both of these systems are predicted to be near half-metals as described in Table \ref{tab:near_half_metallic_ferromagnets}. Cr$_2$CoSn has a majority gap after 9 states and a pseudogap in minority near 12 states.

Cr$_2$Ni(Al,Ga,In): Cr$_2$NiAl has gaps after 9 and 14 states in majority and after 12 states in minority.  It uses the gap after 12 states in minority to generate a half metal as described in Table \ref{tab:half_metallic_ferromagnets}.  Cr$_2$NiGa has gaps after 9  states in majority and after 14 states in minority.  It also has a pseudogap near 12 states in minority. Cr$_2$NiIn has a gap after 9 states in majority and a pseudogap near 12 states in minority. 

Mn2Fe(Al,Ga,In): Mn$_2$FeAl and Mn$_2$FeGa both have gaps after 9 states in majority and after 12 states in minority.  Both place the Fermi energy slightly below the minority gap to generate near-half-metals as described in Table \ref{tab:near_half_metallic_ferromagnets}.
In Mn$_2$FeIn, the minority gap after 12 states becomes a pseudogap.   It also has minority gaps after 9 states and after 14 states.

\textbf{26 electron systems:} V$_2$Cu(P,As,Sb), V$_2$Zn(Si,GeSn), Cr$_2$Co(P,As,Sb), Cr$_2$Ni(Si,Ge,Sn), Cr$_2$Cu(Al,Ga,In), Mn$_2$Fe(Si,GeSn), Mn$_2$Co(Al,Ga,As). 
Many of the 26 electron systems show gaps after 9 states, some show gaps after 12 states and a few show gaps after 14 states. None of these systems can generate the 8 $\mu_B$ moment needed to generate a half-metal using the gap after 9 states. Several, however, can generate the  moment of 2$\mu_B$ needed to generate a half-metal or near half metal by using a minority gap after 12 states. A moment of 2$\mu_B$ would also be consistent with a half-metal taking advantage of a a majority gap after 14 states. In principle, a majority gap after 14 states and a minority gap after 12 combined with a moment of 2$\mu_B$ could lead to a magnetic semiconductor. Mn$_2$CoAl comes close to this situation.

V$_2$Cu(P,As,Sb) have gaps in both spin channels after 9 states and are non-magnetic. V$_2$CuSb also has pseudogaps in both spin-channels after 14 states. 

V$_2$Zn(Si,Ge,Sn) have only pseudogaps near 9 states and are also non-magnetic. None of the 26 electron systems with $X$=V are predicted to be half-metals or near half metals.

Cr$_2$Co(P,As,Sb) have majority gaps after 9 states and minority gaps after 12 states. Cr$_2$CoP also has a small minority gap after 9 states that becomes a pseudopgap for Cr$_2$CoAs and Cr$_2$CoSb. Cr$_2$CoP is a half metal with a moment of 2$\mu_B$. Cr$_2$CoAs and Cr$_2$CoSb are near half metals with moments of 2.0047 and 2.0016$\mu_B$ respectively, however they are omitted from Table \ref{tab:near_half_metallic_ferromagnets} because of their positive formation energy.

Cr$_2$Ni(Si,Ge,Sn) are remarkable for the large gaps after 9 states in the majority channel. Cr$_2$NiSi and Cr$_2$NiGe also have smaller gaps after 9 states in the minority channel.  Cr$_2$NiSi\ and Cr$_2$NiGe also have gaps  after 12 states in minority. The Fermi energy falls just above these gaps after 12 states giving moments of 1.89 and 1.94$\mu_B$ respectively. Cr$_2$NiGe also has a minority gap after 14 states. Cr$_2$NiSn has a single gap after 9 states in the majority channel and a relatively small moment (0.83$\mu_B$).
Cr$_2$NiSi\ and Cr$_2$NiGe satisfy our criteria for near half metals and are listed in Table \ref{tab:near_half_metallic_ferromagnets}.

Cr$_2$Cu(Al,Ga,In) all show gaps after 9 states in both spin channels. In addition Cr$_2$CuAl has gaps after 14 states in both channels while Cr$_2$CuGa has gaps after 14 states in the minority channel. For all cases, however, the net moments are too small to place the Fermi energy near any of the gaps. These three systems are ferrimagnets with relatively large anti-aligned moments on the two Cr atoms.

Mn$_2$Fe(Si,Ge,Sn)all have minority gaps after 12 states. Mn$_2$FeSi and Mn$_2$FeSn also have pseudogaps after 9 states in majority while Mn$_2$FeGe has a gap after 9 states in minority.  The moments (2.008$\mu_B$, 2.016 $\mu_B$ and 1.999$\mu_B$) are the right size to make all three of these systems near half-metals (Table \ref{tab:near_half_metallic_ferromagnets}); however the formation of  Mn$_2$FeSn is positive.  
 
Mn$_2$Co(Al,Ga,In) tend to have minority gaps (Al,Ga) or pseudogaps (In) after 12 states and their net moments tend to be large enough to take advantage of these gaps to make near half-metals.  Thus Mn$_2$CoAl is a half-metal (Table \ref{tab:half_metallic_ferromagnets})  and Mn$_2$CoGa is a near half-metal(Table \ref{tab:near_half_metallic_ferromagnets}). 

 In addition to the gap in the minority channel after 12 states that Mn$_2$CoAl utilizes to make itself a half-metal, it also has a slightly negative gap in the majority channel, \textit{i.e.} a very slight overlap between bands 14 and 15.
The maximum of band 14 at the $\Gamma$ point is only slightly higher than the minimum of band 15 at the  $X$ point. This situation has led some to refer this system as a ``spin-gapless semiconductor"\cite{PhysRevLett.100.156404, Galanakis2013SGS}. In our calculation and also that of Ouardi \textit{et al.} \cite{PhysRevLett.110.100401} it is a semi-metal rather than a semiconductor.

\textbf{27 electron systems:} The 21 systems in our database comprise: V$_2$Zn(P,As,Sb), Cr$_2$Ni(P,As,Sb), Cr$_2$Cu(Si,Ge,Sn), Cr$_2$Zn(Al,Ga,In), Mn$_2$Fe(P,As,Sb), Mn$_2$Co(Si,Ge,Sn), Mn$_2$Ni(Al,Ga,As).  A system with 27 electrons could take advantage of a minority gap after 12 states to make a half metal with a moment of 3$\mu_B$ or a majority gap after 14 states to to make a half-metal with a moment of 1$\mu_B$. 

V$_2$Zn(P,As,Sb): All three have gaps after 9 states in both spin channels. V$_2$ZnP is non-magnetic while V$_2$ZnAs and V$_2$ZnSb have relatively small moments on the V sites with ferrimagnetic alignment.Because their moments are small and there are no gaps after 12 or 14 states, these systems are very far from being half metals.
  
Cr$_2$Ni(P,As,Sb): These three systems  all generate large gaps after 9 states in the majority channel, and gaps   after 12 states and after 14 states in the minority channel.  The gaps after 14 states cannot be used to generate a half-metal because they are in the minority channel. The gaps after 12 states could be used to make half-metals or near half-metals but the net moments are too small (2.03, 1.69 and 1.29 $\mu_B$ respectively) for the Fermi energy to be near the gap.   

Cr$_2$Cu(Si,Ge,Sn):\ These three systems  generate gaps after 9 states in both channels and gaps (Si and Ge) or pseudogaps (Sn) after 14 states in the minority channel . There are no gaps after 12 states. The moments are too small to take advantage of the gaps after 9 states and the gaps after 14 states are in the wrong channel to be useful for making half-metals.

Cr$_2$Zn(Al,Ga,In): These systems  only generate gaps after 14 states; Cr$_2$ZnAl in both channels, Cr$_2$ZnGa in the minority and for Cr$_2$ZnIn there are no gaps. The one opportunity for making a half-metal is the majority channel of Cr$_2$ZnAl, but the net moment is somewhat smaller (0.706 $\mu_B$) than the 1.0$\mu_B$ needed to make a half-metal.
Pseudogaps near 9 states can be identified in both majority and minority of all three.

Mn$_2$Fe(P,As,Sb): These systems    all generate gaps after 12 states in the minority channel and  the moments have the correct value (3$\mu_B$)to generate  half-metals (Table \ref{tab:half_metallic_ferromagnets}).    The predicted half-metal, Mn$_2$FeSb is omitted from  Table \ref{tab:half_metallic_ferromagnets} because it is predicted to have positive formation energy. These systems are also predicted to have pseudogaps around 9 states in both majority and minority. Mn$_2$FeP has a small gap after 9 states in minority. 

Mn$_2$Co(Si,Ge,Sn): These systems    all generate gaps after 12 states in the minority channel.  For  Mn$_2$CoSi and Mn$_2$CoGe  the moment is large enough (3$\mu_B$)to generate a half-metal (Table \ref{tab:half_metallic_ferromagnets}).
The moment of Mn$_2$CoSn is too small to generate a gap.  There are also gaps after 9 states in majority for Mn$_2$CoSi and Mn$_2$CoGe.  This gap becomes a pseudogap for Mn$_2$CoSn.

Mn$_2$Ni(Al,Ga,As): No gaps are predicted for these three systems. Mn$_2$NiAl and Mn$_2$NiGa
have pseudogaps around 14 states.  

\textbf{28 electron systems:}  The 18 systems with 28 valence electrons in our dataset comprise: Cr$_2$Cu(P,As,Sb), Cr$_2$Zn(Si,Ge,Sn), Mn$_2$Co(P,As,Sb), Mn$_2$Ni(Si,Ge,Sb), Mn$_2$Cu(Al,Ga,In), and Fe$_2$Co(Al,Ga,In).  In principle, these  systems could take advantage of gaps after 14 electrons to make zero moment half-metals or semiconductors, however we found no examples of such an electronic structure in a system with negative formation energy.  They can also take advantage of a minority gap after 12 states to make half metals with moments of 4$\mu_B$ per formula unit.  We found three examples of this type of electronic structure.

Two systems 
Cr$_2$Cu(P,As,Sb):
These systems all have sizable gaps in majority after 9 states and smaller gaps in minority after 9 states. Cr$_2$CuP has a pseudogap after 14 states in minority and Cr$_2$CuSb
has a minority pseudogap after 14 states.  The moments are small.

Cr$_2$Zn(Si,Ge,Sn):
These systems have pseudogaps near 9 states per formula unit in both spin channels. Cr$_2$ZnSi and Cr$_2$ZnGe have gaps after 14 electrons. Cr$_2$ZnSi has gaps in both channels so that it is predicted to be a ferrimagnetic  semiconductor with zero net moment and two different gap widths. The smaller gap is a direct gap at the $L$ point and is on the order of 0.02eV so that $XA$ phase Cr$_2$ZnSi could be called a ``spin gapless semiconductor". Cr$_2$ZnGe only has a gap after 14 states in one spin channel. Its net moment is almost zero (0.0004$\mu_B$) so that it is predicted to be a near-half-metal. Unfortunately, both of these interesting systems are predicted to have positive formation energy and hull distances in excess of 0.3 eV/atom.

Mn$_2$Co(P,As,Sb): Mn$_2$CoP and Mn$_2$CoAs have gaps in both channels after 9 states.  All three systems have gaps after 12 states in minority. All three  have the requisite moment (4$\mu_B$) to generate half metals and are listed in Table \ref{tab:half_metallic_ferromagnets}.  We include details about calculated $XA$- Mn$_2$CoSb in Table \ref{tab:half_metallic_ferromagnets} even though its calculated formation energy is positive because of the controversy over the experimentally observed phase as discussed in Section \ref{FE and HD}.

Mn$_2$Ni(Si,Ge,Sb):
These systems all have minority gaps after 12 states. Mn$_2$NiSi and Mn$_2$NiGe also have majority pseudogaps near 9 states.  The moments on these systems are all too small to take advantage of the minority gap after 12 states.

Mn$_2$Cu(Al,Ga,In):
One can identify pseudogaps near 9 states in majority in Mn$_2$CuAl and Mn$_2$CuGa along with a tiny gap after 9 states in minority in Mn$_2$CuGa.  One can also identify pseudogaps near 14 states in both majority and minority in Mn$_2$CuAl.

Fe$_2$Co(Al,Ga,In):
These systems have minority gaps after 9 states.  Pseudogaps near 12 minority states can also be identified above the Fermi energy in all three systems.  Even if there were a minority gap after 12 states, the moments in these systems would be too large to take advantage of it.

\textbf{29 electron systems:}  Our database contained 18 systems with 29 valence electrons per formula unit: Cr$_2$Zn(P,As,Sb), Mn$_2$Ni(P,As,Sb), Mn$_2$Cu(Si,Ge,Sn), Mn$_2$Zn(Al,Ga,In), Fe$_2$Co(Si,Ge,Sn), and Fe$_2$Ni(Al,Ga,In).    In principle, a system with 29 electrons per formula unit could utilize a minority gap after 9, 12, or 14 states together with moments of 9$\mu_B$, 5$\mu_B$, or 1$\mu_B$   respectively, to make a half-metal. In practice we only find one 29 electron half-metal, Mn$_2$CuSi, with a moment of 1$\mu_B$. 

Cr$_2$Zn(P,As,Sb):  These systems all showed gaps after 9 states and 15 states in the minority channel.  Cr$_2$ZnP also has a gap after 9 states in the majority channel and a pseudogap after 14 states in minority.  Cr$_2$ZnAs also has a pseudogap near 9 states and a gap after 14 states in majority.  In principle the  gap after 14 states in the majority could be utilized to make a half-metal, but the moment is too small (0.47$\mu_B$).
Cr$_2$ZnSb  has in addition to the minority gaps after 9 and 15 states, a pseudogap near 9 states in majority. The gap after 15 states is unusual and can be traced to a singly degenerate state at $\Gamma$ that is usually band 15 or band 17 dropping below a triply degenerate state.

Mn$_2$Ni(P,As,Sb): These three systems  have gaps (Mn$_2$NiP) or pseudogaps (Mn$_2$Ni(As,Sb)) near 12 states in the minority channel, but the moments are all much less than the 5$\mu_B$ that would be necessary to make a half-metal.
In addition, pseudogaps near 9 states in majority and 14 states in minority can be identified for Mn$_2$NiP.

Mn$_2$Cu(Si,Ge,Sn):\ all have gaps after 9 states in both the majority and minority channels.  Mn$_2$CuSi also has a minority pseudogap near 12 states and a minority gap after 14 states. The system uses the minority gap after 14 states to make a half-metal (Table \ref{tab:half_metallic_ferromagnets}).

Mn$_2$Zn(Al,Ga,In):\  all have small moments and no gaps. Pseudogaps in both channels near 9 states can be identified.

Fe$_2$Co(Si,Ge,Sn):\ all have moments near 5$\mu_B$.  However instead of gaps after 12 states in minority there are pseudogaps.  All three have gaps after 9 states in minority.  Fe$_2$CoSi also has a pseudogap near 9 states in majority.

Fe$_2$Ni(Al,Ga,In):\ also all have moments near  5$\mu_B$, but the only gaps are after 9 states in the minority channel. Fe$_2$NiAl has a minority pseudogap near 12 states and majority pseudogaps near 14 and 16 states. Fe$_2$NiGa has majority pseudogaps near 9 states and 16 states.   

\textbf{30 electron systems:}
The 15 systems in our database with 30 electrons are: Mn$_2$Cu(P,As,Sb), Mn$_2$Zn(Si,Ge,Sn), Fe$_2$Co(P,As,Sb), Fe$_2$Ni(Si,Ge,Sn), and Fe$_2$Cu(Al,Ga,In). In principle gaps after 14, 12, or 9 states could be used to make half-metals with moments of 2$\mu_B$ , 6$\mu_B$  or 12$\mu_B$ respectively.  In practice we find one negative-formation-energy near-half metal with a moment near 2$\mu_B$.  

Mn$_2$Cu(P,As,Sb): These all have gaps after 9 states in both majority and minority.  Mn$_2$CuP also has a gap after 14 states in the minority channel.  The moment is nearly 2$\mu_B$ which makes it a near-half-metal.

Mn$_2$Zn(Si,Ge,Sn): These systems have pseudogaps in both spin channels near 9 states.  Mn$_2$ZnSi also has a gap after 14 electrons in the minority channel and a spin moment of exactly 2$\mu_B$ which makes it a half-metal.  Unfortunately, its calculated formation energy is positive.  Mn$_2$ZnGe and Mn$_2$ZnSn do not have gaps in the DOS at the experimental lattice constant.  If, however, their lattices are significantly contracted, the gap after 14 states reappears and the Fermi energy falls into the gap.    

Fe$_2$Co(P,As,Sb): These\ all have gaps after 9 electrons in the minority channel and pseudogaps near 12 states also in the minority channel.  The moments are large enough (5.76, 5.95, 5.93 respectively) to place the Fermi energy in the vicinity of the 12 state pseudogap, but because there is not really a gap, these are not near half-metals.

Fe$_2$Ni(Si,Ge,Sn): These all\ have gaps after 9 states in the minority.  Fe$_2$NiSi and Fe$_2$NiGe also have majority gaps after 9 states.  Fe$_2$NiSi also has a minority gap after 12 states while Fe$_2$NiGe has a minority pseudogap after 12 states.  The moments are relatively large (4.78$\mu_B$, 5.09$\mu_B$, 5.23$\mu_B$ respectively) but not large enough (6$\mu_B$ would have been required) to make them half-metals or near half-metals.

Fe$_2$Cu(Al,Ga,In): These\ all have relatively large minority gaps after 9 states and  small majority gaps after 14 states (Fe$_2$CuAl and Fe$_2$CuGa) or after 16 states (Fe$_2$CuIn). Fe$_2$CuIn also has a pseudogap after 14 states in the majority.  The majority gaps after 14 states cannot be used to make a half metal and although the majority gap after 16 could be used, the moment is too large, 4.84 rather than 2.

\textbf{31 electron systems:}  The 15 systems in our database with 31 electrons comprise Mn$_2$Zn(P,As,Sb), Fe$_2$Ni(P,As,Sb), Fe$_2$Cu(Si,Ge,Sn), Fe$_2$Zn(Al,Ga,In) and Co$_2$Ni(Al,Ga,In).   Compounds with 31 electrons could, in principle, take advantage of a minority gap after 14, 12 or 9 states with a moment of 3$\mu_B$, 7$\mu_B$, or 13$\mu_B$ respectively.   In addition, we find one example of a gap after 15 states which would allow a half-metal with a moment of 1$\mu_B$.  In practice, we find no 31-electron half-metals or near half-metals with negative formation energy.

Mn$_2$Zn(P,As,Sb): All three systems have gaps after 9 states in minority.   Mn$_2$ZnP also has a gap after 9 states in majority while Mn$_2$ZnAs has a pseudogap near 9 states in majority. Mn$_2$ZnP has a pseudogap near 14 states in minority. Mn$_2$ZnAs has a tiny minority  gap after 15 states and a moment (1.001$\mu_B$) that places the Fermi energy very near this gap.  Unfortunately, its formation energy is positive.  Mn$_2$ZnSb has a similar electronic structure and moment, but the tiny gap in Mn$_2$ZnAs is replaced by a pseudogap in Mn$_2$ZnSb.

Fe$_2$Ni(P,As,Sb): These all have minority gaps after 9 states. Fe$_2$NiP also has a minority gap after 12 states while Fe$_2$NiAs and Fe$_2$NiSb have pseudogaps after 12 states.
 The moments are too small to take advantage of the gap or pseudogaps after 12 states.
 
Fe$_2$Cu(Si,Ge,Sn): These all have gaps after 9 states in both spin channels, with the minority gap being larger than the majority gap.  In addition, Fe$_2$CuSi and Fe$_2$CuSn have narrow gaps after 14 states in the majority.  None of these gaps are near the Fermi energy. 

Fe$_2$Zn(Al,Ga,In): These systems have gaps (Al)\ or pseudogaps (Ga,In) at 14 states in minority.  The moments are to large to place the Fermi energy near these gaps or pseudogaps. 

Co$_2$Ni(Al,Ga,In): These systems have numerous gaps and pseudogaps.  All three have narrow, deep, pseudogaps near 14 states in minority and have moments (2.99, 3.02, 3.13)$\mu_B$ that place the Fermi energy in the pseudogaps.  They are not   ``near half-metals" according to our definition.  In addition, all three have minority pseudogaps near 12 states.  They all also  have gaps after 9 states in minority. Co$_2$NiAl has a gap after 14 states in majority.  Co$_2$NiGa has a gap after 9 states in majority and pseudogaps near 14 and 16 states in majority.

\textbf{32 electron systems:} The 12 systems with 32 electrons in our database  include Fe$_2$Cu(P,As,Sb), Fe$_2$Zn(Si,Ge,Sn), Co$_2$Ni(Si,Ge,Sn), and Co$_2$Cu(Al,Ga,In).  

Fe$_2$Cu(P,As,Sb): These systems all have gaps after 9 states in both spin channels  and pseudogaps after 14 states in the minority channel.  Their moments are all greater than 4$\mu_B$, and the moment of Fe$_2$CuP is 4.018$\mu_B$, quite close the the value of 4 needed to generate a half-metal.  Unfortunately, the Fermi energy is on the edge of a pseudogap rather than a gap.   

Fe$_2$Zn(Si,Ge,Sn): The 31-electron Fe$_2$Zn$Z$ $XA$ compounds have pseudogaps rather than gaps near 9 states in both spin channels, but similar to the 31-electron Fe$_2$Cu$Z$ systems, they have  pseudogaps  near 14 states in the minority channel.  The pseudogap near 14 states in Fe$_2$ZnSi is remarkable in that the DOS plots appear to show a gap after 14 states.  Its moment (3.982 $\mu_B$) is close enough to 4 that we would classify Fe$_2$ZnSi as a near half metal and include it in Table \ref{NearHalfMetals} because it also has a negative formation energy.
 However, bands 14 and 15 cross between $X$ and $W$, spoiling the gap.  

Co$_2$Ni(Si,Ge,Sn):
These all have
gaps after nine states in both spin channels and pseudogaps near 12 states in minority.  Their moments are too small (2.51$\mu_B$, 2.56$\mu_B$, 2.70$\mu_B$) to place the Fermi energy near the pseudogaps.
  None are half-metals or near half metals.

Co$_2$Cu(Al,Ga,In)
These all have gaps after 9 states in the minority channel.
None are half-metals or near half metals.

\textbf{33 electron systems:}  The 12 systems with 33 electrons in our database include Fe$_2$Zn(P,As,Sb), Co$_2$Ni(P,As,Sb), Co$_2$Cu(Si,Ge,Sn), and Co$_2$Zn(Al,Ga,In).  These systems would need a moment of 5$\mu_B$ to utilize a gap after 14 states to make a half metal.  They have neither the gaps nor the moment.

Fe$_2$Zn(P,As,Sb): These systems have either gaps (P,As) or pseudogaps after 9 states in both spin channels. 

Co$_2$Ni(P,As,Sb):  These systems all have gaps after 9 states in both spin channels.  Pseudogaps near 12 states in minority can also be identified for all three systems. 

Co$_2$Cu(Si,Ge,Sn):  These systems also have gaps after 9 states in both channels.  Two (Si,Ge) have pseudogaps after 12 states in both spin channels.

Co$_2$Zn(Al,Ga,In): No gaps or convincing pseudogaps were observed in these systems.

\textbf{34 electron systems: }The 9 systems in our database with 34 electrons include Co$_2$Cu(P,As,Sb), Co$_2$Zn(Si,Ge,Sn) and Ni$_2$Cu(Al,Ga,In).  

Co$_2$Cu(P,As,Sb): These systems have gaps after 9 states in both channels. Co$_2$CuP\ also has a majority gap after 12 states which becomes a pseudogap for the other two compounds.  The moments are not nearly large enough to place the Fermi energy near the gaps or pseudogaps. 

Co$_2$Zn(Si,Ge,Sn): These systems all have pseudogaps near 9 states in both spin channels and also a pseudogap near 12 states in the minority channel. 

Ni$_2$Cu(Al,Ga,In): These systems are non-magnetic and have no gaps.  Ni$_2$CuAl and Ni$_2$CuGa have pseudogaps near 12 states in both channels.  Ni$_2$CuGa and Ni$_2$CuIn have pseudogaps after 9 states in both channels.

\textbf{35 electron systems:} The 9 systems in our database with 35 electrons include Co$_2$Zn(P,As,Sb), Ni$_2$Cu(Si,Ge,Sn) and Ni$_2$Zn(Al,Ga,In).  A system with 35 electrons would need a moment of 7$\mu_B$ to take advantage of a gap after 14 states in minority.  Such moments would be unlikely in these systems because the d-bands are nearly filled for Co and especially for Ni.  

Co$_2$Zn(P,As,Sb): These systems are magnetic with moments relatively small moments, (2.4$\mu_B$, 2.5$\mu_B$, 2.7$\mu_B$).  
Co$_2$ZnP\ and Co$_2$ZnAs  have gaps after 9 states in both spin channels while Co$_2$ZnSb has pseudogaps near 9 states in both spin channels.  Pseudogaps near 14 states can be identified in all three systems.    

Ni$_2$Cu(Si,Ge,Sn): These\    are all non-magnetic and have gaps after 9 states in each of the identical spin channels. Pseudogaps near 12 states can be identified in all three as well as near 14 states in Ni$_2$CuSn. \ 
 
\textbf{36 electron systems:} The 6 systems in our database with 36 electrons include Ni$_2$Cu(P,As,Sb) and Ni$_2$Zn(Si,Ge,Sn).   None of these systems are magnetic.

Ni$_2$Cu(P,As,Sb): These have gaps after 9 states and pseudogaps near 12 states and 14 states.

Ni$_2$Zn(Si,Ge,Sn):  These have pseudogaps near 14 states.

\textbf{37 electron systems:} The 6 systems in our database with 37 electrons are Ni$_2$Zn(P,As,Sb) and Cu$_2$Zn(Al,Ga,In). None of these 6 systems has a magnetic moment and therefore cannot be half-metals.

Ni$_2$Zn(P,As,Sb):
Ni$_2$ZnP and Ni$_2$ZnAs have gaps after 9 states. Ni$_2$ZnSb has a pseudogap near 9 states.  Pseudogaps after 14 states can be identified for all three.

Cu$_2$Zn(Al,Ga,In): These systems have no gaps.

\textbf{38 electron systems:}  The three systems in our database with 38 electrons are Cu$_2$Zn(Si,Ge,Sn).  They have neither moments nor gaps.
 
\textbf{39 electron systems:}  The three systems in our database with 39 electrons are Cu$_2$Zn(P,As,Sb).  They have neither moments nor gaps. 

\subsubsection{Moment vs. Electron Count}

 Figure \ref{slaterpauling} shows the total spin magnetic moment per formula unit ($M_{tot}$) plotted versus the total valence electron number per formula unit ($N_{V}$) for 254 $XA$ compounds with negative formation energy.    In this plot, the half-metallic  phases  fall along one of the lines, $M_{tot}=|N_{V}-18|$ , $M_{tot}=|N_{V}-24|$, or  $M_{tot}=|N_{V}-28|$ depending on whether the system is placing the Fermi energy in a gap after 9, 12, or 14 states respectively.
 The half-metals are listed in Table \ref{tab:half_metallic_ferromagnets} along with their calculated properties.   Table \ref{tab:near_half_metallic_ferromagnets} is a similar list of the near-half-metals.  These tables are divided into three sections according to whether the gap leading to half-metallicity is located after 9, 12 or 14 states.

   The half-metals and near-half-metals found along the $M_{tot}=|N_{V}-18|$ line with a gap after  9 states per formula unit in one of the spin channels primarily have $X$=Sc,  or Ti, while those found along the  $M_{tot}=|N_{V}-24|$ lines generally have $X$=Cr or Mn.    Systems with $X$=V also fall mainly along the  $M_{tot}=|N_{V}-18|$ line, but there is one near-half-metal, (V$_2$CoSi) that uses the gap after 12 states.   We find only one half-metal (Mn$_2$CuSi) and one near half-metal (Mn$_2$CuP) along the $M_{tot}=|N_{V}-28|$ line. The   systems that appear to be on that line
at $N_V=30$ and $N_V=31$ have pseudogaps rather than gaps near the Fermi energy.

Some systems appear on or near two lines.  Systems with $N_V=21$ can in principle have gaps after both 9 states and 12 states.  Ti$_2$CoSi is a half-metal with the Fermi energy in the minority gap after 9 states, but the Fermi energy also falls in a pseudogap at 12 states in majority. The near-half-metals Ti$_2$Fe(P,As,Sb) all have gaps after 9 states in minority and after 12 states in majority, however the moment is not large enough to place the Ferni energy in the gaps.  In each case it falls just above the minority gap and just below the majority gap by an equal number of states.
 Similarly, for $N_V=26$ one can have systems on both the $M_{tot}=|N_{V}-24|$ and the  $M_{tot}=|N_{V}-28|$ lines.  Thus Mn$_2$CoAl has a minority gap after 12 states and a pseudogap at 14 majority states.    

 Although we predicted that Mn$_{2}$CoSb would have a positive formation energy (0.042 eV),  we  found  references indicating that it can be fabricated in the $XA$ phase \cite{PhysRevB.77.014424,Dai2006533} using melt-spinning. For this reason Mn$_{2}$CoSb was included in Fig. \ref{slaterpauling} and Table \ref{tab:half_metallic_ferromagnets}.
 As discussed in Sec. \ref{FE and HD}, we speculate that the experimental samples may be $L2_{1}B$ phase rather than $XA$.


For almost all of the half-metals and near-half-metals that follow the Slater-Pauling rule ($M_{tot}=|N_{V}-24|$), the spin moments on the $X_{1}$ atoms are antiparallel to those on $X_{2}$ atoms. If one uses the local moments  given in Table \ref{tab:half_metallic_ferromagnets} to estimate the number of majority and minority $s-d$ electrons on the transition metal sites one finds that for the Slater-Pauling  half-metals with minority gap, there are approximately 4 minority electrons on the $X_1$ and $Y$ sites and approximately 2 on the $X_2$ sites.  This implies that the transition metal nearest neighbors always differ in number of electrons by approximately 2.  The implied contrast in the positions of the $d$-states leads to the hybridization gap in the minority channel.  Since the average of 4 and 2 is 3, the gap should fall after 3 states per atom\cite{butler2011rational}.

\subsubsection{Hull Distances}


\begin{table*}[htbp]
        \caption{DFT-calculated properties of 51 half-metals in the inverse-Heusler phase that have negative formation energy. Successive columns present: number of valence electrons per formula unit, $N_{V}$, calculated lattice constant, $a$, total spin moment per f.u., $M_{tot}$, local moments for atoms on the  $X_{1}$, $X_{2}$, $Y$, and $Z$ sites: $m(X_{1})$, $m(X_{2})$, $m(Y)$ and $m(Z)$, formation energy $\Delta E_f$, distance from the convex hull $\Delta E_{\rm HD}$, formation energy in $L2_{1}$ phase $\Delta E_f (L2_{1})$, band gap $E_{g}$, gap type, experimental reports of compounds with composition \textit{\textit{X$_{2}$YZ}}, and experimental reports of corresponding $Y_2XZ$ full-Heusler compounds, if any.}
         \begin{tabular}{|c|c|c|c|c|c|c|c|c|c|c|c|c|c|}
    \toprule
   \textit{X$_2$YZ}   & $N_{V}$ & $a$ & $M_{tot}$ & $m(X_{1})$ & $m(X_{2})$ & $m(Y)$ & $m(Z)$    & $
  \Delta E_f$ & $\Delta E_{\rm HD} $ &$ \Delta E_f(L2_{1})$  & $E_{g}$  & Experimental & $Y_{2}XZ$  \\
      &  & ({\AA}) &  \multicolumn{5}{c|}{($\mu_{B}$)}  & \multicolumn{3}{c|}{(eV/atom)} & (eV)  &   reports   & reports \\
    \hline
    \midrule
    Sc$_{2}$VIn  & 14 & 6.938 & -4 & -0.072 & -0.509 & -2.675 & 0.041 &  -0.013 & 0.276 & -0.100 & 0.504   & & \\
    Sc$_{2}$VSi  & 15 & 6.53 & -3 & 0.058 & -0.452 & -2.228 & 0.073  &  -0.125 & 0.388 & -0.196 & 0.499   & & \\
    Sc$_{2}$VGe & 15 & 6.61 & -3 & 0.084 &  -0.424 &  -2.274 & 0.070 &  -0.118 & 0.446  & -0.203 &  0.480   & & \\
    Sc$_{2}$VSn & 15 & 6.87 & -3 & 0.144 &  -0.327 &  -2.388 & 0.058 &  -0.157 & 0.323 &-0.210 & 0.518    & & \\
    Sc$_{2}$CrAl & 15 & 6.67 & -3 & 0.295 &  -0.064&  -3.133 & 0.082 &  -0.039 & 0.225 & -0.096 & 0.684   & & \\
    Sc$_{2}$CrGa & 15 & 6.62 & -3 & 0.284 &  -0.050&  -3.121 & 0.067 &  -0.065 & 0.283 &-0.158 & 0.684   & & \\
    Sc$_{2}$CrIn & 15 & 6.86 & -3 & 0.337 &  0.056 &   -3.302 & 0.049 &  -0.003 & 0.286 & -0.069 & 0.646   & & \\
    Sc$_{2}$CrSi & 16 & 6.43 & -2 & 0.380 &  0.020 &   -2.454 & 0.079 &  -0.120 & 0.393 & -0.161 & 0.632   & & \\
    Sc$_{2}$CrGe & 16 & 6.52 & -2 & 0.429 &  0.112 &   -2.613 & 0.065 &  -0.110 & 0.453 &-0.184  & 0.646  & & \\
    Sc$_{2}$CrSn & 16 & 6.79 & -2 & 0.505 &  0.234 &   -2.878 & 0.051 &  -0.136 & 0.344 & -0.180 & 0.631   & & \\
    Sc$_{2}$MnSi & 17 & 6.37 & -1 & 0.465 &  0.443 &   -2.218 & 0.062 &  -0.200 & 0.350 & -0.285  &  0.518   & & \\
    Sc$_{2}$MnGe & 17 & 6.461 & -1 & 0.536 & 0.554 &  -2.455 & 0.041 &  -0.188 & 0.410 & -0.314 & 0.428   & & \\
    Sc$_{2}$MnSn & 17 & 6.75 & -1 & 0.628 & 0.654 &  -2.833 & 0.031 &  -0.202 & 0.312 & -0.287 &  0.266  & & \\
    Ti$_{2}$VSi & 17 & 6.173 & -1 & 0.575 & -0.028 &  -1.476 & 0.007 &  -0.316 & 0.217 & -0.379 & 0.219   & & \\
    Ti$_{2}$VGe & 17 & 6.262 & -1 & 0.688 & 0.094 &  -1.706 & 0.008 &  -0.246 & 0.240 & -0.316 & 0.285   & & \\
    Ti$_{2}$VSn & 17 & 6.523 & -1 & 0.879 &  0.257 & -2.052  & 0.004 &  -0.166 & 0.139& -0.189 & 0.423   & & \\
    Sc$_{2}$CoSi & 19 & 6.28 & 1 & 0.426 &  0.173 & 0.220  & 0.053 &  -0.268 & 0.406 & -0.506 &0.67 &   & \\
    Sc$_{2}$CoGe & 19 & 6.367 & 1 & 0.432 &  0.221 & 0.165  & 0.029&  -0.248 & 0.425 & -0.521 & 0.591    & & \\
    Sc$_{2}$CoSn & 19 & 6.627 & 1 & 0.421 &  0.206 & 0.182  & 0.010 &  -0.239 & 0.345 & -0.421 & 0.594    & & \\
    Ti$_{2}$MnSi & 19 & 6.024 & 1 & 1.055 &  0.652 & -0.920  & 0.043 &  -0.394 & 0.231 & -0.409 & 0.461   & & \\
    Ti$_{2}$MnGe & 19 & 6.123 & 1 & 1.204 &  0.844 & -1.314  & 0.024 & -0.295 & 0.265 & -0.346& 0.517   & & \\
    Ti$_{2}$MnSn & 19 & 6.39 & 1 & 1.41 & 1.106 & -1.916 & 0.017 &  -0.188 & 0.188 & -0.178 & 0.593   &                                       & \\
    Ti$_{2}$FeAl & 19 & 6.12 & 1 & 1.013 &  0.709 & -0.964   &   0.031 & -0.256 & 0.187& -0.279 & 0.536  & & \\
    Ti$_{2}$FeGa & 19 & 6.122 & 1 & 1.005 &  0.757 & -0.966   &   0.009 & -0.265 & 0.212 & -0.344 & 0.57   & & \\
    Ti$_{2}$FeIn & 19 & 6.37 & 1 & 1.208 &   1.064 & -1.673   &   0.004 & -0.045 & 0.274 & -0.072 & 0.532   & & \\
    Ti$_{2}$FeSi & 20 & 5.99 & 2 & 1.257 &   0.621 & -0.101   &   0.041 & -0.395 & 0.295 & -0.447 &0.628    & & \\
    Ti$_{2}$FeGe & 20 & 6.07 & 2 & 1.282 &   0.677 & -0.186   &   0.011 & -0.297 & 0.307 & -0.394 & 0.622    & & \\
    Ti$_{2}$FeSn & 20 & 6.339 & 2 & 1.356 &   0.801 & -0.509  &   0.001 & -0.185 & 0.256& -0.227& 0.617   & & \\
    Ti$_{2}$CoAl & 20 & 6.13 & 2 & 1.252 &   0.705 & -0.204  &   0.020 & -0.287 & 0.185& -0.293 & 0.684   & & \\
    Ti$_{2}$CoGa & 20 & 6.11 & 2 & 1.230 &   0.773 & -0.211  &   -0.005 &-0.295 & 0.178 & -0.363 & 0.689   & & \\
    Ti$_{2}$CoIn & 20 & 6.351 & 2 & 1.245 & 0.837 & -0.381 &-0.009 & -0.089 & 0.195 & -0.086 & 0.613   &  &  \\ 
    V$_{2}$CrGe & 20 & 5.94 & 2 & 1.167 & -0.415 & 1.158  & 0.003 & -0.118  & 0.151 & -0.120 & 0.086   &                      &\\
    V$_{2}$MnAl & 20 &5.922 & 2 & 1.354 & -0.309 & 0.870 & 0.016  & -0.148 & 0.097 & -0.072 & 0.142  &              &\cite{Mn2VAl}\\
    V$_{2}$MnGa & 20 & 5.924 & 2 & 1.229 & -0.416 & 1.138 & -0.007 & -0.118 & 0.123& -0.070 & 0.124 &                &\cite{Buschow198190}\\
    Ti$_{2}$CoSi & 21 & 6.02 & 3 & 1.528 &   0.789 & 0.394  &   0.059 &-0.387 & 0.307 & -0.437 & 0.798   & & \\
    Ti$_{2}$CoGe & 21 & 6.10 & 3 & 1.518 &   0.812 & 0.396  &   0.018 &-0.299 & 0.303 & -0.384 & 0.779  & & \\
    Ti$_{2}$CoSn & 21 & 6.35 & 3 & 1.495 &   0.802 & 0.373  &  -0.004 &-0.217 & 0.220 & -0.197 & 0.727   & & \\
    Ti$_{2}$NiAl & 21 & 6.19 & 3 & 1.524 &   0.982 & 0.149  &  0.028 &-0.285& 0.200 & -0.277 & 0.46   & & \\
    Ti$_{2}$NiGa & 21 & 6.17 & 3 &  1.491 &   1.056 & 0.158  &  -0.003 &-0.289& 0.193 & -0.364 & 0.522   & & \\
    Ti$_{2}$NiIn & 21 & 6.4 & 3 &  1.452 &   1.058 & 0.135  & -0.013  &-0.127& 0.216 & -0.096 & 0.418   & & \\
    \hline
    Cr$_{2}$MnAl & 22 & 5.831 & $-2$ & -1.835 & 1.536 & -1.678 & -0.021  &-0.086 &  0.048 & 0.002 & 0.204  &    &\\
    Cr$_{2}$NiAl & 25 & 5.7 & 1 & -1.239 & 1.828 & 0.452 & -0.057 & -0.071 & 0.236 & 0.148 & 0.142   &                            &  \cite{Buschow19831}\\
    Cr$_{2}$CoP & 26 & 5.619 & 2     & -0.794 & 1.912 & 0.854 &-0.012  & -0.22  & 0.217& -0.159 & 0.48   &              &\\
    Mn$_{2}$CoAl & 26 & 5.735 &  2     & -1.617 & 2.701 & 0.958 & -0.061   & -0.27  & 0.036& -0.141 & 0.38  &    XA\cite{PhysRevB.77.014424,PhysRevLett.110.100401}       & \cite{Co2MnAl} \\ 
    Mn$_{2}$FeP & 27 & 5.558 & 3     & -0.452 &  2.632 & 0.733 & 0.024   & -0.352  &0.096 & -0.292 & 0.252  &                          & \\
    Mn$_{2}$FeAs & 27 & 5.72 & 3     & -0.732 & 2.772  & 0.885 &0.022 & -0.071  & 0.086 & 0.014 & 0.337  &  & \\
    Mn$_{2}$CoSi & 27 & 5.621 & 3 & -0.548 & 2.670 & 0.849 & -0.012  & -0.365 & 0.018 & -0.177 & 0.513 &   &\cite{doi:10.1143/JPSJ.63.1881,PhysRevB.79.100405,1347-4065-48-8R-083002}\\
    Mn$_{2}$CoGe & 27 & 5.75 & 3    & -0.804 & 2.863 & 0.903 &0.005  & -0.153  & 0.03 & -0.011 & 0.323    &    XA\cite{PhysRevB.77.014424}  &\cite{Sobczak1976}\\
    Mn$_{2}$CoP & 28 & 5.581 & 4     & 0.091 & 2.793 & 1.017 & 0.032  & -0.333 & 0.216 & -0.232 & 0.508    &                 &\\
    Mn$_{2}$CoAs & 28 &5.738 & 4     & -0.007 & 2.916 & 1.013 & 0.025  & -0.085 & 0.08 & 0.081 & 0.295    &                 &\\
    Mn$_{2}$CoSb & 28 &5.985 & 4     & -0.041 &  3.093 & 0.938 & -0.003  & 0.042 & 0.133 & 0.193 & 0.499    &        XA\cite{PhysRevB.77.014424,Dai2006533}        &\cite{Co2MnSb,Buschow19831}\\
     \hline
    Mn$_{2}$CuSi & 29 & 5.762 & 1 &  -1.876 &   2.838 & 0.003  & 0.018  &-0.101& 0.177 & -0.091 & 0.318   & & \\
    \bottomrule
    \end{tabular}%

        \label{tab:half_metallic_ferromagnets}%
\end{table*}%

Using our calculated $XA$ formation energies, and the formation energies of all the other phases in the OQMD database, we calculated the hull distance $\Delta E_{\rm HD}$ for  the  negative formation energy half-metallic inverse-Heusler compounds listed in Table \ref{tab:half_metallic_ferromagnets} and the near-half-metals listed in Table \ref{tab:near_half_metallic_ferromagnets}.  Our calculations predict all the Sc, Ti-, and V-based half-metals lie significantly above the convex hull and have a $\Delta E_{\rm HD}$ that exceeds our empirical criterion  of  $\Delta E_{\rm HD}\le 0.052 $eV/atom for having a reasonable probability of synthesizability.  We predict 4 half-metals (Cr$_{2}$MnAl, Mn$_{2}$CoAl, Mn$_{2}$CoSi, Mn$_{2}$CoGe) to lie close enough to the convex hull to meet our criterion. Two of these compounds  Mn$_{2}$CoAl\cite{PhysRevB.77.014424,PhysRevLett.110.100401} and Mn$_{2}$CoGe\cite{PhysRevB.77.014424} have been experimentally observed by X-ray diffraction (XRD) measurements to exist in the $XA$ phase. We also found 6 near-half-metals (Cr$_{2}$MnGa, Mn$_{2}$FeAl, Mn$_{2}$FeGa, Mn$_{2}$FeSi, Mn$_{2}$FeGe) that meet our criterion for stability, three of these (Mn$_{2}$FeGa, Mn$_{2}$FeSi, Mn$_{2}$FeGe) have a calculated hull distance of zero meaning that no phase or combination of phases in the OQMD database had lower energy.

\begin{table*}[htbp]
        \centering
        \begin{threeparttable}
         \caption{\label{NearHalfMetals}DFT-calculated properties of 50 near half-metallic $X_{2}YZ$ inverse-Heusler compounds with negative formation energy. Successive columns present: number of valence electrons per formula unit $N_{V}$, calculated lattice constant ($a_{calc}$), total spin moment ($M_{tot}$) per f.u., local moments for atoms on the $X_{1}$, $X_{2}$, $Y$ and $Z$ sites: $m(X_{1})$, $m(X_{2})$, $m(Y)$ and $m(Z)$, formation energy ($\Delta E_{f}$), distance from the convex hull $\Delta E_{\rm HD}$, formation energy in $L2_{1}$ phase $\Delta E_f (L2_{1})$,  experimental reports of compounds with composition $X_{2}YZ$, and experimental reports of corresponding $Y_{2}XZ$ full-Heusler compounds.}
            \begin{tabular}{|c|c|c|c|c|c|c|c|c|c|c|c|c|c|}
    \toprule
      $X_{2}YZ$   & $N_{V}$ & $a_{calc}$ & $M_{tot}$ & $m(X_{1})$ & $m(X_{2})$ & $m(Y)$ & $m(Z)$    & $\Delta E_{f}$ & $\Delta E_{\rm HD}$ & $ \Delta E_f(L2_{1})$ & Experimental     & Rep.  \\
      &  & $(\AA)$ &  ($\mu_{B}$) & &  &  &    & \multicolumn{3}{c|}{(eV/atom)} &   records   &  Y$_{2}$XZ\\
    \hline
    \midrule
    Sc$_{2}$VAl     & 14    & 6.756 &  -3.8779    & -0.077 & -0.508 & -2.588 &0.040 & -0.033 & 0.231  & -0.134 &    & \\
    Sc$_{2}$VGa   & 14    & 6.708 & -3.9989    & -0.114 & -0.579 & -2.613 & 0.050 &  -0.064 & 0.285  &-0.184 &              & \\
    Sc$_{2}$MnAl   & 16    & 6.59 & -1.9983   & 0.429 & 0.231 & -2.903 & 0.074 &  -0.122 & 0.174 & -0.160  &              & \\
    Sc$_{2}$MnGa   & 16    & 6.54 & -1.9999    & 0.425 & 0.285 & -2.919 & 0.052 &  -0.159 & 0.229  & -0.243 &              & \\
    Sc$_{2}$MnIn   & 16    & 6.80 & -1.9969    & 0.489 & 0.398 & -3.203 & 0.038 &  -0.085 & 0.221 &-0.138   &              & \\
    Sc$_{2}$VSb   & 16    & 6.78 & -2.0011    & 0.319 & -0.188& -1.949 & 0.049 &  -0.150& 0.439  &-0.173 &              & \\
     Ti$_{2}$VAl   & 16   & 6.32 & -1.9633    & 0.380 & -0.579 & -1.541 & 0.016 &  -0.120 & 0.153  & -0.179 &              & \\
     Ti$_{2}$VGa   & 16   & 6.27 & -1.9529    & 0.314 & -0.567 & -1.499 & 0.022 &  -0.145 & 0.189  & -0.219  &              & \\
    Sc$_{2}$CrP   & 17    & 6.303 & -1.0043    & 0.491 & 0.349 & -2.051 & 0.063 &  -0.166 & 0.699 &-0.296  &              & \\
    Sc$_{2}$CrSb   & 17    & 6.74 & -1.0001   & 0.660 & 0.526 & -2.598 & 0.046 &  -0.097 &0.492 &-0.178 &              & \\
    Sc$_{2}$FeAl   & 17    & 6.47 & -0.9071   & 0.248 & 0.216 & -1.539 & 0.029 & -0.1323 & 0.216 &-0.244  &              & \\
    Sc$_{2}$FeGa   & 17   & 6.43 &-0.9488    & 0.264 &  0.277 & -1.656 & 0.015 &  -0.1735 & 0.267 &-0.345   &              & \\
    Sc$_{2}$FeIn   & 17   & 6.69 & -0.9714   & 0.344 & 0.412& -2.042 & 0.013 &  -0.069 & 0.290   & -0.189 &              & \\
    Ti$_{2}$CrAl   & 17   & 6.297 & -0.8021  & 1.032 & 0.775& -2.723 & 0.019 &  -0.1556 & 0.141 &-0.197   &              & \\
    Ti$_{2}$CrGa  & 17   &6.266 & -0.8659   & 1.008 & 0.785 &-2.747 & 0.011 &  -0.1781 & 0.166 &-0.245  &              & \\
    Ti$_{2}$CrIn   & 17   & 6.51 & -0.8538    & 1.141& 0.939 &  -3.078 & 0.005 &  -0.002 & 0.149 & 0.022 &              & \\
    Sc$_{2}$MnSb   & 18   & 6.685 & 0.0047    & -0.656 &-0.712 & 1.984 & 0.018 & -0.1449 & 0.444 & -0.296  &              & \\
    Ti$_{2}$VP   & 18   & 6.06 & 0.0033    &  -0.941 &-0.365 & 1.371 & -0.013 &  -0.4397 & 0.442  & -0.495 &              & \\
    Sc$_{2}$NiAl  & 19   & 6.502 & 0.9649    &  0.413& 0.217 & 0.120 &  0.031 &  -0.2413 & 0.249 &-0.425   &              & \\
    Sc$_{2}$NiGa  & 19   & 6.453 & 0.9738  & 0.423 & 0.272 & 0.081 & 0.015 &  -0.2817 & 0.278 &-0.544  &              & \\
    Sc$_{2}$NiIn  & 19   & 6.69 & 0.9839  &  0.414 & 0.271 & 0.078 & 0.006 &  -0.1993 & 0.309   & -0.375 &              & \\
   Ti$_{2}$CrAs   & 19   & 6.11 & 0.8077 & 1.015 & 0.535 & -0.941 & 0.023 & -0.2035 & 0.444 &  -0.255 &  & \\
   Ti$_{2}$CrSb   & 19    & 6.38 &  0.965  & 1.301 & 0.775 & -1.441 &0.019 & -0.111 & 0.279  & -0.084 &    &  \\ 
   Sc$_{2}$CoSb   & 20    & 6.64 &  1.9999  & 0.656 & 0.299 & 0.664 & 0.055 & -0.138 & 0.528 &-0.394   &    &  \\ 
   Sc$_{2}$NiSi  & 20 & 6.38 & 1.8015 & 0.689 & 0.459 & 0.220 & 0.093 & -0.203 & 0.527 & -0.591 & & \\
   Sc$_{2}$NiGe  & 20 & 6.46 & 1.8123 & 0.691 & 0.517 & 0.184 & 0.050 & -0.200 & 0.540 & -0.621 & & \\
   Sc$_{2}$NiSn  & 20 & 6.72 & 1.8438 &  0.690 & 0.523 &  0.164 & 0.019 & -0.228 & 0.430 & -0.516 & & \\
   Ti$_{2}$MnSb   & 20    & 6.323 &  1.9891  & 1.318 & 0.564 & -0.231 &0.009 & -0.1596 & 0.277& -0.181   &    &  \\
   Ti$_{2}$FeAs   & 21    & 6.07&  2.9395 & 1.424 & 0.469 & 0.780 &0.036 & -0.2374 & 0.475 &  -0.391 &    &  \\     
    V$_{2}$FeGa     & 21    & 5.91  & 2.8487    & 1.803 & -0.231& 1.200 & -0.001  & -0.146 & 0.091  & -0.047 &              &\cite{Buschow19831}\\
    Ti$_{2}$NiSn  & 22 & 6.427 & 3.8602 & 1.604 & 1.307 & 0.451 & 0.001 & -0.189 & 0.294 & -0.236 & & \\
    \hline
    Ti$_{2}$FeSb   & 21    & 6.31 &  2.9885 & 1.449 & 0.441 & 0.802 &0.009 &-0.1583 & 0.319&  -0.209 &    &  \\
    V$_{2}$FeAl    & 21   & 5.914  & 2.911   & 1.862 & -0.197& 1.132  & 0.026 & -0.191 & 0.088 &  -0.023 &                      & \cite{0953-8984-20-4-045212}\\
    \hline
    V$_{2}$MnGe & 21 & 5.92 & -2.8136 & -1.627 & 0.516 & -1.610 & 0.005 & -0.193 & 0.115 & -0.110 & & \\
    V$_{2}$MnSn & 21 & 6.20 & -2.9604 & -1.690 & 0.944 & -2.148 & 0.014 & -0.002 & 0.188 & 0.179 & & \\
    V$_{2}$FeGe & 22 & 5.891 & -1.9779 & -1.415 & 0.434 & -0.944 & -0.004 & -0.165 & 0.146 & -0.129 & & \\
    Cr$_{2}$MnGa  & 22    & 5.832 & -1.993  & -1.880 & 1.614 & -1.716 & -0.003 & -0.004 & 0.049  & 0.038 &  &                 \\
    Cr$_{2}$MnGe   & 23    & 5.8 & -0.9919  & -1.478 & 1.610 & - 1.147 & 0.010  &-0.007 & 0.122  & -0.008 &  &     \\
    V$_{2}$CoSi     & 23 & 5.78 & -0.997  & -0.754 & 0.182 & -0.374 & -0.029  &-0.291 & 0.199  &-0.193 &     &\\
    Cr$_{2}$CoAl   & 24    & 5.79 &0.0095& 1.892 & -1.698 &   -0.307 & 0.086 &-0.08 & 0.227   & 0.126  & & \cite{Carbonari1996313}\\
    Cr$_{2}$CoSi   & 25    & 5.657 &1.0086& -0.949 & 1.440 &   0.547 & 0.053 &-0.2187 & 0.139  & -0.022&    & \\
    Cr$_{2}$CoGe   & 25    & 5.792 &1.0225& -1.593 & 2.034 &  0.619 & 0.043 &-0.0171 & 0.109 & 0.155 &    & \\
    
    Mn$_{2}$FeAl      & 25    & 5.75         & 1.0009     & -1.856 & 2.714 & 0.142 & -0.025 & -0.158 &  0.008 &  -0.086  &                          & \cite{Buschow198190}\\
    Mn$_{2}$FeGa     & 25    & 5.79   & 1.0386     & -2.185 & 2.888 & 0.315 &-0.004 & -0.075 & 0.018 & -0.038 &             &\\
    Cr$_{2}$NiSi & 26   & 5.705 & 1.8915  & -0.857 & 2.130 &  0.573 & -0.023 & -0.195 & 0.201 & -0.035 & &\\
    Cr$_{2}$NiGe & 26   &5.82   & 1.9414  & -1.132 & 2.434 & 0.573  & -0.008 & -0.012 & 0.176 & 0.124 &            &\\
    Mn$_{2}$CoGa    & 26  & 5.76    &  2.0045    & -1.820 & 2.879 & 0.955 & -0.030   & -0.167 & 0  &-0.087   &    XA\cite{PhysRevB.77.014424}    &  \cite{Varaprasad20092702} \\
    Mn$_{2}$FeSi    & 26    & 5.6006     & 2.0077     & -0.798 &  2.407 & 0.371 & -0.015   & -0.362 & 0 & -0.259   &                          & \cite{Buschow19831}\\
    Mn$_{2}$FeGe & 26    & 5.72       & 2.0135    & -1.243 & 2.667  & 0.556 &-0.004 & -0.14 & 0   & -0.035 &         & \\
\hline
    Mn$_{2}$CuP  & 30   & 5.716       & 1.9946   & -1.242 & 3.042  & 0.057 & 0.083 & -0.048 & 0.399  &-0.091  &         & \\
    \bottomrule
    \end{tabular}%

        \label{tab:near_half_metallic_ferromagnets}%
        \end{threeparttable}
\end{table*}%

We also collected  references reporting  fabrication of  corresponding $Y_{2}XZ$ in the $L2_{1}$ phase. The presence of these phases indicates that it may be difficult to distinguish a mixture of $X_{2}YZ$ in the $XA$ phase and $Y_{2}XZ$ in the $L2_{1}$ phase.  Since they have the same XRD reflection peaks, it is difficult to distinguish the composition from the XRD patterns, which may lead to a wrong conclusion about the composition of the sample.

\section{\label{Summary}Conclusion}

We have performed extensive first-principle calculations on 405 inverse-Heusler compounds $X_{2}YZ$, where X=Sc, Ti, V, Cr, Mn, Fe, Co, Ni, or Cu; Y=Ti, V, Cr, Mn, Fe, Co, Ni, Cu, or Zn; and $Z$=Al, Ga, In, Si, Ge, Sn, P, As, or Sb with a goal of identifying materials of interest for spintronic applications.  We identified 14 semiconductors with negative formation energy. These are listed in Table \ref{tab:SP_semiconductors}. We identified many half-metals and near half-metals.  Those half-metals and near-half-metals that we identified and  that have negative formation energy are listed in Tables \ref{tab:half_metallic_ferromagnets} and \ref{tab:near_half_metallic_ferromagnets}. respectively.

We have identified four half-metals and six near-half-metals that meet a criterion for stability that we estimated based on a study of the hull distances of known $XA$ phases.  Seven of these ten predicted phases have not (to our knowledge) been observed experimentally.   

In establishing our stability criterion, we analyzed the experimental and theoretical data for several reported $XA$ phases (Mn$_{2}$CoIn, Mn$_{2}$CoSb, Mn$_2$CoSn, Mn$_2$RuSn)  and tentatively concluded that they are not $XA$ but are more likely a disordered phase such as $L2_1B$. We recommend more extensive studies of these materials. 

The ten half-metallic or near-half-metallic phases that we identified as reasonably likely to be stable, are all Slater-Pauling phases,\textit{ i.e.} they have a gap in the density of states after three states per atom in one of the two spin channels, and the Fermi energy is in or near this gap. Many half-metallic and near-half-metallic phases based on a gap after 9 states per formula unit (2.25 states per atom) were identified, but we estimate that other phases or combination of phases are likely to be more stable in equilibrium.


\begin{acknowledgments}
The authors acknowledge support from the National Science Foundation through grants DMREF-1235230 and DMREF-1235396. JH and CW (OQMD stability and SQS calculations) acknowledge support via ONR STTR N00014-13-P-1056. The authors also acknowledge Advanced Research Computing Services at the University of Virginia and High Performance Computing staff from the Center for Materials for Information Technology at the University of Alabama for providing technical support that has contributed to the results in this paper. The computational work was done using the High Performance Computing Cluster at the Center for Materials for Information Technology, University of Alabama, resources of the National Energy Research Scientific Computing Center (NERSC), a DOE Office of Science User Facility supported by the Office of Science of the U.S. Department of Energy under Contract No. DE-AC02-05CH11231 and the Rivanna high-performance cluster at the University of Virginia.
\end{acknowledgments}

\appendix

\bibliography{inverseheusler}

\begin{thebibliography}{86}%
\makeatletter
\providecommand \@ifxundefined [1]{%
 \@ifx{#1\undefined}
}%
\providecommand \@ifnum [1]{%
 \ifnum #1\expandafter \@firstoftwo
 \else \expandafter \@secondoftwo
 \fi
}%
\providecommand \@ifx [1]{%
 \ifx #1\expandafter \@firstoftwo
 \else \expandafter \@secondoftwo
 \fi
}%
\providecommand \natexlab [1]{#1}%
\providecommand \enquote  [1]{``#1''}%
\providecommand \bibnamefont  [1]{#1}%
\providecommand \bibfnamefont [1]{#1}%
\providecommand \citenamefont [1]{#1}%
\providecommand \href@noop [0]{\@secondoftwo}%
\providecommand \href [0]{\begingroup \@sanitize@url \@href}%
\providecommand \@href[1]{\@@startlink{#1}\@@href}%
\providecommand \@@href[1]{\endgroup#1\@@endlink}%
\providecommand \@sanitize@url [0]{\catcode `\\12\catcode `\$12\catcode
  `\&12\catcode `\#12\catcode `\^12\catcode `\_12\catcode `\%12\relax}%
\providecommand \@@startlink[1]{}%
\providecommand \@@endlink[0]{}%
\providecommand \url  [0]{\begingroup\@sanitize@url \@url }%
\providecommand \@url [1]{\endgroup\@href {#1}{\urlprefix }}%
\providecommand \urlprefix  [0]{URL }%
\providecommand \Eprint [0]{\href }%
\providecommand \doibase [0]{http://dx.doi.org/}%
\providecommand \selectlanguage [0]{\@gobble}%
\providecommand \bibinfo  [0]{\@secondoftwo}%
\providecommand \bibfield  [0]{\@secondoftwo}%
\providecommand \translation [1]{[#1]}%
\providecommand \BibitemOpen [0]{}%
\providecommand \bibitemStop [0]{}%
\providecommand \bibitemNoStop [0]{.\EOS\space}%
\providecommand \EOS [0]{\spacefactor3000\relax}%
\providecommand \BibitemShut  [1]{\csname bibitem#1\endcsname}%
\let\auto@bib@innerbib\@empty
\bibitem [{\citenamefont {\ifmmode \check{Z}\else
  \v{Z}\fi{}uti\ifmmode~\acute{c}\else \'{c}\fi{}}\ \emph
  {et~al.}(2004)\citenamefont {\ifmmode \check{Z}\else
  \v{Z}\fi{}uti\ifmmode~\acute{c}\else \'{c}\fi{}}, \citenamefont {Fabian},\
  and\ \citenamefont {Das~Sarma}}]{RevModPhys.76.323}%
  \BibitemOpen
  \bibfield  {author} {\bibinfo {author} {\bibfnamefont {I.}~\bibnamefont
  {\ifmmode \check{Z}\else \v{Z}\fi{}uti\ifmmode~\acute{c}\else \'{c}\fi{}}},
  \bibinfo {author} {\bibfnamefont {J.}~\bibnamefont {Fabian}}, \ and\ \bibinfo
  {author} {\bibfnamefont {S.}~\bibnamefont {Das~Sarma}},\ }\href {\doibase
  10.1103/RevModPhys.76.323} {\bibfield  {journal} {\bibinfo  {journal} {Rev.
  Mod. Phys.}\ }\textbf {\bibinfo {volume} {76}},\ \bibinfo {pages} {323}
  (\bibinfo {year} {2004})}\BibitemShut {NoStop}%
\bibitem [{\citenamefont {Hirohata}\ and\ \citenamefont
  {Takanashi}(2014)}]{Takanashi2014Future}%
  \BibitemOpen
  \bibfield  {author} {\bibinfo {author} {\bibfnamefont {A.}~\bibnamefont
  {Hirohata}}\ and\ \bibinfo {author} {\bibfnamefont {K.}~\bibnamefont
  {Takanashi}},\ }\href {http://stacks.iop.org/0022-3727/47/i=19/a=193001}
  {\bibfield  {journal} {\bibinfo  {journal} {Journal of Physics D: Applied
  Physics}\ }\textbf {\bibinfo {volume} {47}},\ \bibinfo {pages} {193001}
  (\bibinfo {year} {2014})}\BibitemShut {NoStop}%
\bibitem [{\citenamefont {Ando}(2015)}]{Ando2015spintronics}%
  \BibitemOpen
  \bibfield  {author} {\bibinfo {author} {\bibfnamefont {Y.}~\bibnamefont
  {Ando}},\ }\href {http://stacks.iop.org/1347-4065/54/i=7/a=070101} {\bibfield
   {journal} {\bibinfo  {journal} {Japanese Journal of Applied Physics}\
  }\textbf {\bibinfo {volume} {54}},\ \bibinfo {pages} {070101} (\bibinfo
  {year} {2015})}\BibitemShut {NoStop}%
\bibitem [{\citenamefont {Felser}\ \emph {et~al.}(2007)\citenamefont {Felser},
  \citenamefont {Fecher},\ and\ \citenamefont {Balke}}]{ANIE:ANIE200601815}%
  \BibitemOpen
  \bibfield  {author} {\bibinfo {author} {\bibfnamefont {C.}~\bibnamefont
  {Felser}}, \bibinfo {author} {\bibfnamefont {G.}~\bibnamefont {Fecher}}, \
  and\ \bibinfo {author} {\bibfnamefont {B.}~\bibnamefont {Balke}},\ }\href
  {\doibase 10.1002/anie.200601815} {\bibfield  {journal} {\bibinfo  {journal}
  {Angewandte Chemie International Edition}\ }\textbf {\bibinfo {volume}
  {46}},\ \bibinfo {pages} {668} (\bibinfo {year} {2007})}\BibitemShut
  {NoStop}%
\bibitem [{\citenamefont {Katsnelson}\ \emph {et~al.}(2008)\citenamefont
  {Katsnelson}, \citenamefont {Irkhin}, \citenamefont {Chioncel}, \citenamefont
  {Lichtenstein},\ and\ \citenamefont {de~Groot}}]{RevModPhys.80.315}%
  \BibitemOpen
  \bibfield  {author} {\bibinfo {author} {\bibfnamefont {M.~I.}\ \bibnamefont
  {Katsnelson}}, \bibinfo {author} {\bibfnamefont {V.~Y.}\ \bibnamefont
  {Irkhin}}, \bibinfo {author} {\bibfnamefont {L.}~\bibnamefont {Chioncel}},
  \bibinfo {author} {\bibfnamefont {A.~I.}\ \bibnamefont {Lichtenstein}}, \
  and\ \bibinfo {author} {\bibfnamefont {R.~A.}\ \bibnamefont {de~Groot}},\
  }\href {\doibase 10.1103/RevModPhys.80.315} {\bibfield  {journal} {\bibinfo
  {journal} {Rev. Mod. Phys.}\ }\textbf {\bibinfo {volume} {80}},\ \bibinfo
  {pages} {315} (\bibinfo {year} {2008})}\BibitemShut {NoStop}%
\bibitem [{\citenamefont {de~Groot}\ \emph {et~al.}(1983)\citenamefont
  {de~Groot}, \citenamefont {Mueller}, \citenamefont {Engen},\ and\
  \citenamefont {Buschow}}]{PhysRevLett.50.2024}%
  \BibitemOpen
  \bibfield  {author} {\bibinfo {author} {\bibfnamefont {R.~A.}\ \bibnamefont
  {de~Groot}}, \bibinfo {author} {\bibfnamefont {F.~M.}\ \bibnamefont
  {Mueller}}, \bibinfo {author} {\bibfnamefont {P.~G.~v.}\ \bibnamefont
  {Engen}}, \ and\ \bibinfo {author} {\bibfnamefont {K.~H.~J.}\ \bibnamefont
  {Buschow}},\ }\href {\doibase 10.1103/PhysRevLett.50.2024} {\bibfield
  {journal} {\bibinfo  {journal} {Phys. Rev. Lett.}\ }\textbf {\bibinfo
  {volume} {50}},\ \bibinfo {pages} {2024} (\bibinfo {year}
  {1983})}\BibitemShut {NoStop}%
\bibitem [{\citenamefont {Galanakis}\ \emph
  {et~al.}(2002{\natexlab{a}})\citenamefont {Galanakis}, \citenamefont
  {Dederichs},\ and\ \citenamefont {Papanikolaou}}]{PhysRevB.66.134428}%
  \BibitemOpen
  \bibfield  {author} {\bibinfo {author} {\bibfnamefont {I.}~\bibnamefont
  {Galanakis}}, \bibinfo {author} {\bibfnamefont {P.~H.}\ \bibnamefont
  {Dederichs}}, \ and\ \bibinfo {author} {\bibfnamefont {N.}~\bibnamefont
  {Papanikolaou}},\ }\href {\doibase 10.1103/PhysRevB.66.134428} {\bibfield
  {journal} {\bibinfo  {journal} {Phys. Rev. B}\ }\textbf {\bibinfo {volume}
  {66}},\ \bibinfo {pages} {134428} (\bibinfo {year}
  {2002}{\natexlab{a}})}\BibitemShut {NoStop}%
\bibitem [{\citenamefont {Galanakis}\ \emph
  {et~al.}(2002{\natexlab{b}})\citenamefont {Galanakis}, \citenamefont
  {Dederichs},\ and\ \citenamefont {Papanikolaou}}]{PhysRevB.66.174429}%
  \BibitemOpen
  \bibfield  {author} {\bibinfo {author} {\bibfnamefont {I.}~\bibnamefont
  {Galanakis}}, \bibinfo {author} {\bibfnamefont {P.~H.}\ \bibnamefont
  {Dederichs}}, \ and\ \bibinfo {author} {\bibfnamefont {N.}~\bibnamefont
  {Papanikolaou}},\ }\href {\doibase 10.1103/PhysRevB.66.174429} {\bibfield
  {journal} {\bibinfo  {journal} {Phys. Rev. B}\ }\textbf {\bibinfo {volume}
  {66}},\ \bibinfo {pages} {174429} (\bibinfo {year}
  {2002}{\natexlab{b}})}\BibitemShut {NoStop}%
\bibitem [{\citenamefont {Kandpal}\ \emph {et~al.}(2006)\citenamefont
  {Kandpal}, \citenamefont {Felser},\ and\ \citenamefont
  {Seshadri}}]{kandpal2006covalent}%
  \BibitemOpen
  \bibfield  {author} {\bibinfo {author} {\bibfnamefont {H.~C.}\ \bibnamefont
  {Kandpal}}, \bibinfo {author} {\bibfnamefont {C.}~\bibnamefont {Felser}}, \
  and\ \bibinfo {author} {\bibfnamefont {R.}~\bibnamefont {Seshadri}},\ }\href
  {http://stacks.iop.org/0022-3727/39/i=5/a=S02} {\bibfield  {journal}
  {\bibinfo  {journal} {Journal of Physics D: Applied Physics}\ }\textbf
  {\bibinfo {volume} {39}},\ \bibinfo {pages} {776} (\bibinfo {year}
  {2006})}\BibitemShut {NoStop}%
\bibitem [{\citenamefont {Galanakis}\ \emph {et~al.}(2006)\citenamefont
  {Galanakis}, \citenamefont {Mavropoulos},\ and\ \citenamefont
  {Dederichs}}]{galanakis2006electronic}%
  \BibitemOpen
  \bibfield  {author} {\bibinfo {author} {\bibfnamefont {I.}~\bibnamefont
  {Galanakis}}, \bibinfo {author} {\bibfnamefont {P.}~\bibnamefont
  {Mavropoulos}}, \ and\ \bibinfo {author} {\bibfnamefont {P.~H.}\ \bibnamefont
  {Dederichs}},\ }\href {http://stacks.iop.org/0022-3727/39/i=5/a=S01}
  {\bibfield  {journal} {\bibinfo  {journal} {Journal of Physics D: Applied
  Physics}\ }\textbf {\bibinfo {volume} {39}},\ \bibinfo {pages} {765}
  (\bibinfo {year} {2006})}\BibitemShut {NoStop}%
\bibitem [{\citenamefont {Galanakis}(2004)}]{0953-8984-16-18-010}%
  \BibitemOpen
  \bibfield  {author} {\bibinfo {author} {\bibfnamefont {I.}~\bibnamefont
  {Galanakis}},\ }\href {http://stacks.iop.org/0953-8984/16/i=18/a=010}
  {\bibfield  {journal} {\bibinfo  {journal} {Journal of Physics: Condensed
  Matter}\ }\textbf {\bibinfo {volume} {16}},\ \bibinfo {pages} {3089}
  (\bibinfo {year} {2004})}\BibitemShut {NoStop}%
\bibitem [{\citenamefont {Skaftouros}\ \emph
  {et~al.}(2013{\natexlab{a}})\citenamefont {Skaftouros}, \citenamefont
  {\"Ozdo\ifmmode~\breve{g}\else \u{g}\fi{}an}, \citenamefont {\ifmmode
  \mbox{\c{S}}\else \c{S}\fi{}a\ifmmode \mbox{\c{s}}\else \c{s}\fi{}\ifmmode
  \imath \else \i \fi{}o\ifmmode~\breve{g}\else \u{g}\fi{}lu},\ and\
  \citenamefont {Galanakis}}]{PhysRevB.87.024420}%
  \BibitemOpen
  \bibfield  {author} {\bibinfo {author} {\bibfnamefont {S.}~\bibnamefont
  {Skaftouros}}, \bibinfo {author} {\bibfnamefont {K.}~\bibnamefont
  {\"Ozdo\ifmmode~\breve{g}\else \u{g}\fi{}an}}, \bibinfo {author}
  {\bibfnamefont {E.}~\bibnamefont {\ifmmode \mbox{\c{S}}\else
  \c{S}\fi{}a\ifmmode \mbox{\c{s}}\else \c{s}\fi{}\ifmmode \imath \else \i
  \fi{}o\ifmmode~\breve{g}\else \u{g}\fi{}lu}}, \ and\ \bibinfo {author}
  {\bibfnamefont {I.}~\bibnamefont {Galanakis}},\ }\href {\doibase
  10.1103/PhysRevB.87.024420} {\bibfield  {journal} {\bibinfo  {journal} {Phys.
  Rev. B}\ }\textbf {\bibinfo {volume} {87}},\ \bibinfo {pages} {024420}
  (\bibinfo {year} {2013}{\natexlab{a}})}\BibitemShut {NoStop}%
\bibitem [{\citenamefont {K{\"u}bler}(2006)}]{0953-8984-18-43-003}%
  \BibitemOpen
  \bibfield  {author} {\bibinfo {author} {\bibfnamefont {J.}~\bibnamefont
  {K{\"u}bler}},\ }\href {http://stacks.iop.org/0953-8984/18/i=43/a=003}
  {\bibfield  {journal} {\bibinfo  {journal} {Journal of Physics: Condensed
  Matter}\ }\textbf {\bibinfo {volume} {18}},\ \bibinfo {pages} {9795}
  (\bibinfo {year} {2006})}\BibitemShut {NoStop}%
\bibitem [{\citenamefont {K\"ubler}\ \emph {et~al.}(2007)\citenamefont
  {K\"ubler}, \citenamefont {Fecher},\ and\ \citenamefont
  {Felser}}]{PhysRevB.76.024414}%
  \BibitemOpen
  \bibfield  {author} {\bibinfo {author} {\bibfnamefont {J.}~\bibnamefont
  {K\"ubler}}, \bibinfo {author} {\bibfnamefont {G.~H.}\ \bibnamefont
  {Fecher}}, \ and\ \bibinfo {author} {\bibfnamefont {C.}~\bibnamefont
  {Felser}},\ }\href {\doibase 10.1103/PhysRevB.76.024414} {\bibfield
  {journal} {\bibinfo  {journal} {Phys. Rev. B}\ }\textbf {\bibinfo {volume}
  {76}},\ \bibinfo {pages} {024414} (\bibinfo {year} {2007})}\BibitemShut
  {NoStop}%
\bibitem [{\citenamefont {Blum}\ \emph {et~al.}(2009)\citenamefont {Blum},
  \citenamefont {Jenkins}, \citenamefont {Barth}, \citenamefont {Felser},
  \citenamefont {Wurmehl}, \citenamefont {Friemel}, \citenamefont {Hess},
  \citenamefont {Behr}, \citenamefont {Büchner}, \citenamefont {Reller},
  \citenamefont {Riegg}, \citenamefont {Ebbinghaus}, \citenamefont {Ellis},
  \citenamefont {Jacobs}, \citenamefont {Kohlhepp},\ and\ \citenamefont
  {Swagten}}]{Blum2009Co2FeSi}%
  \BibitemOpen
  \bibfield  {author} {\bibinfo {author} {\bibfnamefont {C.~G.~F.}\
  \bibnamefont {Blum}}, \bibinfo {author} {\bibfnamefont {C.~A.}\ \bibnamefont
  {Jenkins}}, \bibinfo {author} {\bibfnamefont {J.}~\bibnamefont {Barth}},
  \bibinfo {author} {\bibfnamefont {C.}~\bibnamefont {Felser}}, \bibinfo
  {author} {\bibfnamefont {S.}~\bibnamefont {Wurmehl}}, \bibinfo {author}
  {\bibfnamefont {G.}~\bibnamefont {Friemel}}, \bibinfo {author} {\bibfnamefont
  {C.}~\bibnamefont {Hess}}, \bibinfo {author} {\bibfnamefont {G.}~\bibnamefont
  {Behr}}, \bibinfo {author} {\bibfnamefont {B.}~\bibnamefont {Büchner}},
  \bibinfo {author} {\bibfnamefont {A.}~\bibnamefont {Reller}}, \bibinfo
  {author} {\bibfnamefont {S.}~\bibnamefont {Riegg}}, \bibinfo {author}
  {\bibfnamefont {S.~G.}\ \bibnamefont {Ebbinghaus}}, \bibinfo {author}
  {\bibfnamefont {T.}~\bibnamefont {Ellis}}, \bibinfo {author} {\bibfnamefont
  {P.~J.}\ \bibnamefont {Jacobs}}, \bibinfo {author} {\bibfnamefont {J.~T.}\
  \bibnamefont {Kohlhepp}}, \ and\ \bibinfo {author} {\bibfnamefont {H.~J.~M.}\
  \bibnamefont {Swagten}},\ }\href {\doibase 10.1063/1.3242370} {\bibfield
  {journal} {\bibinfo  {journal} {Applied Physics Letters}\ }\textbf {\bibinfo
  {volume} {95}},\ \bibinfo {eid} {161903} (\bibinfo {year}
  {2009})}\BibitemShut {NoStop}%
\bibitem [{\citenamefont {Felser}\ \emph {et~al.}(2015)\citenamefont {Felser},
  \citenamefont {Wollmann}, \citenamefont {Chadov}, \citenamefont {Fecher},\
  and\ \citenamefont {Parkin}}]{felser2015basics}%
  \BibitemOpen
  \bibfield  {author} {\bibinfo {author} {\bibfnamefont {C.}~\bibnamefont
  {Felser}}, \bibinfo {author} {\bibfnamefont {L.}~\bibnamefont {Wollmann}},
  \bibinfo {author} {\bibfnamefont {S.}~\bibnamefont {Chadov}}, \bibinfo
  {author} {\bibfnamefont {G.~H.}\ \bibnamefont {Fecher}}, \ and\ \bibinfo
  {author} {\bibfnamefont {S.~S.~P.}\ \bibnamefont {Parkin}},\ }\href {\doibase
  10.1063/1.4917387} {\bibfield  {journal} {\bibinfo  {journal} {APL
  Materials}\ }\textbf {\bibinfo {volume} {3}},\ \bibinfo {eid} {041518}
  (\bibinfo {year} {2015})}\BibitemShut {NoStop}%
\bibitem [{\citenamefont {Tsunegi}\ \emph {et~al.}(2009)\citenamefont
  {Tsunegi}, \citenamefont {Sakuraba}, \citenamefont {Oogane}, \citenamefont
  {Naganuma}, \citenamefont {Takanashi},\ and\ \citenamefont
  {Ando}}]{Ando2009MTJs}%
  \BibitemOpen
  \bibfield  {author} {\bibinfo {author} {\bibfnamefont {S.}~\bibnamefont
  {Tsunegi}}, \bibinfo {author} {\bibfnamefont {Y.}~\bibnamefont {Sakuraba}},
  \bibinfo {author} {\bibfnamefont {M.}~\bibnamefont {Oogane}}, \bibinfo
  {author} {\bibfnamefont {H.}~\bibnamefont {Naganuma}}, \bibinfo {author}
  {\bibfnamefont {K.}~\bibnamefont {Takanashi}}, \ and\ \bibinfo {author}
  {\bibfnamefont {Y.}~\bibnamefont {Ando}},\ }\href {\doibase
  10.1063/1.3156858} {\bibfield  {journal} {\bibinfo  {journal} {Applied
  Physics Letters}\ }\textbf {\bibinfo {volume} {94}},\ \bibinfo {eid} {252503}
  (\bibinfo {year} {2009})}\BibitemShut {NoStop}%
\bibitem [{\citenamefont {Tezuka}\ \emph {et~al.}(2009)\citenamefont {Tezuka},
  \citenamefont {Ikeda}, \citenamefont {Mitsuhashi},\ and\ \citenamefont
  {Sugimoto}}]{Sugimoto2009MTJ}%
  \BibitemOpen
  \bibfield  {author} {\bibinfo {author} {\bibfnamefont {N.}~\bibnamefont
  {Tezuka}}, \bibinfo {author} {\bibfnamefont {N.}~\bibnamefont {Ikeda}},
  \bibinfo {author} {\bibfnamefont {F.}~\bibnamefont {Mitsuhashi}}, \ and\
  \bibinfo {author} {\bibfnamefont {S.}~\bibnamefont {Sugimoto}},\ }\href
  {\doibase 10.1063/1.3116717} {\bibfield  {journal} {\bibinfo  {journal}
  {Applied Physics Letters}\ }\textbf {\bibinfo {volume} {94}},\ \bibinfo {eid}
  {162504} (\bibinfo {year} {2009})}\BibitemShut {NoStop}%
\bibitem [{\citenamefont {Liu}\ \emph {et~al.}(2012)\citenamefont {Liu},
  \citenamefont {Honda}, \citenamefont {Taira}, \citenamefont {Matsuda},
  \citenamefont {Arita}, \citenamefont {Uemura},\ and\ \citenamefont
  {Yamamoto}}]{Yamamoto2012MTJ}%
  \BibitemOpen
  \bibfield  {author} {\bibinfo {author} {\bibfnamefont {H.-X.}\ \bibnamefont
  {Liu}}, \bibinfo {author} {\bibfnamefont {Y.}~\bibnamefont {Honda}}, \bibinfo
  {author} {\bibfnamefont {T.}~\bibnamefont {Taira}}, \bibinfo {author}
  {\bibfnamefont {K.-i.}\ \bibnamefont {Matsuda}}, \bibinfo {author}
  {\bibfnamefont {M.}~\bibnamefont {Arita}}, \bibinfo {author} {\bibfnamefont
  {T.}~\bibnamefont {Uemura}}, \ and\ \bibinfo {author} {\bibfnamefont
  {M.}~\bibnamefont {Yamamoto}},\ }\href {\doibase 10.1063/1.4755773}
  {\bibfield  {journal} {\bibinfo  {journal} {Applied Physics Letters}\
  }\textbf {\bibinfo {volume} {101}},\ \bibinfo {eid} {132418} (\bibinfo {year}
  {2012})}\BibitemShut {NoStop}%
\bibitem [{\citenamefont {Liu}\ \emph {et~al.}(2015)\citenamefont {Liu},
  \citenamefont {Kawami}, \citenamefont {Moges}, \citenamefont {Uemura},
  \citenamefont {Yamamoto}, \citenamefont {Shi},\ and\ \citenamefont
  {Voyles}}]{Yamamoto2015MTJ}%
  \BibitemOpen
  \bibfield  {author} {\bibinfo {author} {\bibfnamefont {H.-X.}\ \bibnamefont
  {Liu}}, \bibinfo {author} {\bibfnamefont {T.}~\bibnamefont {Kawami}},
  \bibinfo {author} {\bibfnamefont {K.}~\bibnamefont {Moges}}, \bibinfo
  {author} {\bibfnamefont {T.}~\bibnamefont {Uemura}}, \bibinfo {author}
  {\bibfnamefont {M.}~\bibnamefont {Yamamoto}}, \bibinfo {author}
  {\bibfnamefont {F.}~\bibnamefont {Shi}}, \ and\ \bibinfo {author}
  {\bibfnamefont {P.~M.}\ \bibnamefont {Voyles}},\ }\href
  {http://stacks.iop.org/0022-3727/48/i=16/a=164001} {\bibfield  {journal}
  {\bibinfo  {journal} {Journal of Physics D: Applied Physics}\ }\textbf
  {\bibinfo {volume} {48}},\ \bibinfo {pages} {164001} (\bibinfo {year}
  {2015})}\BibitemShut {NoStop}%
\bibitem [{\citenamefont {Iwase}\ \emph {et~al.}(2009)\citenamefont {Iwase},
  \citenamefont {Sakuraba}, \citenamefont {Bosu}, \citenamefont {Saito},
  \citenamefont {Mitani},\ and\ \citenamefont {Takanashi}}]{Takanashi2009GMR}%
  \BibitemOpen
  \bibfield  {author} {\bibinfo {author} {\bibfnamefont {T.}~\bibnamefont
  {Iwase}}, \bibinfo {author} {\bibfnamefont {Y.}~\bibnamefont {Sakuraba}},
  \bibinfo {author} {\bibfnamefont {S.}~\bibnamefont {Bosu}}, \bibinfo {author}
  {\bibfnamefont {K.}~\bibnamefont {Saito}}, \bibinfo {author} {\bibfnamefont
  {S.}~\bibnamefont {Mitani}}, \ and\ \bibinfo {author} {\bibfnamefont
  {K.}~\bibnamefont {Takanashi}},\ }\href
  {http://stacks.iop.org/1882-0786/2/i=6/a=063003} {\bibfield  {journal}
  {\bibinfo  {journal} {Applied Physics Express}\ }\textbf {\bibinfo {volume}
  {2}},\ \bibinfo {pages} {063003} (\bibinfo {year} {2009})}\BibitemShut
  {NoStop}%
\bibitem [{\citenamefont {Nakatani}\ \emph {et~al.}(2010)\citenamefont
  {Nakatani}, \citenamefont {Furubayashi}, \citenamefont {Kasai}, \citenamefont
  {Sukegawa}, \citenamefont {Takahashi}, \citenamefont {Mitani},\ and\
  \citenamefont {Hono}}]{Hono2010GMR}%
  \BibitemOpen
  \bibfield  {author} {\bibinfo {author} {\bibfnamefont {T.~M.}\ \bibnamefont
  {Nakatani}}, \bibinfo {author} {\bibfnamefont {T.}~\bibnamefont
  {Furubayashi}}, \bibinfo {author} {\bibfnamefont {S.}~\bibnamefont {Kasai}},
  \bibinfo {author} {\bibfnamefont {H.}~\bibnamefont {Sukegawa}}, \bibinfo
  {author} {\bibfnamefont {Y.~K.}\ \bibnamefont {Takahashi}}, \bibinfo {author}
  {\bibfnamefont {S.}~\bibnamefont {Mitani}}, \ and\ \bibinfo {author}
  {\bibfnamefont {K.}~\bibnamefont {Hono}},\ }\href {\doibase
  10.1063/1.3432070} {\bibfield  {journal} {\bibinfo  {journal} {Applied
  Physics Letters}\ }\textbf {\bibinfo {volume} {96}},\ \bibinfo {eid} {212501}
  (\bibinfo {year} {2010})}\BibitemShut {NoStop}%
\bibitem [{\citenamefont {Takahashi}\ \emph {et~al.}(2011)\citenamefont
  {Takahashi}, \citenamefont {Srinivasan}, \citenamefont {Varaprasad},
  \citenamefont {Rajanikanth}, \citenamefont {Hase}, \citenamefont {Nakatani},
  \citenamefont {Kasai}, \citenamefont {Furubayashi},\ and\ \citenamefont
  {Hono}}]{Hono2011GMR}%
  \BibitemOpen
  \bibfield  {author} {\bibinfo {author} {\bibfnamefont {Y.~K.}\ \bibnamefont
  {Takahashi}}, \bibinfo {author} {\bibfnamefont {A.}~\bibnamefont
  {Srinivasan}}, \bibinfo {author} {\bibfnamefont {B.}~\bibnamefont
  {Varaprasad}}, \bibinfo {author} {\bibfnamefont {A.}~\bibnamefont
  {Rajanikanth}}, \bibinfo {author} {\bibfnamefont {N.}~\bibnamefont {Hase}},
  \bibinfo {author} {\bibfnamefont {T.~M.}\ \bibnamefont {Nakatani}}, \bibinfo
  {author} {\bibfnamefont {S.}~\bibnamefont {Kasai}}, \bibinfo {author}
  {\bibfnamefont {T.}~\bibnamefont {Furubayashi}}, \ and\ \bibinfo {author}
  {\bibfnamefont {K.}~\bibnamefont {Hono}},\ }\href {\doibase
  10.1063/1.3576923} {\bibfield  {journal} {\bibinfo  {journal} {Applied
  Physics Letters}\ }\textbf {\bibinfo {volume} {98}},\ \bibinfo {eid} {152501}
  (\bibinfo {year} {2011})}\BibitemShut {NoStop}%
\bibitem [{\citenamefont {Sato}\ \emph {et~al.}(2011)\citenamefont {Sato},
  \citenamefont {Oogane}, \citenamefont {Naganuma},\ and\ \citenamefont
  {Ando}}]{Ando2011GMR}%
  \BibitemOpen
  \bibfield  {author} {\bibinfo {author} {\bibfnamefont {J.}~\bibnamefont
  {Sato}}, \bibinfo {author} {\bibfnamefont {M.}~\bibnamefont {Oogane}},
  \bibinfo {author} {\bibfnamefont {H.}~\bibnamefont {Naganuma}}, \ and\
  \bibinfo {author} {\bibfnamefont {Y.}~\bibnamefont {Ando}},\ }\href
  {http://stacks.iop.org/1882-0786/4/i=11/a=113005} {\bibfield  {journal}
  {\bibinfo  {journal} {Applied Physics Express}\ }\textbf {\bibinfo {volume}
  {4}},\ \bibinfo {pages} {113005} (\bibinfo {year} {2011})}\BibitemShut
  {NoStop}%
\bibitem [{\citenamefont {Chadov}\ \emph {et~al.}(2011)\citenamefont {Chadov},
  \citenamefont {Graf}, \citenamefont {Chadova}, \citenamefont {Dai},
  \citenamefont {Casper}, \citenamefont {Fecher},\ and\ \citenamefont
  {Felser}}]{PhysRevLett.107.047202}%
  \BibitemOpen
  \bibfield  {author} {\bibinfo {author} {\bibfnamefont {S.}~\bibnamefont
  {Chadov}}, \bibinfo {author} {\bibfnamefont {T.}~\bibnamefont {Graf}},
  \bibinfo {author} {\bibfnamefont {K.}~\bibnamefont {Chadova}}, \bibinfo
  {author} {\bibfnamefont {X.}~\bibnamefont {Dai}}, \bibinfo {author}
  {\bibfnamefont {F.}~\bibnamefont {Casper}}, \bibinfo {author} {\bibfnamefont
  {G.~H.}\ \bibnamefont {Fecher}}, \ and\ \bibinfo {author} {\bibfnamefont
  {C.}~\bibnamefont {Felser}},\ }\href {\doibase
  10.1103/PhysRevLett.107.047202} {\bibfield  {journal} {\bibinfo  {journal}
  {Phys. Rev. Lett.}\ }\textbf {\bibinfo {volume} {107}},\ \bibinfo {pages}
  {047202} (\bibinfo {year} {2011})}\BibitemShut {NoStop}%
\bibitem [{\citenamefont {Worledge}\ and\ \citenamefont
  {Geballe}(2000)}]{WorledgeandGeballe}%
  \BibitemOpen
  \bibfield  {author} {\bibinfo {author} {\bibfnamefont {D.~C.}\ \bibnamefont
  {Worledge}}\ and\ \bibinfo {author} {\bibfnamefont {T.~H.}\ \bibnamefont
  {Geballe}},\ }\href {\doibase 10.1063/1.1315619} {\bibfield  {journal}
  {\bibinfo  {journal} {Journal of Applied Physics}\ }\textbf {\bibinfo
  {volume} {88}},\ \bibinfo {pages} {5277} (\bibinfo {year}
  {2000})}\BibitemShut {NoStop}%
\bibitem [{\citenamefont {LeClair}\ \emph {et~al.}(2002)\citenamefont
  {LeClair}, \citenamefont {Ha}, \citenamefont {Swagten}, \citenamefont
  {Kohlhepp}, \citenamefont {van~de Vin},\ and\ \citenamefont
  {de~Jonge}}]{LeClairspinfilter}%
  \BibitemOpen
  \bibfield  {author} {\bibinfo {author} {\bibfnamefont {P.}~\bibnamefont
  {LeClair}}, \bibinfo {author} {\bibfnamefont {J.~K.}\ \bibnamefont {Ha}},
  \bibinfo {author} {\bibfnamefont {H.~J.~M.}\ \bibnamefont {Swagten}},
  \bibinfo {author} {\bibfnamefont {J.~T.}\ \bibnamefont {Kohlhepp}}, \bibinfo
  {author} {\bibfnamefont {C.~H.}\ \bibnamefont {van~de Vin}}, \ and\ \bibinfo
  {author} {\bibfnamefont {W.~J.~M.}\ \bibnamefont {de~Jonge}},\ }\href
  {\doibase 10.1063/1.1436284} {\bibfield  {journal} {\bibinfo  {journal}
  {Applied Physics Letters}\ }\textbf {\bibinfo {volume} {80}},\ \bibinfo
  {pages} {625} (\bibinfo {year} {2002})}\BibitemShut {NoStop}%
\bibitem [{\citenamefont {Ouardi}\ \emph {et~al.}(2013)\citenamefont {Ouardi},
  \citenamefont {Fecher}, \citenamefont {Felser},\ and\ \citenamefont
  {K\"ubler}}]{PhysRevLett.110.100401}%
  \BibitemOpen
  \bibfield  {author} {\bibinfo {author} {\bibfnamefont {S.}~\bibnamefont
  {Ouardi}}, \bibinfo {author} {\bibfnamefont {G.~H.}\ \bibnamefont {Fecher}},
  \bibinfo {author} {\bibfnamefont {C.}~\bibnamefont {Felser}}, \ and\ \bibinfo
  {author} {\bibfnamefont {J.}~\bibnamefont {K\"ubler}},\ }\href {\doibase
  10.1103/PhysRevLett.110.100401} {\bibfield  {journal} {\bibinfo  {journal}
  {Phys. Rev. Lett.}\ }\textbf {\bibinfo {volume} {110}},\ \bibinfo {pages}
  {100401} (\bibinfo {year} {2013})}\BibitemShut {NoStop}%
\bibitem [{\citenamefont {Butler}\ \emph {et~al.}(2016)\citenamefont {Butler},
  \citenamefont {Ghosh} \emph {et~al.}}]{Heusl89:online}%
  \BibitemOpen
  \bibfield  {author} {\bibinfo {author} {\bibfnamefont {W.~H.}\ \bibnamefont
  {Butler}}, \bibinfo {author} {\bibfnamefont {A.~W.}\ \bibnamefont {Ghosh}},
  \emph {et~al.},\ }\href@noop {} {\enquote {\bibinfo {title} {Heuslers
  home},}\ }\bibinfo {howpublished} {\url{http://heusleralloys.mint.ua.edu/}}
  (\bibinfo {year} {2016})\BibitemShut {NoStop}%
\bibitem [{\citenamefont {Ma}\ \emph {et~al.}(2017)\citenamefont {Ma},
  \citenamefont {Hegde}, \citenamefont {Munira}, \citenamefont {Xie},
  \citenamefont {Keshavarz}, \citenamefont {Mildebrath}, \citenamefont
  {Wolverton}, \citenamefont {Ghosh},\ and\ \citenamefont
  {Butler}}]{ma2016computational}%
  \BibitemOpen
  \bibfield  {author} {\bibinfo {author} {\bibfnamefont {J.}~\bibnamefont
  {Ma}}, \bibinfo {author} {\bibfnamefont {V.~I.}\ \bibnamefont {Hegde}},
  \bibinfo {author} {\bibfnamefont {K.}~\bibnamefont {Munira}}, \bibinfo
  {author} {\bibfnamefont {Y.}~\bibnamefont {Xie}}, \bibinfo {author}
  {\bibfnamefont {S.}~\bibnamefont {Keshavarz}}, \bibinfo {author}
  {\bibfnamefont {D.~T.}\ \bibnamefont {Mildebrath}}, \bibinfo {author}
  {\bibfnamefont {C.}~\bibnamefont {Wolverton}}, \bibinfo {author}
  {\bibfnamefont {A.~W.}\ \bibnamefont {Ghosh}}, \ and\ \bibinfo {author}
  {\bibfnamefont {W.~H.}\ \bibnamefont {Butler}},\ }\href {\doibase
  10.1103/PhysRevB.95.024411} {\bibfield  {journal} {\bibinfo  {journal} {Phys.
  Rev. B}\ }\textbf {\bibinfo {volume} {95}},\ \bibinfo {pages} {024411}
  (\bibinfo {year} {2017})}\BibitemShut {NoStop}%
\bibitem [{\citenamefont {Slater}\ and\ \citenamefont
  {Koster}(1954)}]{PhysRev.94.1498}%
  \BibitemOpen
  \bibfield  {author} {\bibinfo {author} {\bibfnamefont {J.~C.}\ \bibnamefont
  {Slater}}\ and\ \bibinfo {author} {\bibfnamefont {G.~F.}\ \bibnamefont
  {Koster}},\ }\href {\doibase 10.1103/PhysRev.94.1498} {\bibfield  {journal}
  {\bibinfo  {journal} {Phys. Rev.}\ }\textbf {\bibinfo {volume} {94}},\
  \bibinfo {pages} {1498} (\bibinfo {year} {1954})}\BibitemShut {NoStop}%
\bibitem [{\citenamefont {Slater}(1937)}]{Slater1937}%
  \BibitemOpen
  \bibfield  {author} {\bibinfo {author} {\bibfnamefont {J.~C.}\ \bibnamefont
  {Slater}},\ }\href {\doibase 10.1063/1.1710311} {\bibfield  {journal}
  {\bibinfo  {journal} {Journal of Applied Physics}\ }\textbf {\bibinfo
  {volume} {8}},\ \bibinfo {pages} {385} (\bibinfo {year} {1937})}\BibitemShut
  {NoStop}%
\bibitem [{\citenamefont {Pauling}(1938)}]{PhysRev.54.899}%
  \BibitemOpen
  \bibfield  {author} {\bibinfo {author} {\bibfnamefont {L.}~\bibnamefont
  {Pauling}},\ }\href {\doibase 10.1103/PhysRev.54.899} {\bibfield  {journal}
  {\bibinfo  {journal} {Phys. Rev.}\ }\textbf {\bibinfo {volume} {54}},\
  \bibinfo {pages} {899} (\bibinfo {year} {1938})}\BibitemShut {NoStop}%
\bibitem [{\citenamefont {Skaftouros}\ \emph
  {et~al.}(2013{\natexlab{b}})\citenamefont {Skaftouros}, \citenamefont
  {{\"O}zdo{\u{g}}an}, \citenamefont {{\c{S}}a{\c{s}}{\i}o{\u{g}}lu},\ and\
  \citenamefont {Galanakis}}]{Galanakis2013SGS}%
  \BibitemOpen
  \bibfield  {author} {\bibinfo {author} {\bibfnamefont {S.}~\bibnamefont
  {Skaftouros}}, \bibinfo {author} {\bibfnamefont {K.}~\bibnamefont
  {{\"O}zdo{\u{g}}an}}, \bibinfo {author} {\bibfnamefont {E.}~\bibnamefont
  {{\c{S}}a{\c{s}}{\i}o{\u{g}}lu}}, \ and\ \bibinfo {author} {\bibfnamefont
  {I.}~\bibnamefont {Galanakis}},\ }\href {\doibase 10.1063/1.4775599}
  {\bibfield  {journal} {\bibinfo  {journal} {Applied Physics Letters}\
  }\textbf {\bibinfo {volume} {102}},\ \bibinfo {eid} {022402} (\bibinfo {year}
  {2013}{\natexlab{b}})}\BibitemShut {NoStop}%
\bibitem [{\citenamefont {Galanakis}\ \emph {et~al.}(2014)\citenamefont
  {Galanakis}, \citenamefont {{\"O}zdo{\u{g}}an}, \citenamefont
  {{\c{S}}a{\c{s}}{\i}o{\u{g}}lu},\ and\ \citenamefont
  {Bl{\"u}gel}}]{Galanakis2014Mn2CoAl}%
  \BibitemOpen
  \bibfield  {author} {\bibinfo {author} {\bibfnamefont {I.}~\bibnamefont
  {Galanakis}}, \bibinfo {author} {\bibfnamefont {K.}~\bibnamefont
  {{\"O}zdo{\u{g}}an}}, \bibinfo {author} {\bibfnamefont {E.}~\bibnamefont
  {{\c{S}}a{\c{s}}{\i}o{\u{g}}lu}}, \ and\ \bibinfo {author} {\bibfnamefont
  {S.}~\bibnamefont {Bl{\"u}gel}},\ }\href {\doibase 10.1063/1.4867917}
  {\bibfield  {journal} {\bibinfo  {journal} {Journal of Applied Physics}\
  }\textbf {\bibinfo {volume} {115}},\ \bibinfo {eid} {093908} (\bibinfo {year}
  {2014})}\BibitemShut {NoStop}%
\bibitem [{\citenamefont {Lakshmi}\ \emph {et~al.}(2002)\citenamefont
  {Lakshmi}, \citenamefont {Pandey},\ and\ \citenamefont
  {Venugopalan}}]{Lakshmi2002Mn2CoSn}%
  \BibitemOpen
  \bibfield  {author} {\bibinfo {author} {\bibfnamefont {N.}~\bibnamefont
  {Lakshmi}}, \bibinfo {author} {\bibfnamefont {A.}~\bibnamefont {Pandey}}, \
  and\ \bibinfo {author} {\bibfnamefont {K.}~\bibnamefont {Venugopalan}},\
  }\href {\doibase 10.1007/BF02704123} {\bibfield  {journal} {\bibinfo
  {journal} {Bulletin of Materials Science}\ }\textbf {\bibinfo {volume}
  {25}},\ \bibinfo {pages} {309} (\bibinfo {year} {2002})}\BibitemShut
  {NoStop}%
\bibitem [{\citenamefont {Meinert}\ \emph
  {et~al.}(2011{\natexlab{a}})\citenamefont {Meinert}, \citenamefont
  {Schmalhorst}, \citenamefont {Klewe}, \citenamefont {Reiss}, \citenamefont
  {Arenholz}, \citenamefont {B\"ohnert},\ and\ \citenamefont
  {Nielsch}}]{PhysRevB.84.132405}%
  \BibitemOpen
  \bibfield  {author} {\bibinfo {author} {\bibfnamefont {M.}~\bibnamefont
  {Meinert}}, \bibinfo {author} {\bibfnamefont {J.-M.}\ \bibnamefont
  {Schmalhorst}}, \bibinfo {author} {\bibfnamefont {C.}~\bibnamefont {Klewe}},
  \bibinfo {author} {\bibfnamefont {G.}~\bibnamefont {Reiss}}, \bibinfo
  {author} {\bibfnamefont {E.}~\bibnamefont {Arenholz}}, \bibinfo {author}
  {\bibfnamefont {T.}~\bibnamefont {B\"ohnert}}, \ and\ \bibinfo {author}
  {\bibfnamefont {K.}~\bibnamefont {Nielsch}},\ }\href {\doibase
  10.1103/PhysRevB.84.132405} {\bibfield  {journal} {\bibinfo  {journal} {Phys.
  Rev. B}\ }\textbf {\bibinfo {volume} {84}},\ \bibinfo {pages} {132405}
  (\bibinfo {year} {2011}{\natexlab{a}})}\BibitemShut {NoStop}%
\bibitem [{\citenamefont {Winterlik}\ \emph {et~al.}(2011)\citenamefont
  {Winterlik}, \citenamefont {Fecher}, \citenamefont {Balke}, \citenamefont
  {Graf}, \citenamefont {Alijani}, \citenamefont {Ksenofontov}, \citenamefont
  {Jenkins}, \citenamefont {Meshcheriakova}, \citenamefont {Felser},
  \citenamefont {Liu}, \citenamefont {Ueda}, \citenamefont {Kobayashi},
  \citenamefont {Nakamura},\ and\ \citenamefont
  {W\'ojcik}}]{PhysRevB.83.174448}%
  \BibitemOpen
  \bibfield  {author} {\bibinfo {author} {\bibfnamefont {J.}~\bibnamefont
  {Winterlik}}, \bibinfo {author} {\bibfnamefont {G.~H.}\ \bibnamefont
  {Fecher}}, \bibinfo {author} {\bibfnamefont {B.}~\bibnamefont {Balke}},
  \bibinfo {author} {\bibfnamefont {T.}~\bibnamefont {Graf}}, \bibinfo {author}
  {\bibfnamefont {V.}~\bibnamefont {Alijani}}, \bibinfo {author} {\bibfnamefont
  {V.}~\bibnamefont {Ksenofontov}}, \bibinfo {author} {\bibfnamefont {C.~A.}\
  \bibnamefont {Jenkins}}, \bibinfo {author} {\bibfnamefont {O.}~\bibnamefont
  {Meshcheriakova}}, \bibinfo {author} {\bibfnamefont {C.}~\bibnamefont
  {Felser}}, \bibinfo {author} {\bibfnamefont {G.}~\bibnamefont {Liu}},
  \bibinfo {author} {\bibfnamefont {S.}~\bibnamefont {Ueda}}, \bibinfo {author}
  {\bibfnamefont {K.}~\bibnamefont {Kobayashi}}, \bibinfo {author}
  {\bibfnamefont {T.}~\bibnamefont {Nakamura}}, \ and\ \bibinfo {author}
  {\bibfnamefont {M.}~\bibnamefont {W\'ojcik}},\ }\href {\doibase
  10.1103/PhysRevB.83.174448} {\bibfield  {journal} {\bibinfo  {journal} {Phys.
  Rev. B}\ }\textbf {\bibinfo {volume} {83}},\ \bibinfo {pages} {174448}
  (\bibinfo {year} {2011})}\BibitemShut {NoStop}%
\bibitem [{\citenamefont {Xin}\ \emph {et~al.}(2017)\citenamefont {Xin},
  \citenamefont {Hao}, \citenamefont {Ma}, \citenamefont {Luo}, \citenamefont
  {Meng}, \citenamefont {Liu}, \citenamefont {Liu},\ and\ \citenamefont
  {Wu}}]{Xin201710}%
  \BibitemOpen
  \bibfield  {author} {\bibinfo {author} {\bibfnamefont {Y.}~\bibnamefont
  {Xin}}, \bibinfo {author} {\bibfnamefont {H.}~\bibnamefont {Hao}}, \bibinfo
  {author} {\bibfnamefont {Y.}~\bibnamefont {Ma}}, \bibinfo {author}
  {\bibfnamefont {H.}~\bibnamefont {Luo}}, \bibinfo {author} {\bibfnamefont
  {F.}~\bibnamefont {Meng}}, \bibinfo {author} {\bibfnamefont {H.}~\bibnamefont
  {Liu}}, \bibinfo {author} {\bibfnamefont {E.}~\bibnamefont {Liu}}, \ and\
  \bibinfo {author} {\bibfnamefont {G.}~\bibnamefont {Wu}},\ }\href {\doibase
  10.1016/j.intermet.2016.10.001} {\bibfield  {journal} {\bibinfo  {journal}
  {Intermetallics}\ }\textbf {\bibinfo {volume} {80}},\ \bibinfo {pages} {10 }
  (\bibinfo {year} {2017})}\BibitemShut {NoStop}%
\bibitem [{\citenamefont {Liu}\ \emph {et~al.}(2008)\citenamefont {Liu},
  \citenamefont {Dai}, \citenamefont {Liu}, \citenamefont {Chen}, \citenamefont
  {Li}, \citenamefont {Xiao},\ and\ \citenamefont {Wu}}]{PhysRevB.77.014424}%
  \BibitemOpen
  \bibfield  {author} {\bibinfo {author} {\bibfnamefont {G.~D.}\ \bibnamefont
  {Liu}}, \bibinfo {author} {\bibfnamefont {X.~F.}\ \bibnamefont {Dai}},
  \bibinfo {author} {\bibfnamefont {H.~Y.}\ \bibnamefont {Liu}}, \bibinfo
  {author} {\bibfnamefont {J.~L.}\ \bibnamefont {Chen}}, \bibinfo {author}
  {\bibfnamefont {Y.~X.}\ \bibnamefont {Li}}, \bibinfo {author} {\bibfnamefont
  {G.}~\bibnamefont {Xiao}}, \ and\ \bibinfo {author} {\bibfnamefont {G.~H.}\
  \bibnamefont {Wu}},\ }\href {\doibase 10.1103/PhysRevB.77.014424} {\bibfield
  {journal} {\bibinfo  {journal} {Phys. Rev. B}\ }\textbf {\bibinfo {volume}
  {77}},\ \bibinfo {pages} {014424} (\bibinfo {year} {2008})}\BibitemShut
  {NoStop}%
\bibitem [{\citenamefont {Ahmadian}\ and\ \citenamefont
  {Salary}(2014)}]{Ahmadian2014243}%
  \BibitemOpen
  \bibfield  {author} {\bibinfo {author} {\bibfnamefont {F.}~\bibnamefont
  {Ahmadian}}\ and\ \bibinfo {author} {\bibfnamefont {A.}~\bibnamefont
  {Salary}},\ }\href {\doibase 10.1016/j.intermet.2013.11.021} {\bibfield
  {journal} {\bibinfo  {journal} {Intermetallics}\ }\textbf {\bibinfo {volume}
  {46}},\ \bibinfo {pages} {243 } (\bibinfo {year} {2014})}\BibitemShut
  {NoStop}%
\bibitem [{\citenamefont {Jakobsson}\ \emph {et~al.}(2015)\citenamefont
  {Jakobsson}, \citenamefont {Mavropoulos}, \citenamefont {\ifmmode
  \mbox{\c{S}}\else \c{S}\fi{}a\ifmmode \mbox{\c{s}}\else \c{s}\fi{}\ifmmode
  \imath \else \i \fi{}o\ifmmode~\breve{g}\else \u{g}\fi{}lu}, \citenamefont
  {Bl\"ugel}, \citenamefont {Le\ifmmode \check{z}\else
  \v{z}\fi{}ai\ifmmode~\acute{c}\else \'{c}\fi{}}, \citenamefont {Sanyal},\
  and\ \citenamefont {Galanakis}}]{PhysRevB.91.174439}%
  \BibitemOpen
  \bibfield  {author} {\bibinfo {author} {\bibfnamefont {A.}~\bibnamefont
  {Jakobsson}}, \bibinfo {author} {\bibfnamefont {P.}~\bibnamefont
  {Mavropoulos}}, \bibinfo {author} {\bibfnamefont {E.}~\bibnamefont {\ifmmode
  \mbox{\c{S}}\else \c{S}\fi{}a\ifmmode \mbox{\c{s}}\else \c{s}\fi{}\ifmmode
  \imath \else \i \fi{}o\ifmmode~\breve{g}\else \u{g}\fi{}lu}}, \bibinfo
  {author} {\bibfnamefont {S.}~\bibnamefont {Bl\"ugel}}, \bibinfo {author}
  {\bibfnamefont {M.}~\bibnamefont {Le\ifmmode \check{z}\else
  \v{z}\fi{}ai\ifmmode~\acute{c}\else \'{c}\fi{}}}, \bibinfo {author}
  {\bibfnamefont {B.}~\bibnamefont {Sanyal}}, \ and\ \bibinfo {author}
  {\bibfnamefont {I.}~\bibnamefont {Galanakis}},\ }\href {\doibase
  10.1103/PhysRevB.91.174439} {\bibfield  {journal} {\bibinfo  {journal} {Phys.
  Rev. B}\ }\textbf {\bibinfo {volume} {91}},\ \bibinfo {pages} {174439}
  (\bibinfo {year} {2015})}\BibitemShut {NoStop}%
\bibitem [{\citenamefont {Saal}\ \emph {et~al.}(2013)\citenamefont {Saal},
  \citenamefont {Kirklin}, \citenamefont {Aykol}, \citenamefont {Meredig},\
  and\ \citenamefont {Wolverton}}]{raey}%
  \BibitemOpen
  \bibfield  {author} {\bibinfo {author} {\bibfnamefont {J.}~\bibnamefont
  {Saal}}, \bibinfo {author} {\bibfnamefont {S.}~\bibnamefont {Kirklin}},
  \bibinfo {author} {\bibfnamefont {M.}~\bibnamefont {Aykol}}, \bibinfo
  {author} {\bibfnamefont {B.}~\bibnamefont {Meredig}}, \ and\ \bibinfo
  {author} {\bibfnamefont {C.}~\bibnamefont {Wolverton}},\ }\href {\doibase
  10.1007/s11837-013-0755-4} {\bibfield  {journal} {\bibinfo  {journal} {JOM}\
  }\textbf {\bibinfo {volume} {65}},\ \bibinfo {pages} {1501} (\bibinfo {year}
  {2013})}\BibitemShut {NoStop}%
\bibitem [{\citenamefont {Kirklin}\ \emph {et~al.}(2015)\citenamefont
  {Kirklin}, \citenamefont {Saal}, \citenamefont {Meredig}, \citenamefont
  {Thompson}, \citenamefont {Doak}, \citenamefont {Aykol}, \citenamefont
  {R{\"u}hl},\ and\ \citenamefont {Wolverton}}]{oqmd_npj_2015}%
  \BibitemOpen
  \bibfield  {author} {\bibinfo {author} {\bibfnamefont {S.}~\bibnamefont
  {Kirklin}}, \bibinfo {author} {\bibfnamefont {J.~E.}\ \bibnamefont {Saal}},
  \bibinfo {author} {\bibfnamefont {B.}~\bibnamefont {Meredig}}, \bibinfo
  {author} {\bibfnamefont {A.}~\bibnamefont {Thompson}}, \bibinfo {author}
  {\bibfnamefont {J.~W.}\ \bibnamefont {Doak}}, \bibinfo {author}
  {\bibfnamefont {M.}~\bibnamefont {Aykol}}, \bibinfo {author} {\bibfnamefont
  {S.}~\bibnamefont {R{\"u}hl}}, \ and\ \bibinfo {author} {\bibfnamefont
  {C.}~\bibnamefont {Wolverton}},\ }\href {\doibase
  10.1038/npjcompumats.2015.10} {\bibfield  {journal} {\bibinfo  {journal} {npj
  Computational Materials}\ }\textbf {\bibinfo {volume} {1}},\ \bibinfo {pages}
  {15010} (\bibinfo {year} {2015})}\BibitemShut {NoStop}%
\bibitem [{\citenamefont {Kresse}\ and\ \citenamefont
  {Furthmüller}(1996)}]{Kresse199615}%
  \BibitemOpen
  \bibfield  {author} {\bibinfo {author} {\bibfnamefont {G.}~\bibnamefont
  {Kresse}}\ and\ \bibinfo {author} {\bibfnamefont {J.}~\bibnamefont
  {Furthmüller}},\ }\href {\doibase dx.doi.org/10.1016/0927-0256(96)00008-0}
  {\bibfield  {journal} {\bibinfo  {journal} {Computational Materials Science}\
  }\textbf {\bibinfo {volume} {6}},\ \bibinfo {pages} {15 } (\bibinfo {year}
  {1996})}\BibitemShut {NoStop}%
\bibitem [{\citenamefont {Bl\"ochl}(1994)}]{PhysRevB.50.17953}%
  \BibitemOpen
  \bibfield  {author} {\bibinfo {author} {\bibfnamefont {P.~E.}\ \bibnamefont
  {Bl\"ochl}},\ }\href {\doibase 10.1103/PhysRevB.50.17953} {\bibfield
  {journal} {\bibinfo  {journal} {Phys. Rev. B}\ }\textbf {\bibinfo {volume}
  {50}},\ \bibinfo {pages} {17953} (\bibinfo {year} {1994})}\BibitemShut
  {NoStop}%
\bibitem [{\citenamefont {Perdew}\ \emph {et~al.}(1996)\citenamefont {Perdew},
  \citenamefont {Burke},\ and\ \citenamefont
  {Ernzerhof}}]{PhysRevLett.77.3865}%
  \BibitemOpen
  \bibfield  {author} {\bibinfo {author} {\bibfnamefont {J.~P.}\ \bibnamefont
  {Perdew}}, \bibinfo {author} {\bibfnamefont {K.}~\bibnamefont {Burke}}, \
  and\ \bibinfo {author} {\bibfnamefont {M.}~\bibnamefont {Ernzerhof}},\ }\href
  {\doibase 10.1103/PhysRevLett.77.3865} {\bibfield  {journal} {\bibinfo
  {journal} {Phys. Rev. Lett.}\ }\textbf {\bibinfo {volume} {77}},\ \bibinfo
  {pages} {3865} (\bibinfo {year} {1996})}\BibitemShut {NoStop}%
\bibitem [{\citenamefont {Monkhorst}\ and\ \citenamefont
  {Pack}(1976)}]{PhysRevB.13.5188}%
  \BibitemOpen
  \bibfield  {author} {\bibinfo {author} {\bibfnamefont {H.~J.}\ \bibnamefont
  {Monkhorst}}\ and\ \bibinfo {author} {\bibfnamefont {J.~D.}\ \bibnamefont
  {Pack}},\ }\href {\doibase 10.1103/PhysRevB.13.5188} {\bibfield  {journal}
  {\bibinfo  {journal} {Phys. Rev. B}\ }\textbf {\bibinfo {volume} {13}},\
  \bibinfo {pages} {5188} (\bibinfo {year} {1976})}\BibitemShut {NoStop}%
\bibitem [{\citenamefont {Bl\"ochl}\ \emph {et~al.}(1994)\citenamefont
  {Bl\"ochl}, \citenamefont {Jepsen},\ and\ \citenamefont
  {Andersen}}]{PhysRevB.49.16223}%
  \BibitemOpen
  \bibfield  {author} {\bibinfo {author} {\bibfnamefont {P.~E.}\ \bibnamefont
  {Bl\"ochl}}, \bibinfo {author} {\bibfnamefont {O.}~\bibnamefont {Jepsen}}, \
  and\ \bibinfo {author} {\bibfnamefont {O.~K.}\ \bibnamefont {Andersen}},\
  }\href {\doibase 10.1103/PhysRevB.49.16223} {\bibfield  {journal} {\bibinfo
  {journal} {Phys. Rev. B}\ }\textbf {\bibinfo {volume} {49}},\ \bibinfo
  {pages} {16223} (\bibinfo {year} {1994})}\BibitemShut {NoStop}%
\bibitem [{\citenamefont {Vosko}\ \emph {et~al.}(1980)\citenamefont {Vosko},
  \citenamefont {Wilk},\ and\ \citenamefont {Nusair}}]{vosko1980accurate}%
  \BibitemOpen
  \bibfield  {author} {\bibinfo {author} {\bibfnamefont {S.~H.}\ \bibnamefont
  {Vosko}}, \bibinfo {author} {\bibfnamefont {L.}~\bibnamefont {Wilk}}, \ and\
  \bibinfo {author} {\bibfnamefont {M.}~\bibnamefont {Nusair}},\ }\href
  {\doibase 10.1139/p80-159} {\bibfield  {journal} {\bibinfo  {journal}
  {Canadian Journal of Physics}\ }\textbf {\bibinfo {volume} {58}},\ \bibinfo
  {pages} {1200} (\bibinfo {year} {1980})}\BibitemShut {NoStop}%
\bibitem [{\citenamefont {…zdog÷an}\ and\ \citenamefont
  {Galanakis}(2009)}]{Ozdogan2009L34}%
  \BibitemOpen
  \bibfield  {author} {\bibinfo {author} {\bibfnamefont {K.}~\bibnamefont
  {…zdog÷an}}\ and\ \bibinfo {author} {\bibfnamefont {I.}~\bibnamefont
  {Galanakis}},\ }\href {\doibase dx.doi.org/10.1016/j.jmmm.2009.01.006}
  {\bibfield  {journal} {\bibinfo  {journal} {Journal of Magnetism and Magnetic
  Materials}\ }\textbf {\bibinfo {volume} {321}},\ \bibinfo {pages} {L34 }
  (\bibinfo {year} {2009})}\BibitemShut {NoStop}%
\bibitem [{\citenamefont {Luo}\ \emph {et~al.}(2008)\citenamefont {Luo},
  \citenamefont {Zhu}, \citenamefont {Ma}, \citenamefont {Xu}, \citenamefont
  {Zhu}, \citenamefont {Jiang}, \citenamefont {Xu},\ and\ \citenamefont
  {Wu}}]{AlessthanB-1}%
  \BibitemOpen
  \bibfield  {author} {\bibinfo {author} {\bibfnamefont {H.}~\bibnamefont
  {Luo}}, \bibinfo {author} {\bibfnamefont {Z.}~\bibnamefont {Zhu}}, \bibinfo
  {author} {\bibfnamefont {L.}~\bibnamefont {Ma}}, \bibinfo {author}
  {\bibfnamefont {S.}~\bibnamefont {Xu}}, \bibinfo {author} {\bibfnamefont
  {X.}~\bibnamefont {Zhu}}, \bibinfo {author} {\bibfnamefont {C.}~\bibnamefont
  {Jiang}}, \bibinfo {author} {\bibfnamefont {H.}~\bibnamefont {Xu}}, \ and\
  \bibinfo {author} {\bibfnamefont {G.}~\bibnamefont {Wu}},\ }\href
  {http://stacks.iop.org/0022-3727/41/i=5/a=055010} {\bibfield  {journal}
  {\bibinfo  {journal} {Journal of Physics D: Applied Physics}\ }\textbf
  {\bibinfo {volume} {41}},\ \bibinfo {pages} {055010} (\bibinfo {year}
  {2008})}\BibitemShut {NoStop}%
\bibitem [{\citenamefont {Meinert}\ \emph
  {et~al.}(2011{\natexlab{b}})\citenamefont {Meinert}, \citenamefont
  {Schmalhorst},\ and\ \citenamefont {Reiss}}]{AlessthanB-2}%
  \BibitemOpen
  \bibfield  {author} {\bibinfo {author} {\bibfnamefont {M.}~\bibnamefont
  {Meinert}}, \bibinfo {author} {\bibfnamefont {J.-M.}\ \bibnamefont
  {Schmalhorst}}, \ and\ \bibinfo {author} {\bibfnamefont {G.}~\bibnamefont
  {Reiss}},\ }\href {http://stacks.iop.org/0953-8984/23/i=11/a=116005}
  {\bibfield  {journal} {\bibinfo  {journal} {Journal of Physics: Condensed
  Matter}\ }\textbf {\bibinfo {volume} {23}},\ \bibinfo {pages} {116005}
  (\bibinfo {year} {2011}{\natexlab{b}})}\BibitemShut {NoStop}%
\bibitem [{\citenamefont {Xu}\ \emph {et~al.}(2011)\citenamefont {Xu},
  \citenamefont {Zhang},\ and\ \citenamefont {Yan}}]{AlessthanB-3}%
  \BibitemOpen
  \bibfield  {author} {\bibinfo {author} {\bibfnamefont {B.}~\bibnamefont
  {Xu}}, \bibinfo {author} {\bibfnamefont {M.}~\bibnamefont {Zhang}}, \ and\
  \bibinfo {author} {\bibfnamefont {H.}~\bibnamefont {Yan}},\ }\href {\doibase
  10.1002/pssb.201147020} {\bibfield  {journal} {\bibinfo  {journal} {physica
  status solidi b}\ }\textbf {\bibinfo {volume} {248}},\ \bibinfo {pages}
  {2870} (\bibinfo {year} {2011})}\BibitemShut {NoStop}%
\bibitem [{\citenamefont {Pugaczowa-Michalska}(2012)}]{AlessthanB-4}%
  \BibitemOpen
  \bibfield  {author} {\bibinfo {author} {\bibfnamefont {M.}~\bibnamefont
  {Pugaczowa-Michalska}},\ }\href {\doibase
  dx.doi.org/10.1016/j.intermet.2012.01.004} {\bibfield  {journal} {\bibinfo
  {journal} {Intermetallics}\ }\textbf {\bibinfo {volume} {24}},\ \bibinfo
  {pages} {128 } (\bibinfo {year} {2012})}\BibitemShut {NoStop}%
\bibitem [{\citenamefont {Kervan}\ and\ \citenamefont
  {Kervan}(2011)}]{AlessthanB-5}%
  \BibitemOpen
  \bibfield  {author} {\bibinfo {author} {\bibfnamefont {N.}~\bibnamefont
  {Kervan}}\ and\ \bibinfo {author} {\bibfnamefont {S.}~\bibnamefont
  {Kervan}},\ }\href {\doibase dx.doi.org/10.1016/j.jpcs.2011.08.017}
  {\bibfield  {journal} {\bibinfo  {journal} {Journal of Physics and Chemistry
  of Solids}\ }\textbf {\bibinfo {volume} {72}},\ \bibinfo {pages} {1358 }
  (\bibinfo {year} {2011})}\BibitemShut {NoStop}%
\bibitem [{\citenamefont {Klaer}\ \emph {et~al.}(2011)\citenamefont {Klaer},
  \citenamefont {Jenkins}, \citenamefont {Alijani}, \citenamefont {Winterlik},
  \citenamefont {Balke}, \citenamefont {Felser},\ and\ \citenamefont
  {Elmers}}]{Klaer}%
  \BibitemOpen
  \bibfield  {author} {\bibinfo {author} {\bibfnamefont {P.}~\bibnamefont
  {Klaer}}, \bibinfo {author} {\bibfnamefont {C.~A.}\ \bibnamefont {Jenkins}},
  \bibinfo {author} {\bibfnamefont {V.}~\bibnamefont {Alijani}}, \bibinfo
  {author} {\bibfnamefont {J.}~\bibnamefont {Winterlik}}, \bibinfo {author}
  {\bibfnamefont {B.}~\bibnamefont {Balke}}, \bibinfo {author} {\bibfnamefont
  {C.}~\bibnamefont {Felser}}, \ and\ \bibinfo {author} {\bibfnamefont {H.~J.}\
  \bibnamefont {Elmers}},\ }\href {\doibase 10.1063/1.3592802} {\bibfield
  {journal} {\bibinfo  {journal} {Applied Physics Letters}\ }\textbf {\bibinfo
  {volume} {98}} (\bibinfo {year} {2011}),\ 10.1063/1.3592802}\BibitemShut
  {NoStop}%
\bibitem [{\citenamefont {Alijani}\ \emph {et~al.}(2011)\citenamefont
  {Alijani}, \citenamefont {Winterlik}, \citenamefont {Fecher},\ and\
  \citenamefont {Felser}}]{Alijani}%
  \BibitemOpen
  \bibfield  {author} {\bibinfo {author} {\bibfnamefont {V.}~\bibnamefont
  {Alijani}}, \bibinfo {author} {\bibfnamefont {J.}~\bibnamefont {Winterlik}},
  \bibinfo {author} {\bibfnamefont {G.~H.}\ \bibnamefont {Fecher}}, \ and\
  \bibinfo {author} {\bibfnamefont {C.}~\bibnamefont {Felser}},\ }\href
  {\doibase 10.1063/1.3665260} {\bibfield  {journal} {\bibinfo  {journal}
  {Applied Physics Letters}\ }\textbf {\bibinfo {volume} {99}} (\bibinfo {year}
  {2011}),\ 10.1063/1.3665260}\BibitemShut {NoStop}%
\bibitem [{\citenamefont {Meshcheriakova}\ \emph {et~al.}(2014)\citenamefont
  {Meshcheriakova}, \citenamefont {Chadov}, \citenamefont {Nayak},
  \citenamefont {R\"o\ss{}ler}, \citenamefont {K\"ubler}, \citenamefont
  {Andr\'e}, \citenamefont {Tsirlin}, \citenamefont {Kiss}, \citenamefont
  {Hausdorf}, \citenamefont {Kalache}, \citenamefont {Schnelle}, \citenamefont
  {Nicklas},\ and\ \citenamefont {Felser}}]{PhysRevLett.113.087203_Mn2RhSn}%
  \BibitemOpen
  \bibfield  {author} {\bibinfo {author} {\bibfnamefont {O.}~\bibnamefont
  {Meshcheriakova}}, \bibinfo {author} {\bibfnamefont {S.}~\bibnamefont
  {Chadov}}, \bibinfo {author} {\bibfnamefont {A.~K.}\ \bibnamefont {Nayak}},
  \bibinfo {author} {\bibfnamefont {U.~K.}\ \bibnamefont {R\"o\ss{}ler}},
  \bibinfo {author} {\bibfnamefont {J.}~\bibnamefont {K\"ubler}}, \bibinfo
  {author} {\bibfnamefont {G.}~\bibnamefont {Andr\'e}}, \bibinfo {author}
  {\bibfnamefont {A.~A.}\ \bibnamefont {Tsirlin}}, \bibinfo {author}
  {\bibfnamefont {J.}~\bibnamefont {Kiss}}, \bibinfo {author} {\bibfnamefont
  {S.}~\bibnamefont {Hausdorf}}, \bibinfo {author} {\bibfnamefont
  {A.}~\bibnamefont {Kalache}}, \bibinfo {author} {\bibfnamefont
  {W.}~\bibnamefont {Schnelle}}, \bibinfo {author} {\bibfnamefont
  {M.}~\bibnamefont {Nicklas}}, \ and\ \bibinfo {author} {\bibfnamefont
  {C.}~\bibnamefont {Felser}},\ }\href {\doibase
  10.1103/PhysRevLett.113.087203} {\bibfield  {journal} {\bibinfo  {journal}
  {Phys. Rev. Lett.}\ }\textbf {\bibinfo {volume} {113}},\ \bibinfo {pages}
  {087203} (\bibinfo {year} {2014})}\BibitemShut {NoStop}%
\bibitem [{\citenamefont {Zunger}\ \emph {et~al.}(1990)\citenamefont {Zunger},
  \citenamefont {Wei}, \citenamefont {Ferreira},\ and\ \citenamefont
  {Bernard}}]{PhysRevLett.65.353}%
  \BibitemOpen
  \bibfield  {author} {\bibinfo {author} {\bibfnamefont {A.}~\bibnamefont
  {Zunger}}, \bibinfo {author} {\bibfnamefont {S.-H.}\ \bibnamefont {Wei}},
  \bibinfo {author} {\bibfnamefont {L.~G.}\ \bibnamefont {Ferreira}}, \ and\
  \bibinfo {author} {\bibfnamefont {J.~E.}\ \bibnamefont {Bernard}},\ }\href
  {\doibase 10.1103/PhysRevLett.65.353} {\bibfield  {journal} {\bibinfo
  {journal} {Phys. Rev. Lett.}\ }\textbf {\bibinfo {volume} {65}},\ \bibinfo
  {pages} {353} (\bibinfo {year} {1990})}\BibitemShut {NoStop}%
\bibitem [{\citenamefont {Van~de Walle}\ \emph {et~al.}(2013)\citenamefont
  {Van~de Walle}, \citenamefont {Tiwary}, \citenamefont {De~Jong},
  \citenamefont {Olmsted}, \citenamefont {Asta}, \citenamefont {Dick},
  \citenamefont {Shin}, \citenamefont {Wang}, \citenamefont {Chen},\ and\
  \citenamefont {Liu}}]{van2013efficient}%
  \BibitemOpen
  \bibfield  {author} {\bibinfo {author} {\bibfnamefont {A.}~\bibnamefont
  {Van~de Walle}}, \bibinfo {author} {\bibfnamefont {P.}~\bibnamefont
  {Tiwary}}, \bibinfo {author} {\bibfnamefont {M.}~\bibnamefont {De~Jong}},
  \bibinfo {author} {\bibfnamefont {D.}~\bibnamefont {Olmsted}}, \bibinfo
  {author} {\bibfnamefont {M.}~\bibnamefont {Asta}}, \bibinfo {author}
  {\bibfnamefont {A.}~\bibnamefont {Dick}}, \bibinfo {author} {\bibfnamefont
  {D.}~\bibnamefont {Shin}}, \bibinfo {author} {\bibfnamefont {Y.}~\bibnamefont
  {Wang}}, \bibinfo {author} {\bibfnamefont {L.-Q.}\ \bibnamefont {Chen}}, \
  and\ \bibinfo {author} {\bibfnamefont {Z.-K.}\ \bibnamefont {Liu}},\
  }\href@noop {} {\bibfield  {journal} {\bibinfo  {journal} {Calphad}\ }\textbf
  {\bibinfo {volume} {42}},\ \bibinfo {pages} {13} (\bibinfo {year}
  {2013})}\BibitemShut {NoStop}%
\bibitem [{\citenamefont {van~de Walle}(2009)}]{van2009multicomponent}%
  \BibitemOpen
  \bibfield  {author} {\bibinfo {author} {\bibfnamefont {A.}~\bibnamefont
  {van~de Walle}},\ }\href@noop {} {\bibfield  {journal} {\bibinfo  {journal}
  {Calphad}\ }\textbf {\bibinfo {volume} {33}},\ \bibinfo {pages} {266}
  (\bibinfo {year} {2009})}\BibitemShut {NoStop}%
\bibitem [{\citenamefont {R.~Akbarzadeh}\ \emph {et~al.}(2007)\citenamefont
  {R.~Akbarzadeh}, \citenamefont {Ozolins},\ and\ \citenamefont
  {Wolverton}}]{ADMA:ADMA200700843}%
  \BibitemOpen
  \bibfield  {author} {\bibinfo {author} {\bibfnamefont {A.}~\bibnamefont
  {R.~Akbarzadeh}}, \bibinfo {author} {\bibfnamefont {V.}~\bibnamefont
  {Ozolins}}, \ and\ \bibinfo {author} {\bibfnamefont {C.}~\bibnamefont
  {Wolverton}},\ }\href {\doibase 10.1002/adma.200700843} {\bibfield  {journal}
  {\bibinfo  {journal} {Advanced Materials}\ }\textbf {\bibinfo {volume}
  {19}},\ \bibinfo {pages} {3233} (\bibinfo {year} {2007})}\BibitemShut
  {NoStop}%
\bibitem [{\citenamefont {Kirklin}\ \emph {et~al.}(2013)\citenamefont
  {Kirklin}, \citenamefont {Meredig},\ and\ \citenamefont
  {Wolverton}}]{AENM:AENM201200593}%
  \BibitemOpen
  \bibfield  {author} {\bibinfo {author} {\bibfnamefont {S.}~\bibnamefont
  {Kirklin}}, \bibinfo {author} {\bibfnamefont {B.}~\bibnamefont {Meredig}}, \
  and\ \bibinfo {author} {\bibfnamefont {C.}~\bibnamefont {Wolverton}},\ }\href
  {\doibase 10.1002/aenm.201200593} {\bibfield  {journal} {\bibinfo  {journal}
  {Advanced Energy Materials}\ }\textbf {\bibinfo {volume} {3}},\ \bibinfo
  {pages} {252} (\bibinfo {year} {2013})}\BibitemShut {NoStop}%
\bibitem [{Note1()}]{Note1}%
  \BibitemOpen
  \bibinfo {note} {\protect \url {http://oqmd.org/analysis/gclp}}\BibitemShut
  {NoStop}%
\bibitem [{\citenamefont {Yin}\ \emph {et~al.}(2015)\citenamefont {Yin},
  \citenamefont {Nash},\ and\ \citenamefont {Chen}}]{yin2015enthalpies}%
  \BibitemOpen
  \bibfield  {author} {\bibinfo {author} {\bibfnamefont {M.}~\bibnamefont
  {Yin}}, \bibinfo {author} {\bibfnamefont {P.}~\bibnamefont {Nash}}, \ and\
  \bibinfo {author} {\bibfnamefont {S.}~\bibnamefont {Chen}},\ }\href {\doibase
  https://doi.org/10.1016/j.intermet.2014.10.001} {\bibfield  {journal}
  {\bibinfo  {journal} {Intermetallics}\ }\textbf {\bibinfo {volume} {57}},\
  \bibinfo {pages} {34 } (\bibinfo {year} {2015})}\BibitemShut {NoStop}%
\bibitem [{\citenamefont {Kreiner}\ \emph {et~al.}(2014)\citenamefont
  {Kreiner}, \citenamefont {Kalache}, \citenamefont {Hausdorf}, \citenamefont
  {Alijani}, \citenamefont {Qian}, \citenamefont {Shan}, \citenamefont
  {Burkhardt}, \citenamefont {Ouardi},\ and\ \citenamefont
  {Felser}}]{kreiner2014new}%
  \BibitemOpen
  \bibfield  {author} {\bibinfo {author} {\bibfnamefont {G.}~\bibnamefont
  {Kreiner}}, \bibinfo {author} {\bibfnamefont {A.}~\bibnamefont {Kalache}},
  \bibinfo {author} {\bibfnamefont {S.}~\bibnamefont {Hausdorf}}, \bibinfo
  {author} {\bibfnamefont {V.}~\bibnamefont {Alijani}}, \bibinfo {author}
  {\bibfnamefont {J.-F.}\ \bibnamefont {Qian}}, \bibinfo {author}
  {\bibfnamefont {G.}~\bibnamefont {Shan}}, \bibinfo {author} {\bibfnamefont
  {U.}~\bibnamefont {Burkhardt}}, \bibinfo {author} {\bibfnamefont
  {S.}~\bibnamefont {Ouardi}}, \ and\ \bibinfo {author} {\bibfnamefont
  {C.}~\bibnamefont {Felser}},\ }\href {\doibase 10.1002/zaac.201300665}
  {\bibfield  {journal} {\bibinfo  {journal} {Zeitschrift fŸr anorganische und
  allgemeine Chemie}\ }\textbf {\bibinfo {volume} {640}},\ \bibinfo {pages}
  {738} (\bibinfo {year} {2014})}\BibitemShut {NoStop}%
\bibitem [{\citenamefont {Luo}\ \emph {et~al.}(2009)\citenamefont {Luo},
  \citenamefont {Liu}, \citenamefont {Feng}, \citenamefont {Li}, \citenamefont
  {Ma}, \citenamefont {Wu}, \citenamefont {Zhu}, \citenamefont {Jiang},\ and\
  \citenamefont {Xu}}]{luo2009effect}%
  \BibitemOpen
  \bibfield  {author} {\bibinfo {author} {\bibfnamefont {H.}~\bibnamefont
  {Luo}}, \bibinfo {author} {\bibfnamefont {G.}~\bibnamefont {Liu}}, \bibinfo
  {author} {\bibfnamefont {Z.}~\bibnamefont {Feng}}, \bibinfo {author}
  {\bibfnamefont {Y.}~\bibnamefont {Li}}, \bibinfo {author} {\bibfnamefont
  {L.}~\bibnamefont {Ma}}, \bibinfo {author} {\bibfnamefont {G.}~\bibnamefont
  {Wu}}, \bibinfo {author} {\bibfnamefont {X.}~\bibnamefont {Zhu}}, \bibinfo
  {author} {\bibfnamefont {C.}~\bibnamefont {Jiang}}, \ and\ \bibinfo {author}
  {\bibfnamefont {H.}~\bibnamefont {Xu}},\ }\href {\doibase
  https://doi.org/10.1016/j.jmmm.2009.08.002} {\bibfield  {journal} {\bibinfo
  {journal} {Journal of Magnetism and Magnetic Materials}\ }\textbf {\bibinfo
  {volume} {321}},\ \bibinfo {pages} {4063 } (\bibinfo {year}
  {2009})}\BibitemShut {NoStop}%
\bibitem [{\citenamefont {Bakkar}\ and\ \citenamefont
  {Mazumdar}(2017)}]{DipanjanMn2CoGa}%
  \BibitemOpen
  \bibfield  {author} {\bibinfo {author} {\bibfnamefont {S.}~\bibnamefont
  {Bakkar}}\ and\ \bibinfo {author} {\bibfnamefont {D.}~\bibnamefont
  {Mazumdar}},\ }\href@noop {} {} (\bibinfo {year} {2017}),\ \bibinfo {note}
  {unpublished}\BibitemShut {NoStop}%
\bibitem [{\citenamefont {Dai}\ \emph {et~al.}(2006)\citenamefont {Dai},
  \citenamefont {Liu}, \citenamefont {Chen}, \citenamefont {Chen},\ and\
  \citenamefont {Wu}}]{Dai2006533}%
  \BibitemOpen
  \bibfield  {author} {\bibinfo {author} {\bibfnamefont {X.}~\bibnamefont
  {Dai}}, \bibinfo {author} {\bibfnamefont {G.}~\bibnamefont {Liu}}, \bibinfo
  {author} {\bibfnamefont {L.}~\bibnamefont {Chen}}, \bibinfo {author}
  {\bibfnamefont {J.}~\bibnamefont {Chen}}, \ and\ \bibinfo {author}
  {\bibfnamefont {G.}~\bibnamefont {Wu}},\ }\href {\doibase
  dx.doi.org/10.1016/j.ssc.2006.09.030} {\bibfield  {journal} {\bibinfo
  {journal} {Solid State Communications}\ }\textbf {\bibinfo {volume} {140}},\
  \bibinfo {pages} {533 } (\bibinfo {year} {2006})}\BibitemShut {NoStop}%
\bibitem [{\citenamefont {Endo}\ \emph {et~al.}(2012)\citenamefont {Endo},
  \citenamefont {Kanomata}, \citenamefont {Nishihara},\ and\ \citenamefont
  {Ziebeck}}]{endo2012magnetic}%
  \BibitemOpen
  \bibfield  {author} {\bibinfo {author} {\bibfnamefont {K.}~\bibnamefont
  {Endo}}, \bibinfo {author} {\bibfnamefont {T.}~\bibnamefont {Kanomata}},
  \bibinfo {author} {\bibfnamefont {H.}~\bibnamefont {Nishihara}}, \ and\
  \bibinfo {author} {\bibfnamefont {K.}~\bibnamefont {Ziebeck}},\ }\href
  {\doibase 10.1016/j.jallcom.2011.08.090} {\bibfield  {journal} {\bibinfo
  {journal} {Journal of Alloys and Compounds}\ }\textbf {\bibinfo {volume}
  {510}},\ \bibinfo {pages} {1 } (\bibinfo {year} {2012})}\BibitemShut
  {NoStop}%
\bibitem [{\citenamefont {Yang}\ \emph {et~al.}(2015)\citenamefont {Yang},
  \citenamefont {Liu}, \citenamefont {Luo}, \citenamefont {Meng}, \citenamefont
  {Liu}, \citenamefont {Liu}, \citenamefont {Wang},\ and\ \citenamefont
  {Wu}}]{YANG2015247}%
  \BibitemOpen
  \bibfield  {author} {\bibinfo {author} {\bibfnamefont {L.}~\bibnamefont
  {Yang}}, \bibinfo {author} {\bibfnamefont {B.}~\bibnamefont {Liu}}, \bibinfo
  {author} {\bibfnamefont {H.}~\bibnamefont {Luo}}, \bibinfo {author}
  {\bibfnamefont {F.}~\bibnamefont {Meng}}, \bibinfo {author} {\bibfnamefont
  {H.}~\bibnamefont {Liu}}, \bibinfo {author} {\bibfnamefont {E.}~\bibnamefont
  {Liu}}, \bibinfo {author} {\bibfnamefont {W.}~\bibnamefont {Wang}}, \ and\
  \bibinfo {author} {\bibfnamefont {G.}~\bibnamefont {Wu}},\ }\href {\doibase
  http://dx.doi.org/10.1016/j.jmmm.2015.01.081} {\bibfield  {journal} {\bibinfo
   {journal} {Journal of Magnetism and Magnetic Materials}\ }\textbf {\bibinfo
  {volume} {382}},\ \bibinfo {pages} {247 } (\bibinfo {year}
  {2015})}\BibitemShut {NoStop}%
\bibitem [{\citenamefont {Wang}(2008)}]{PhysRevLett.100.156404}%
  \BibitemOpen
  \bibfield  {author} {\bibinfo {author} {\bibfnamefont {X.~L.}\ \bibnamefont
  {Wang}},\ }\href {\doibase 10.1103/PhysRevLett.100.156404} {\bibfield
  {journal} {\bibinfo  {journal} {Phys. Rev. Lett.}\ }\textbf {\bibinfo
  {volume} {100}},\ \bibinfo {pages} {156404} (\bibinfo {year}
  {2008})}\BibitemShut {NoStop}%
\bibitem [{\citenamefont {Butler}\ \emph {et~al.}(2011)\citenamefont {Butler},
  \citenamefont {Mewes}, \citenamefont {Liu},\ and\ \citenamefont
  {Xu}}]{butler2011rational}%
  \BibitemOpen
  \bibfield  {author} {\bibinfo {author} {\bibfnamefont {W.~H.}\ \bibnamefont
  {Butler}}, \bibinfo {author} {\bibfnamefont {C.~K.}\ \bibnamefont {Mewes}},
  \bibinfo {author} {\bibfnamefont {C.}~\bibnamefont {Liu}}, \ and\ \bibinfo
  {author} {\bibfnamefont {T.}~\bibnamefont {Xu}},\ }\href@noop {} {\bibfield
  {journal} {\bibinfo  {journal} {arXiv preprint arXiv:1103.3855}\ } (\bibinfo
  {year} {2011})}\BibitemShut {NoStop}%
\bibitem [{\citenamefont {Nakamichi}\ and\ \citenamefont
  {Itoh}(1975)}]{Mn2VAl}%
  \BibitemOpen
  \bibfield  {author} {\bibinfo {author} {\bibfnamefont {T.}~\bibnamefont
  {Nakamichi}}\ and\ \bibinfo {author} {\bibfnamefont {H.}~\bibnamefont
  {Itoh}},\ }\href {\doibase 10.1143/JPSJ.38.1781} {\bibfield  {journal}
  {\bibinfo  {journal} {Journal of the Physical Society of Japan}\ }\textbf
  {\bibinfo {volume} {38}},\ \bibinfo {pages} {1781} (\bibinfo {year}
  {1975})}\BibitemShut {NoStop}%
\bibitem [{\citenamefont {Buschow}\ and\ \citenamefont {van
  Engen}(1981)}]{Buschow198190}%
  \BibitemOpen
  \bibfield  {author} {\bibinfo {author} {\bibfnamefont {K.}~\bibnamefont
  {Buschow}}\ and\ \bibinfo {author} {\bibfnamefont {P.}~\bibnamefont {van
  Engen}},\ }\href {\doibase dx.doi.org/10.1016/0304-8853(81)90151-7}
  {\bibfield  {journal} {\bibinfo  {journal} {Journal of Magnetism and Magnetic
  Materials}\ }\textbf {\bibinfo {volume} {25}},\ \bibinfo {pages} {90 }
  (\bibinfo {year} {1981})}\BibitemShut {NoStop}%
\bibitem [{\citenamefont {Buschow}\ \emph {et~al.}(1983)\citenamefont
  {Buschow}, \citenamefont {van Engen},\ and\ \citenamefont
  {Jongebreur}}]{Buschow19831}%
  \BibitemOpen
  \bibfield  {author} {\bibinfo {author} {\bibfnamefont {K.}~\bibnamefont
  {Buschow}}, \bibinfo {author} {\bibfnamefont {P.}~\bibnamefont {van Engen}},
  \ and\ \bibinfo {author} {\bibfnamefont {R.}~\bibnamefont {Jongebreur}},\
  }\href {\doibase dx.doi.org/10.1016/0304-8853(83)90097-5} {\bibfield
  {journal} {\bibinfo  {journal} {Journal of Magnetism and Magnetic Materials}\
  }\textbf {\bibinfo {volume} {38}},\ \bibinfo {pages} {1 } (\bibinfo {year}
  {1983})}\BibitemShut {NoStop}%
\bibitem [{\citenamefont {Umetsu}\ \emph {et~al.}(2008)\citenamefont {Umetsu},
  \citenamefont {Kobayashi}, \citenamefont {Fujita}, \citenamefont {Kainuma},\
  and\ \citenamefont {Ishida}}]{Co2MnAl}%
  \BibitemOpen
  \bibfield  {author} {\bibinfo {author} {\bibfnamefont {R.~Y.}\ \bibnamefont
  {Umetsu}}, \bibinfo {author} {\bibfnamefont {K.}~\bibnamefont {Kobayashi}},
  \bibinfo {author} {\bibfnamefont {A.}~\bibnamefont {Fujita}}, \bibinfo
  {author} {\bibfnamefont {R.}~\bibnamefont {Kainuma}}, \ and\ \bibinfo
  {author} {\bibfnamefont {K.}~\bibnamefont {Ishida}},\ }\href {\doibase
  10.1063/1.2836677} {\bibfield  {journal} {\bibinfo  {journal} {Journal of
  Applied Physics}\ }\textbf {\bibinfo {volume} {103}} (\bibinfo {year}
  {2008}),\ 10.1063/1.2836677}\BibitemShut {NoStop}%
\bibitem [{\citenamefont {Fujii}\ \emph {et~al.}(1994)\citenamefont {Fujii},
  \citenamefont {Ishida},\ and\ \citenamefont
  {Asano}}]{doi:10.1143/JPSJ.63.1881}%
  \BibitemOpen
  \bibfield  {author} {\bibinfo {author} {\bibfnamefont {S.}~\bibnamefont
  {Fujii}}, \bibinfo {author} {\bibfnamefont {S.}~\bibnamefont {Ishida}}, \
  and\ \bibinfo {author} {\bibfnamefont {S.}~\bibnamefont {Asano}},\ }\href
  {\doibase 10.1143/JPSJ.63.1881} {\bibfield  {journal} {\bibinfo  {journal}
  {Journal of the Physical Society of Japan}\ }\textbf {\bibinfo {volume}
  {63}},\ \bibinfo {pages} {1881} (\bibinfo {year} {1994})}\BibitemShut
  {NoStop}%
\bibitem [{\citenamefont {Miyamoto}\ \emph {et~al.}(2009)\citenamefont
  {Miyamoto}, \citenamefont {Kimura}, \citenamefont {Miura}, \citenamefont
  {Shirai}, \citenamefont {Ye}, \citenamefont {Cui}, \citenamefont {Shimada},
  \citenamefont {Namatame}, \citenamefont {Taniguchi}, \citenamefont {Takeda},
  \citenamefont {Saitoh}, \citenamefont {Ikenaga}, \citenamefont {Ueda},
  \citenamefont {Kobayashi},\ and\ \citenamefont
  {Kanomata}}]{PhysRevB.79.100405}%
  \BibitemOpen
  \bibfield  {author} {\bibinfo {author} {\bibfnamefont {K.}~\bibnamefont
  {Miyamoto}}, \bibinfo {author} {\bibfnamefont {A.}~\bibnamefont {Kimura}},
  \bibinfo {author} {\bibfnamefont {Y.}~\bibnamefont {Miura}}, \bibinfo
  {author} {\bibfnamefont {M.}~\bibnamefont {Shirai}}, \bibinfo {author}
  {\bibfnamefont {M.}~\bibnamefont {Ye}}, \bibinfo {author} {\bibfnamefont
  {Y.}~\bibnamefont {Cui}}, \bibinfo {author} {\bibfnamefont {K.}~\bibnamefont
  {Shimada}}, \bibinfo {author} {\bibfnamefont {H.}~\bibnamefont {Namatame}},
  \bibinfo {author} {\bibfnamefont {M.}~\bibnamefont {Taniguchi}}, \bibinfo
  {author} {\bibfnamefont {Y.}~\bibnamefont {Takeda}}, \bibinfo {author}
  {\bibfnamefont {Y.}~\bibnamefont {Saitoh}}, \bibinfo {author} {\bibfnamefont
  {E.}~\bibnamefont {Ikenaga}}, \bibinfo {author} {\bibfnamefont
  {S.}~\bibnamefont {Ueda}}, \bibinfo {author} {\bibfnamefont {K.}~\bibnamefont
  {Kobayashi}}, \ and\ \bibinfo {author} {\bibfnamefont {T.}~\bibnamefont
  {Kanomata}},\ }\href {\doibase 10.1103/PhysRevB.79.100405} {\bibfield
  {journal} {\bibinfo  {journal} {Phys. Rev. B}\ }\textbf {\bibinfo {volume}
  {79}},\ \bibinfo {pages} {100405} (\bibinfo {year} {2009})}\BibitemShut
  {NoStop}%
\bibitem [{\citenamefont {Nahid}\ \emph {et~al.}(2009)\citenamefont {Nahid},
  \citenamefont {Oogane}, \citenamefont {Naganuma},\ and\ \citenamefont
  {Ando}}]{1347-4065-48-8R-083002}%
  \BibitemOpen
  \bibfield  {author} {\bibinfo {author} {\bibfnamefont {M.~A.~I.}\
  \bibnamefont {Nahid}}, \bibinfo {author} {\bibfnamefont {M.}~\bibnamefont
  {Oogane}}, \bibinfo {author} {\bibfnamefont {H.}~\bibnamefont {Naganuma}}, \
  and\ \bibinfo {author} {\bibfnamefont {Y.}~\bibnamefont {Ando}},\ }\href
  {http://stacks.iop.org/1347-4065/48/i=8R/a=083002} {\bibfield  {journal}
  {\bibinfo  {journal} {Japanese Journal of Applied Physics}\ }\textbf
  {\bibinfo {volume} {48}},\ \bibinfo {pages} {083002} (\bibinfo {year}
  {2009})}\BibitemShut {NoStop}%
\bibitem [{\citenamefont {Sobczak}(1976)}]{Sobczak1976}%
  \BibitemOpen
  \bibfield  {author} {\bibinfo {author} {\bibfnamefont {R.}~\bibnamefont
  {Sobczak}},\ }\href {\doibase 10.1007/BF00904486} {\bibfield  {journal}
  {\bibinfo  {journal} {Monatshefte f{\"u}r Chemie / Chemical Monthly}\
  }\textbf {\bibinfo {volume} {107}},\ \bibinfo {pages} {977} (\bibinfo {year}
  {1976})}\BibitemShut {NoStop}%
\bibitem [{\citenamefont {Paudel}\ \emph {et~al.}(2009)\citenamefont {Paudel},
  \citenamefont {Wolfe}, \citenamefont {Patton}, \citenamefont {Dubenko},
  \citenamefont {Ali}, \citenamefont {Christodoulides},\ and\ \citenamefont
  {Stadler}}]{Co2MnSb}%
  \BibitemOpen
  \bibfield  {author} {\bibinfo {author} {\bibfnamefont {M.~R.}\ \bibnamefont
  {Paudel}}, \bibinfo {author} {\bibfnamefont {C.~S.}\ \bibnamefont {Wolfe}},
  \bibinfo {author} {\bibfnamefont {H.}~\bibnamefont {Patton}}, \bibinfo
  {author} {\bibfnamefont {I.}~\bibnamefont {Dubenko}}, \bibinfo {author}
  {\bibfnamefont {N.}~\bibnamefont {Ali}}, \bibinfo {author} {\bibfnamefont
  {J.~A.}\ \bibnamefont {Christodoulides}}, \ and\ \bibinfo {author}
  {\bibfnamefont {S.}~\bibnamefont {Stadler}},\ }\href {\doibase
  10.1063/1.3054291} {\bibfield  {journal} {\bibinfo  {journal} {Journal of
  Applied Physics}\ }\textbf {\bibinfo {volume} {105}},\ \bibinfo {eid}
  {013716} (\bibinfo {year} {2009}),\ 10.1063/1.3054291}\BibitemShut {NoStop}%
\bibitem [{\citenamefont {Shreder}\ \emph {et~al.}(2008)\citenamefont
  {Shreder}, \citenamefont {Streltsov}, \citenamefont {Svyazhin}, \citenamefont
  {Makhnev}, \citenamefont {Marchenkov}, \citenamefont {Lukoyanov},\ and\
  \citenamefont {Weber}}]{0953-8984-20-4-045212}%
  \BibitemOpen
  \bibfield  {author} {\bibinfo {author} {\bibfnamefont {E.}~\bibnamefont
  {Shreder}}, \bibinfo {author} {\bibfnamefont {S.~V.}\ \bibnamefont
  {Streltsov}}, \bibinfo {author} {\bibfnamefont {A.}~\bibnamefont {Svyazhin}},
  \bibinfo {author} {\bibfnamefont {A.}~\bibnamefont {Makhnev}}, \bibinfo
  {author} {\bibfnamefont {V.~V.}\ \bibnamefont {Marchenkov}}, \bibinfo
  {author} {\bibfnamefont {A.}~\bibnamefont {Lukoyanov}}, \ and\ \bibinfo
  {author} {\bibfnamefont {H.~W.}\ \bibnamefont {Weber}},\ }\href
  {http://stacks.iop.org/0953-8984/20/i=4/a=045212} {\bibfield  {journal}
  {\bibinfo  {journal} {Journal of Physics: Condensed Matter}\ }\textbf
  {\bibinfo {volume} {20}},\ \bibinfo {pages} {045212} (\bibinfo {year}
  {2008})}\BibitemShut {NoStop}%
\bibitem [{\citenamefont {Carbonari}\ \emph {et~al.}(1996)\citenamefont
  {Carbonari}, \citenamefont {Saxena}, \citenamefont {Jr.}, \citenamefont
  {Filho}, \citenamefont {Attili}, \citenamefont {Olzon-Dionysio},\ and\
  \citenamefont {de~Souza}}]{Carbonari1996313}%
  \BibitemOpen
  \bibfield  {author} {\bibinfo {author} {\bibfnamefont {A.}~\bibnamefont
  {Carbonari}}, \bibinfo {author} {\bibfnamefont {R.}~\bibnamefont {Saxena}},
  \bibinfo {author} {\bibfnamefont {W.~P.}\ \bibnamefont {Jr.}}, \bibinfo
  {author} {\bibfnamefont {J.~M.}\ \bibnamefont {Filho}}, \bibinfo {author}
  {\bibfnamefont {R.}~\bibnamefont {Attili}}, \bibinfo {author} {\bibfnamefont
  {M.}~\bibnamefont {Olzon-Dionysio}}, \ and\ \bibinfo {author} {\bibfnamefont
  {S.}~\bibnamefont {de~Souza}},\ }\href {\doibase
  dx.doi.org/10.1016/S0304-8853(96)00338-1} {\bibfield  {journal} {\bibinfo
  {journal} {Journal of Magnetism and Magnetic Materials}\ }\textbf {\bibinfo
  {volume} {163}},\ \bibinfo {pages} {313 } (\bibinfo {year}
  {1996})}\BibitemShut {NoStop}%
\bibitem [{\citenamefont {Varaprasad}\ \emph {et~al.}(2009)\citenamefont
  {Varaprasad}, \citenamefont {Rajanikanth}, \citenamefont {Takahashi},\ and\
  \citenamefont {Hono}}]{Varaprasad20092702}%
  \BibitemOpen
  \bibfield  {author} {\bibinfo {author} {\bibfnamefont {B.}~\bibnamefont
  {Varaprasad}}, \bibinfo {author} {\bibfnamefont {A.}~\bibnamefont
  {Rajanikanth}}, \bibinfo {author} {\bibfnamefont {Y.}~\bibnamefont
  {Takahashi}}, \ and\ \bibinfo {author} {\bibfnamefont {K.}~\bibnamefont
  {Hono}},\ }\href {\doibase dx.doi.org/10.1016/j.actamat.2009.02.024}
  {\bibfield  {journal} {\bibinfo  {journal} {Acta Materialia}\ }\textbf
  {\bibinfo {volume} {57}},\ \bibinfo {pages} {2702 } (\bibinfo {year}
  {2009})}\BibitemShut {NoStop}%
\end{thebibliography}%

\end{document}